\documentclass[prl,aps,twocolumn,preprintnumbers, showpacs, nofootinbib,superscriptaddress,notitlepage]{revtex4-1}
\usepackage{amssymb,amsthm,amsmath}
\usepackage{graphicx}   
\usepackage{color}      
\usepackage{slashed}    
\usepackage{epsfig}
\usepackage{subfigure}  
\usepackage{diagbox}
\usepackage[normalem]{ulem}

\begin{document}


\title{Renormalization of transverse-momentum-dependent parton distribution\\ on the lattice}

\collaboration{\bf{Lattice Parton Collaboration ($\rm {\bf LPC}$)}}

\vspace{1.0cm}

\author{\includegraphics[scale=0.1]{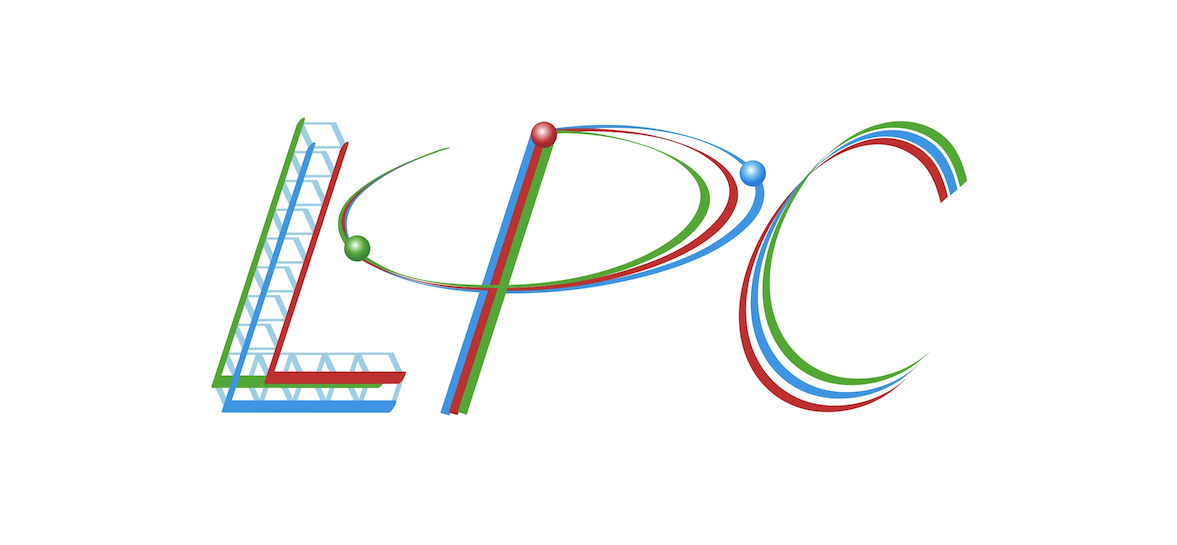}\\
Kuan Zhang}
\affiliation{University of Chinese Academy of Sciences, School of Physical Sciences, Beijing 100049, China}
\affiliation{CAS Key Laboratory of Theoretical Physics, Institute of Theoretical Physics, Chinese Academy of Sciences, Beijing 100190, China}

\author{Xiangdong Ji}
\affiliation{Department of Physics, University of Maryland, College Park, MD 20742, USA}

\author{Yi-Bo Yang}
\email{Corresponding author: ybyang@itp.ac.cn}
\affiliation{University of Chinese Academy of Sciences, School of Physical Sciences, Beijing 100049, China}
\affiliation{CAS Key Laboratory of Theoretical Physics, Institute of Theoretical Physics, Chinese Academy of Sciences, Beijing 100190, China}
\affiliation{School of Fundamental Physics and Mathematical Sciences, Hangzhou Institute for Advanced Study, UCAS, Hangzhou 310024, China}
\affiliation{International Centre for Theoretical Physics Asia-Pacific, Beijing/Hangzhou, China}

\author{Fei Yao}
\affiliation{Center of Advanced Quantum Studies, Department of Physics, Beijing Normal University, Beijing 100875, China}

\author{Jian-Hui Zhang}
\email{Corresponding author: zhangjianhui@bnu.edu.cn}
\affiliation{Center of Advanced Quantum Studies, Department of Physics, Beijing Normal University, Beijing 100875, China}

\date{\today}

\begin{abstract}
To calculate the transverse-momentum-dependent parton distribution functions (TMDPDFs) from lattice QCD, an important goal yet to be realized, it is crucial to establish a viable non-perturbative renormalization approach for linear divergences in the corresponding Euclidean quasi-TMDPDF correlators in large-momentum effective theory. We perform a first systematic study of the renormalization property of the quasi-TMDPDFs 
by calculating the relevant matrix elements in a pion state at 5 lattice spacings ranging from 0.03 fm to 0.12 fm.  We demonstrate that the square root of the Wilson loop combined with the short distance hadron matrix element provides a successful method to remove all ultraviolet divergences of the quasi-TMD operator, and thus provide the necessary justification to perform a continuum limit calculation of TMDPDFs. In contrast, the popular RI/MOM renormalization scheme fails to eliminate all linear divergences.
\end{abstract}

\maketitle

{\it Introduction.} Parton Distribution Functions (PDFs) offer an effective description of quarks and gluons inside a light-travelling hadron~\cite{Ellis:1991qj,Thomas:2001kw}, 
and play an essential role in understanding many processes in high energy and hadron physics. As a natural generalization of the collinear PDFs, the transverse-momentum-dependent (TMD) PDFs also encompass the transverse momentum of partons, and thus provide a useful description of the transverse structure of hadrons. 
They are also crucial inputs for describing multi-scale, noninclusive observables at high-energy colliders such as the LHC~\cite{Constantinou:2020hdm}.
Understanding the TMDPDFs has been an important goal of many experimental facilities around the world, such as COMPASS at CERN, JLab 12 GeV upgrade, RHIC, and in particular, the forthcoming Electron-Ion Collider in the US. Currently, our knowledge of TMDPDFs mainly comes from studies of Drell-Yan and semi-inclusive deep-inelastic scattering processes where the transverse momenta of final state particles are measured. 
In the past, various fittings have been carried out to extract the TMDPDFs from these data (see, e.g., ~\cite{Bacchetta:2017gcc,Scimemi:2017etj,Bertone:2019nxa,Scimemi:2019cmh,Bacchetta:2019sam,Bacchetta:2020gko}). However, calculating the TMDPDFs from first principles has been a challenge, as they are nonperturbative quantities defined in terms of light-cone correlations.

Thanks to the theoretical developments, especially of large-momentum effective theory (LaMET)~\cite{Ji:2013dva,Ji:2020ect}, in the past few years, such calculations have become feasible, but full TMDPDFs from lattice are not yet available. Instead of the standard TMDPDFs involving light-cone Wilson links, LaMET proposes to calculate the quasi-TMDPDF defined by a quark bilinear operator with a staple-shaped Wilson link of finite length along the spacelike direction. The finite link length regulates the so-called pinch-pole singularity associated with infinitely long Wilson lines~\cite{Ji:2020ect}. The singular dependence on such link length is then cancelled by the square root of a Euclidean Wilson loop which, in the mean time, also cancels most of the ultraviolet (UV) divergences (except for the endpoint logarithmic UV divergences which need to be cancelled by other means). The renormalized quasi-TMDPDF can then be factorized into the standard TMDPDF associated with a perturbative hard kernel, a Collins-Soper evolution part and an ``intrinsic soft function", up to power suppressed contributions~\cite{Ji:2019sxk,Ebert:2019okf,Ji:2020ect}. 

Although there have been some preliminary lattice studies of the Collins-Soper evolution kernel and the intrinsic soft function~\cite{LatticeParton:2020uhz,Li:2021wvl,LPC:2022ibr,Shanahan:2020zxr,Shanahan:2021tst,Schlemmer:2021aij}, an important piece of information is still missing in realizing the lattice calculation of TMDPDFs, that is, a systematic study of the nonperturbative renormalization of quasi-TMDPDF operators. This is highly non-trivial, given that both the regularization-independent momentum subtraction (RI/MOM)~\cite{Martinelli:1994ty} and the Wilson line/loop renormalization fail to cancel the power UV divergences in the case of straight-line quasi-PDF operators~\cite{Zhang:2020rsx}. 
It is the purpose of this work to perform such a systematic analysis and to find a viable nonperturbative renormalization approach for the quasi-TMDPDFs, as the hybrid scheme~\cite{Ji:2020brr} with self-renormalization in the case of quasi-PDFs~\cite{Huo:2021rpe}. Indeed, our study shows that the square root of Wilson loop combined with short distance hadron matrix elements is able to eliminate all UV divergences of the quasi-TMDPDF operator and thus ensures a well-defined continuum limit, while the RI/MOM scheme fails in a way similar to that in the quasi-PDF case.

\begin{figure}[!th]
\begin{center}
\includegraphics[width=0.45\textwidth]{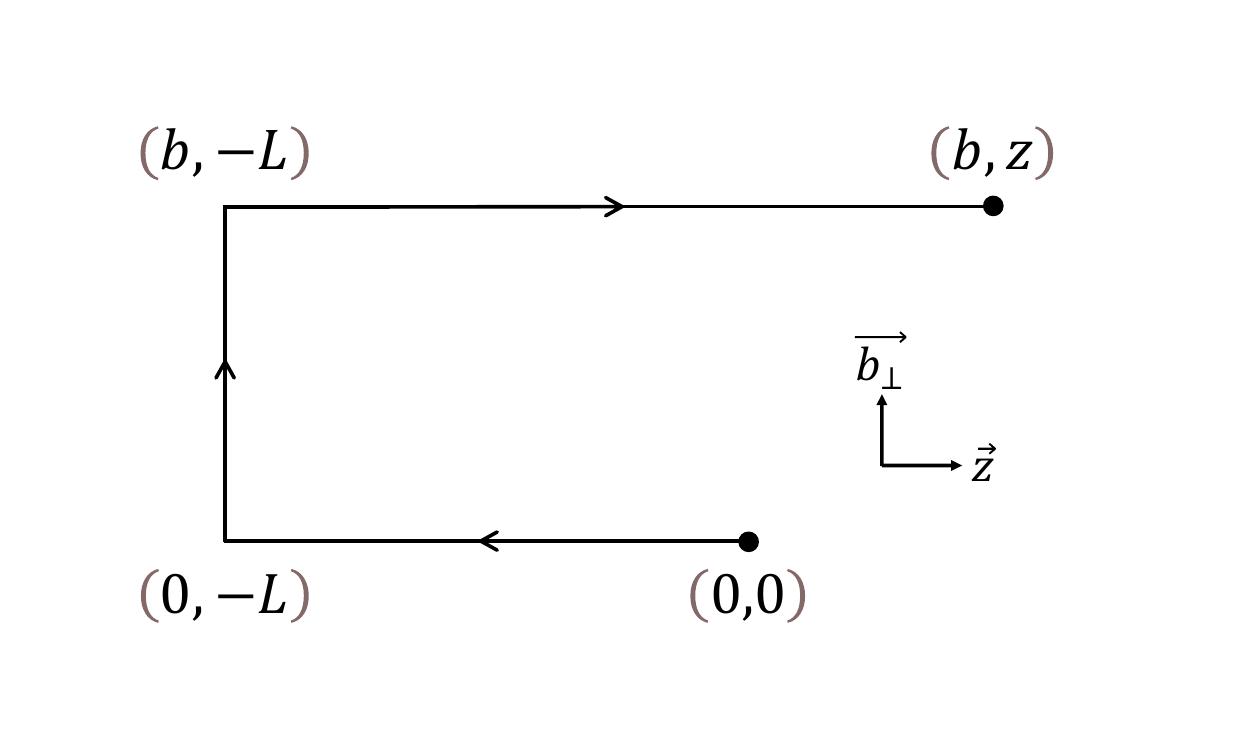}
\caption{Illustration of the Wilson line ${\cal W}(b,z,L)$'s structure needed by the quasi-TMDPDF operator.} \label{fig:staple}
\end{center}
\end{figure}

{\it Theoretical Framework.} In LaMET, the calculation of TMDPDFs starts from the unsubtracted quasi-TMDPDF defined as
\begin{align}
\tilde{h}_{\chi,\gamma_t}(b,z,L,P_z;1/a)&=\langle \chi(P_z)|O_{\gamma_t}(b,z,L)|\chi(P_z)\rangle, \nonumber\\
O_{\Gamma}(b,z,L)&\equiv \bar{\psi}(\vec{0}_{\perp},0) \Gamma {\cal W}(b,z,L) \psi (\vec{b}_{\perp},z),
\end{align}
where we have taken the unpolarized case with $\Gamma=\gamma^t$ as an illustrative example, and assumed a lattice regularization with spacing $a$. $\chi(P_z)$ denotes the hadron state with momentum $P_\mu=(P_0,0,0,P_z)$, the quark bilinear operator $O_\Gamma(b,z,L)$ contains a staple-shaped Wilson link ${\cal W}(b,z,L)$ shown in Fig.~\ref{fig:staple} and defined as
\begin{align}\label{eq:WilsonLink}
{\cal W}(b,z,L)&= {\cal P}{\rm exp} \left[ ig_s\int_{-L}^{z} \textrm{d}s\ \hat{n}_z\cdot  A(b\hat{n}_{\perp}+s\hat{n}_z)\right] \nonumber\\
& \times {\cal P}{\rm exp} \left[ ig_s\int_{0}^{b} \textrm{d}s\ \hat{n}_{\perp}\cdot  A(s\hat{n}_{\perp}-L\hat{n}_z)\right] \nonumber\\
& \times {\cal P}{\rm exp} \left[ ig_s\int_{0}^{-L} \textrm{d}s\ \hat{n}_z\cdot  A(s\hat{n}_z)\right],
\end{align}
where $b=|\vec{b}_{\perp}|$ and $z$ are the separation between the two quark fields along the transverse direction $\hat{n}_{\perp}$ and longitudinal direction $\hat{n}_z$, respectively. $L$ should be large enough to approximate the infinitely long Wilson link in the continuum.

Such a definition suffers from the pinch pole singularity as well as the linear divergence from the Wilson link self-energy, both of which are associated with the length $L$. A more convenient ``subtracted" quasi-TMDPDF is then formed as~\cite{Ji:2019sxk},
\begin{align}\label{eq:quasiTMDcorr}
h_{\chi,\gamma_t}(b,z,P_z;1/a)&=\lim_{L\to\infty}\frac{\tilde{h}_{\chi,\gamma_t}(b,z,L,P_z;1/a)}{\sqrt{Z_E(b,2L+z;1/a)}},
\end{align}
where we have assumed a lattice regularization and included $1/a$ dependence in the matrix elements. $Z_E(b,2L+z)$ is the rectangle Wilson loop with side lengths $b$ and $2L+z$. In $h_{\chi,\gamma_t}(b,z,P_z;1/a)$, the pinch pole singularity, the linear divergences, as well as the cusp divergences are cancelled, but there still remain logarithmic divergences arising from the endpoint of the Wilson links which require further renormalization.

Since the UV divergence is of a multiplicative structure~\cite{Ebert:2019tvc,Green:2020xco} and independent of the external state, the remaining logarithmic divergences of the subtracted quasi-TMDPDF can be removed either by dividing by another subtracted quasi-TMDPDF at zero momentum and short distances $b_0$ and $z_0$, in analogy with the ratio scheme in the quasi-PDF case~\cite{Radyushkin:2017cyf}; or by dividing by an appropriate ratio formed by the quasi-PDF matrix elements at zero momentum and short distances~\cite{Ji:2021uvr}. Here we choose the first option as an illustrative example. In the above renormalization, we work with physical on-shell matrix elements only, and therefore do not suffer from operator mixings due to off-shellness in the RI/MOM scheme. Nevertheless, for non-chiral lattice fermion actions there could be mixings between the quasi-TMDPDF operator with $\Gamma=\gamma^t$ and $\Gamma=\gamma^t\gamma^3$ due to chiral symmetry breaking~\cite{Ji:2021uvr}. However, their numerical impact appears to be negligibly small~\cite{LatticeParton:2020uhz}. Therefore, for the purpose of demonstrating the cancellation of UV divergences, we ignore it here. 

We write the fully renormalized quasi-TMDPDF as
\begin{align}\label{eq:quasiTMDcorrSDR}
h^{{\rm SDR}}_{\chi,\gamma_t}(b,z,P_z;\frac{1}{b_0})=\frac{h_{\chi,\gamma_t}(b,z,P_z;1/a)}{h_{\pi,\gamma_t}(b_0,z_0=0,0,1/a)},
\end{align}
where we have used a superscript $\rm SDR$ to denote the short-distance ratio (SDR) scheme, and chosen the pion matrix element in the denominator (denoted by the subscript $\pi$). For simplicity, we have also chosen $z_0=0$. The singular dependence on $a$ on the r.h.s. of Eq. (\ref{eq:quasiTMDcorrSDR}) has been cancelled, leaving a dependence on the perturbative short scale $b_0$. 
To perform the renormalization, the pion matrix element $h_{\pi,\gamma_t}(b_0,0,0, 1/a)$ needs to be calculated non-perturbatively on the lattice. 
In order to match the renormalized quasi-TMDPDF to the standard TMDPDF, we also need the perturbative results of the above matrix elements, for which we can choose on-shell quark external states. The numerator 
of the r.h.s. of Eq.~(\ref{eq:quasiTMDcorrSDR}) has been calculated previously in dimensional regularization (DR) and $\overline{\rm MS}$ scheme in Ref.~\cite{Ebert:2019okf}, whereas the denominator is independent of $\chi$ and reads~\cite{supplemental} 
\begin{align}\label{eq:hMSbar} 
&h^{\overline{\rm MS}}_{\chi,\gamma_t}(b_0,z_0,0;\mu)=1+\frac{\alpha_s C_F}{2\pi}\bigg\{\frac{1}{2}+3\gamma_E-3\text{ln}2\nonumber\\
&\quad \quad +\frac{3}{2}\text{ln}[\mu^2(b_0^2+z_0^2)]-2\frac{z_0}{b_0}\text{arctan}\frac{z_0}{b_0}\bigg\}+{\cal O}(\alpha_s^2).
\end{align}
With this, we can also convert the SDR result to the $\overline{\rm MS}$ scheme via
\begin{align}\label{eq:quasiTMDcorr1}
\!\! h^{\overline{\rm MS}}_{\chi,\gamma_t}(b,z,P_z;\mu)=h^{\overline{\rm MS}}_{\gamma_t}(b_0,0,0;\mu)h^{{\rm SDR}}_{\chi,\gamma_t}(b,z,P_z;\frac{1}{b_0}),
\end{align}
where $b_0$ dependence cancels. 

Another renormlization option that has been studied in the literature is the RI/MOM renormalization~\cite{Martinelli:1994ty} with the perturbative matching to the $\overline{\text{MS}}$ scheme~\cite{Constantinou:2019vyb,Ebert:2019okf},
\begin{align}\label{eq:quasiTMDcorr_ri}
&h^{\rm MOM}_{\chi,\Gamma}(b,z,P_z;p)=\\
&=\sum_{\Gamma'}[Z^{\rm MOM}(b,z,p;1/a)]^{-1}_{\Gamma\Gamma'} h_{\chi,\Gamma'}(b,z,P_z;1/a)\nonumber\\
&=\lim_{L\to\infty}\sum_{\Gamma'}\tilde{Z}^{\rm MOM}(b,z,L,p;1/a)]^{-1}_{\Gamma\Gamma'} \tilde{h}_{\chi,\Gamma'}(b,z,L,P_z;1/a),\nonumber
\end{align}
\vspace*{-2em}
\begin{align}\label{eq:quasiTMDcorr_MSri}
&h^{\overline{\rm MS}}_{\chi,\gamma_t}(b,z,P_z;\mu)=\\
&\sum_{\Gamma}\mathrm{Tr}[\Gamma h_{q(p),\gamma_t}(b,z,p_z)]^{\overline{\rm MS}}(\mu)
 h^{\rm MOM}_{\chi,\Gamma}(b,z,P_z;p),\nonumber
\end{align}
where $[\tilde{Z}^{\rm MOM}]_{\Gamma\Gamma'}=\mathrm{Tr}[\Gamma'\tilde{h}_{q(p),\Gamma}]$ is the matrix element of ${\cal O}_{\Gamma}$ in the off-shell quark state $q$ with momentum $p$ and projector $\Gamma'$. The $\sqrt{Z_E}$ factor gets cancelled between the non-perturbative renormalization factor $[Z^{\rm MOM}]_{\Gamma\Gamma'}=\mathrm{Tr}[\Gamma'h_{q(p),\Gamma'}]$ and the bare hadron matrix element $h_{\chi}$, but this does not affect the cancellation of singular $L$ dependence between them, thus one can safely take the large-$L$ limit. 
In Eq.~(\ref{eq:quasiTMDcorr_MSri}) the dependence on the four-momentum $p$ on the r.h.s. gets cancelled between the two terms 
up to discretization errors. 

The off-diagonal components of $Z^{\rm MOM}$ can be sizable~\cite{Shanahan:2019zcq} in general, but they turn out to be suppressed with the momentum setup $p_z=p_{\perp}=0$~\cite{LatticeParton:2020uhz}. When the off-diagonal components of $Z^{\rm MOM}$ is negligible, Eq.~(\ref{eq:quasiTMDcorr_ri}) can be simplified into
\begin{align}
h^{\overline{\rm MS}, {\rm MOM}}_{\chi,\gamma_t}(b,z,P_z;\mu)&\simeq \frac{\mathrm{Tr}[\gamma_t h_{q(p),\gamma_t}(b,z,p_z)]^{\overline{\rm MS}}(\mu)}{\mathrm{Tr}[\gamma_th_{q(p),\gamma_t}(b,z,p_z;1/a)]}\nonumber\\
&\times h_{\chi,\gamma_t}(b,z,P_z;1/a).
\end{align}

\begin{table}[htbp]
  \centering
  \begin{tabular}{|c||ccrc|cc|}
\hline
Tag &  $6/g^2$ & $L$ & $T$ & $a(\mathrm{fm})$ & $m^{\textrm{w}}_{q}a$ & $c_{\mathrm{sw}}$ \\
\hline
MILC12 &  3.60 & 24 & 64 & 0.1213(9)    & -0.0695 & 1.0509\\
\hline
MILC09 &  3.78 & 32 & 96 & 0.0882(7)  & -0.0514 & 1.0424 \\
\hline
MILC06 &  4.03 & 48 & 144 & 0.0574(5)  & -0.0398 & 1.0349 \\
\hline
MILC04 &  4.20 & 64 & 192 & 0.0425(5)  & -0.0365 & 1.0314 \\
\hline
MILC03 &  4.37 & 96 & 288 & 0.0318(5)  & -0.0333 & 1.0287 \\
\hline
\end{tabular}
  \caption{Setup of the ensembles, including the bare coupling constant $g$, lattice size $L^3\times T$ and lattice spacing $a$. $m^\textrm{w}_q$ is the bare quark masses. The pion masses in the sea are $\sim$310 MeV and the valence pion mass are in the range of 280-320 MeV.}
  \label{tab:milc}
\end{table}

{\it Simulation setup.} In this work, we use the clover valence quark on the 2+1+1 flavors (degenerate up and down, strange, and charm degrees of freedom) of highly improved staggered quarks (HISQ) and one-loop Symanzik improved~\cite{Hart:2008sq} gauge ensembles from the MILC Collaboration~\cite{Bazavov:2012xda} at five lattice spacings. We tuned the valence light quark mass to be around that in the sea, and the information about the ensembles and parameters we used are listed in Table~\ref{tab:milc}. 

Since the UV divergence shall be independent of the hadron state, as shown in the quasi-PDF case~\cite{Zhang:2020rsx,Huo:2021rpe}, in the following we will concentrate on the pion matrix element $\tilde{h}_{\pi}(b,z,L)\equiv\langle \pi|O_{\gamma_t}(b,z,L)|\pi\rangle$ of the quasi-TMDPDF operator $O_{\gamma_t}(b,z,L)$ as an illustrative example, which can be extracted from the following ratio
\begin{align}\label{eq:ratio}
&R_{\pi}(t_1,b,z,L;a,t_2)\nonumber\\
&\equiv \frac{\langle O_\pi(t_2)\sum_{\vec{x}}O_{\Gamma}(b,z,L;(\vec{x},t_1))O_{\pi}^{\dagger}(0)\rangle}{\langle O_\pi(t_2)O_{\pi}^{\dagger}(0)\rangle}
\nonumber\\
&=\langle \pi|O_{\Gamma}(b,z,L)|\pi\rangle+{\cal O}(e^{-\Delta m t_1},e^{-\Delta m (t_2-t_1)},e^{-\Delta m t_2}),
\end{align}
where $\Delta m$ is the mass gap between the pion and its first excited state which is aournd 1 GeV. 1-step HYP smearing is applied on the staple-shaped link to enhance the signal to noise ratio (SNR), and our previous studies~\cite{Zhang:2020rsx,Huo:2021rpe} on the quasi-PDF operator suggest that such a smearing will not change the renormalization behavior of the results.  The source/sink setup and also the ground state matrix element extraction are similar to that in Refs.~\cite{Zhang:2020rsx,Huo:2021rpe}. 

The square root of the Wilson loop $Z_E(b, 2L+z)$ is expected to cancel the linear divergence in $h_{\pi}$, but its SNR can be very poor at large $b$ and/or $2L+z$. The estimate of $Z_E(b, 2L+z)$ with finite statistics can even lead to a negative central value and make $\sqrt{Z_E(b, 2L+z)}$ ill-defined. But since $Z_E(b, 2L+z)_{\overrightarrow{L\rightarrow \infty}} C(b) e^{-V(b)(2L+z)}$ where $V(b)$ is the QCD static potential, we can fit $Z_E(b, 2L+z)$ by a single exponential term at large $L$ and replace it with the fit result. Such a replacement is essential to access the large $b$ and $z$ region at small lattice spacing. 

For the RI/MOM renormalization, we choose the to-and-fro and transverse gauge links to be along the z and x directions, respectively, and set the external momentum to be $(p_x,p_y,p_z,p_t)=\frac{2\pi}{{L}}(0,5,0,5)$ on all the five ensembles. Such a setup avoids the momentum along the link direction, and thus there is no imaginary part in the quark matrix element projected with $\gamma_t$. $|p|\simeq3$~GeV is, on the one hand, large enough to suppress the IR effect, and on the one hand, not too huge to yield obvious discretization errors at the largest lattice spacing. 

The details of the Wilson loop extrapolation, and $L$ dependence of $h_{\pi,\gamma_t}$ and $h^{\rm MOM}_{\pi,\gamma_t}$ can also be found in the supplemental materials~\cite{supplemental}.

\begin{figure}[!th]
\begin{center}
\includegraphics[width=0.45\textwidth]{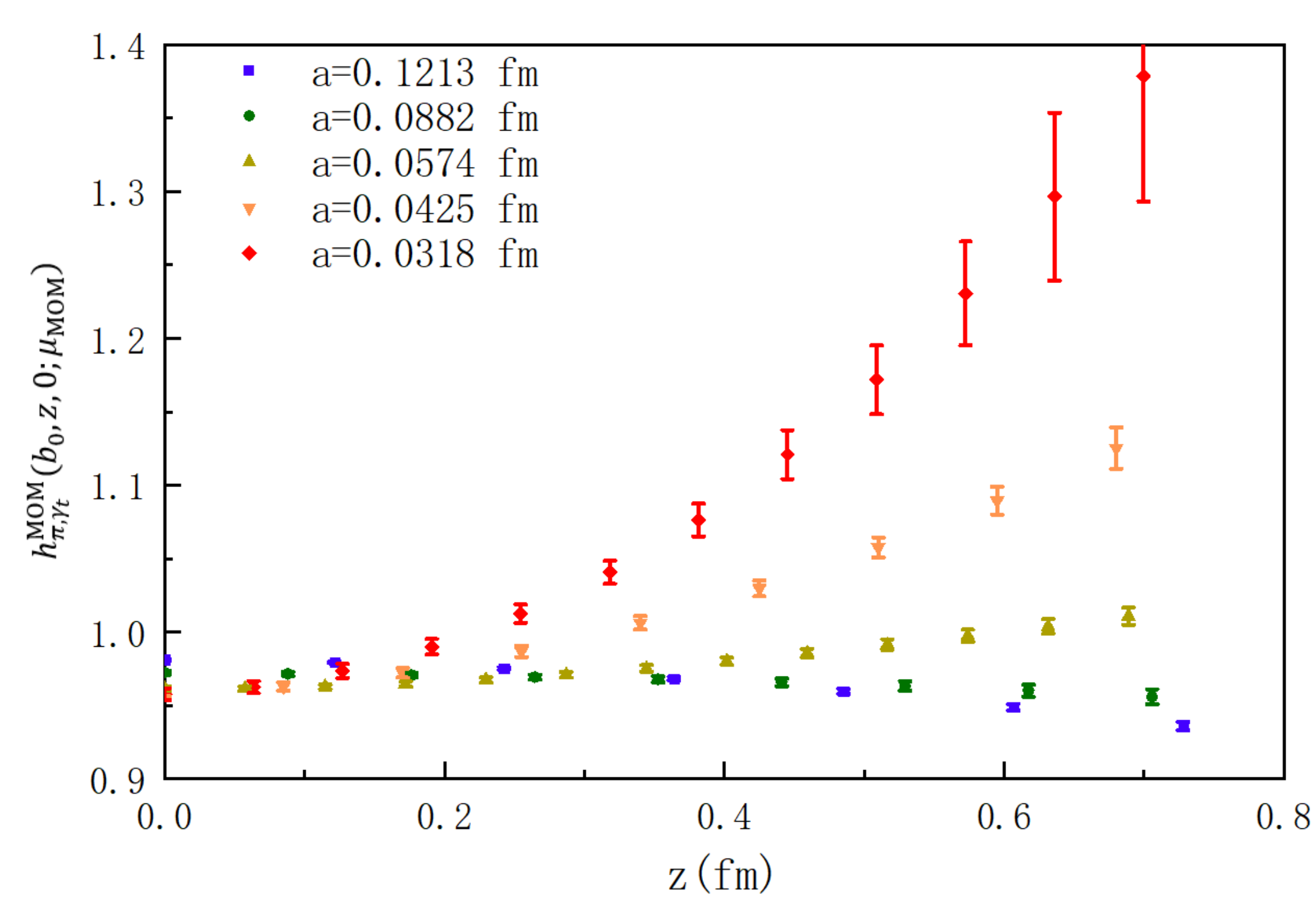}
\caption{The RI/MOM renormalized matrix elements $h^{\rm MOM}_{\pi,\gamma_t}(b_0,z,0;\mu_{\rm MOM})$ defined in Eq.~(\ref{eq:quasiTMDcorr_ri}) using the clover fermion action, with $b_0=0.12~\text{fm}$ and $\mu_{\rm MOM}\simeq 3~\mathrm{GeV}$. The statistical uncertainty comes from bootstrap re-sampling. We use the cubic spline algorithm to interpolate $b$ to the same value for different lattice spacings.}\label{fig:rimom}
\end{center}
\end{figure}

{\it Numerical Results.} With a constant fit in the range $L\ge 0.36~\mathrm{fm}$, one can extract the RI/MOM renormalized matrix element $h^{\rm MOM}_{\pi,\gamma_t}(b,z,0;\mu_{\rm MOM}\simeq 3~\mathrm{GeV})$. Thanks to the cancellation between the $L$ dependence of $\tilde{h}_{\pi}$ and $\tilde{Z}^{\rm MOM}$, the $L$ needed to reach saturation is much smaller than that needed for $h_{\pi}$ defined in Eq.~(\ref{eq:quasiTMDcorr}). We interpolate the results to $b=0.12$~fm with the cubic spline algorithm, and plot the results at different lattice spacings in Fig.~\ref{fig:rimom}. It is clear that the lattice spacing dependence becomes stronger with either larger $z$ or smaller $a$, implying that there are residual linear divergences in $h^{\rm MOM}_{\pi,\gamma_t}$. In other words, the RI/MOM renormalization doesn not cancel all the linear divergences. As suggested by the quasi-PDF study~\cite{Zhang:2020rsx}, the residual linear divergence in $h^{\rm MOM}_{\pi,\gamma_t}$ does not appear at the 1-loop level, and shall be gauge- and action-dependent occurring at higher orders. Since the matching between the RI/MOM and $\overline{\rm MS}$ scheme is calculated under DR in which the linear divergence is absent, 
we expect $h^{\overline{\rm MS}, {\rm MOM}}_{\pi,\gamma_t}$ to inherit the residual linear divergence issue and do not consider it any more in the discussion to follow.

\begin{figure}[!th]
\begin{center}
\includegraphics[width=0.45\textwidth]{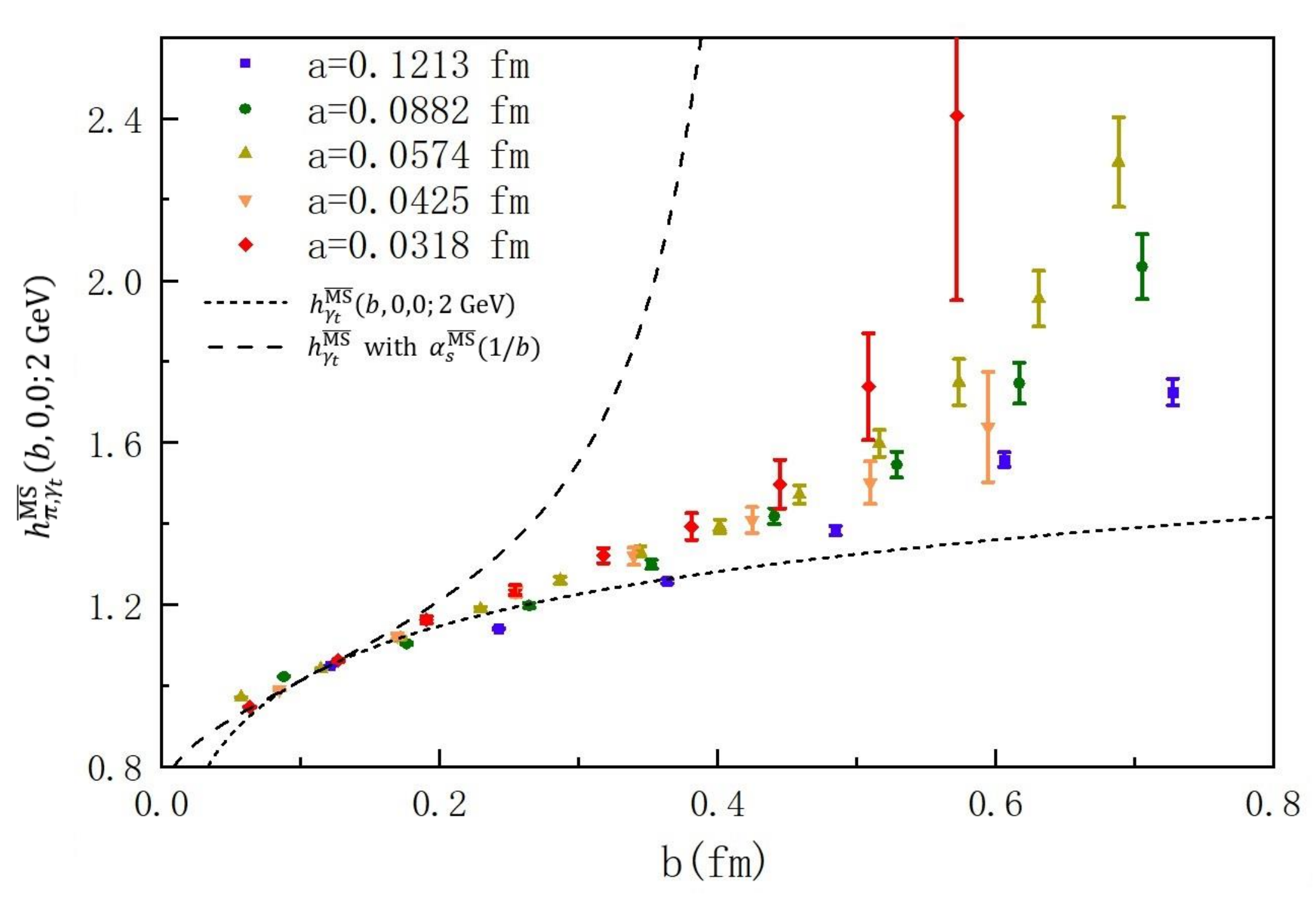}
\caption{The renormalized matrix elements $h^{\overline{\rm MS}}_{\pi,\gamma_t}(b,0,0;2~{\rm GeV})$ defined in Eq.~(\ref{eq:quasiTMDcorr1}) , and the statistical uncertainty comes from bootstrap re-sampling. The dense dashed line is the 1-loop result with $\alpha_s(2\text{ GeV})$ in $\overline{\rm MS}$ scheme. We also show a sparse dashed line for the perturbative result with $\alpha_s(1/b)$ for comparison.}\label{fig:ba_z}
\end{center}
\end{figure}

We also repeat the above calculations with the overlap fermion action, and find that the pion matrix elements are independent of the fermion action within statistical uncertainties, while the quark matrix elements needed by the RI/MOM renormalization can be very sensitive to the fermion action. It is similar to what we observed for the quasi-PDF with straight Wilson link~\cite{Zhang:2020rsx}. Thus, in the main text above we only concentrate on the clover fermion action, and leave the comparison between different fermion actions to the supplemental material~\cite{supplemental}. The discussion on the lattice spacing dependence of the off-diagonal components of $Z^{\rm MOM}$ can also be found there.

\begin{figure}[!th]
\begin{center}
\includegraphics[width=0.45\textwidth]{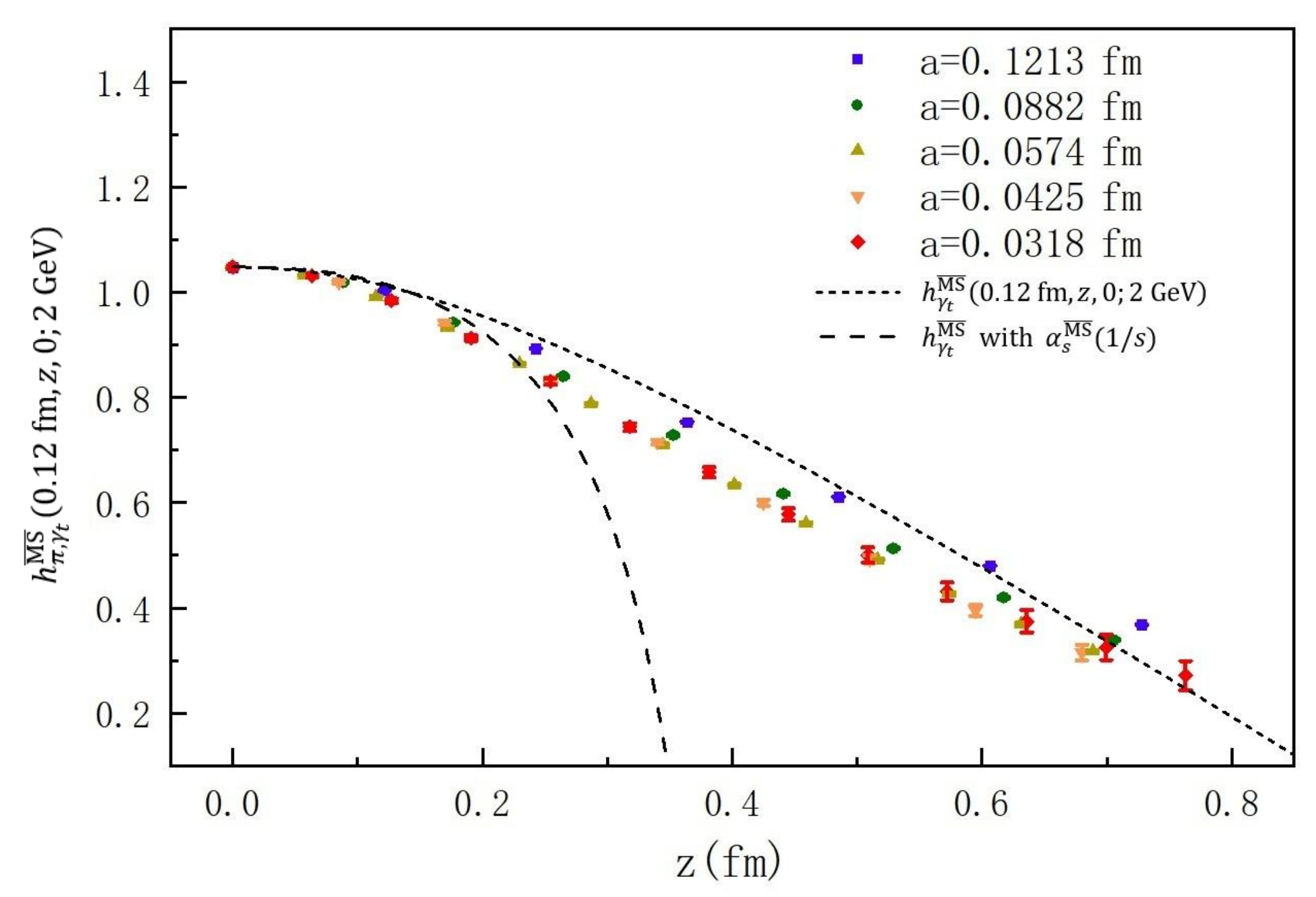}
\includegraphics[width=0.45\textwidth]{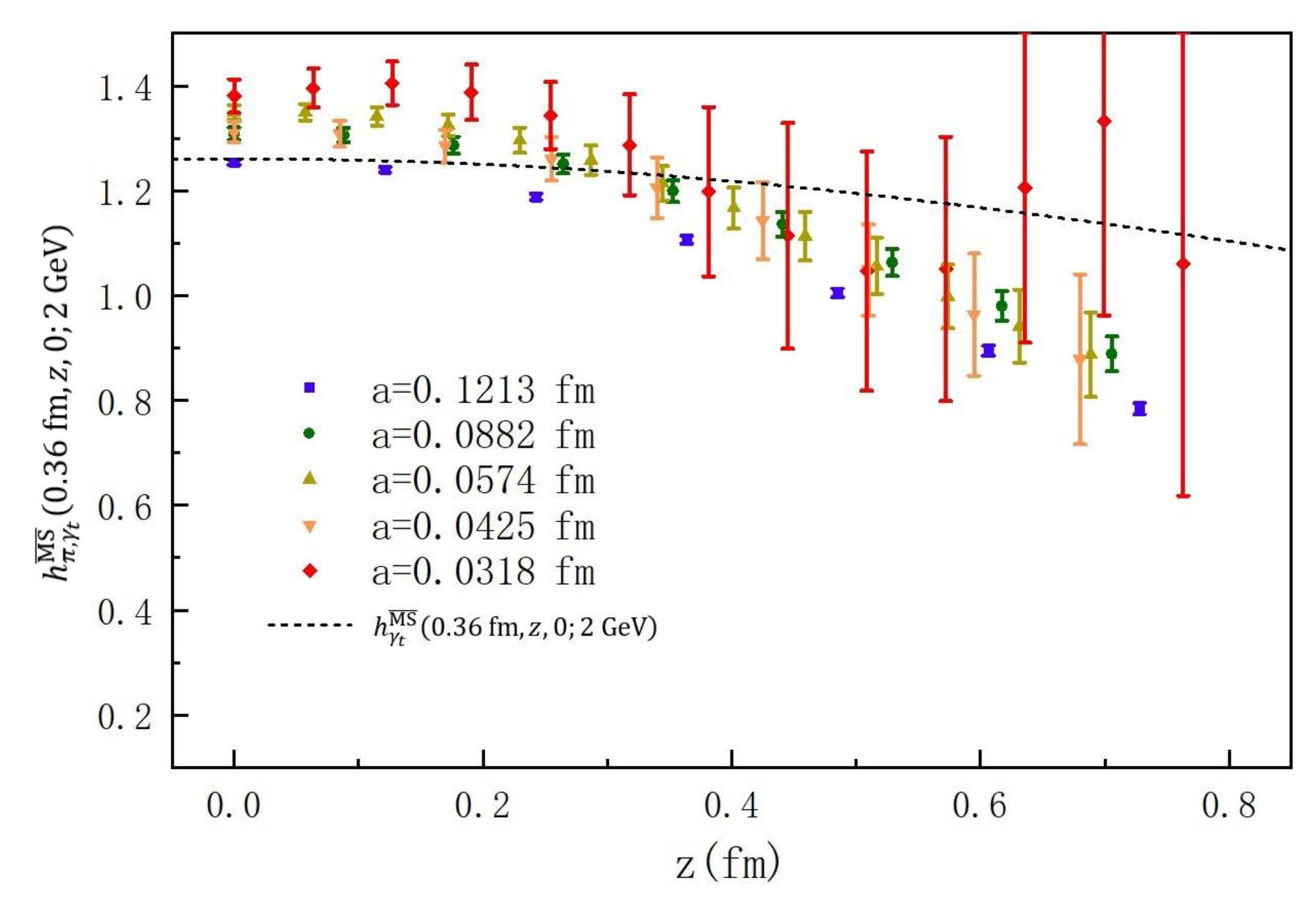}
\caption{The renormalized matrix elements $h^{\overline{\rm MS}}_{\pi,\gamma_t}(b,z,0;2\text{ GeV})$ defined in Eq.~(\ref{eq:quasiTMDcorr1}). The black line is the result in 1-loop level. The upper figure is calculated with $b$=0.12fm, and the lower figure is calculated with $b$=0.36fm. Note that $b$ has been interpolated to the same value but $z$ are kept original. The statistical uncertainty comes from bootstrap re-sampling. The dense dashed line is the 1-loop result with $\alpha_s(2\text{ GeV})$ in $\overline{\rm MS}$ scheme. We also show a sparse dashed line for the perturbative result with $\alpha_s(1/s)$, $s=\sqrt{b^2+z^2}$ for comparison.}\label{fig:renorm_z}
\end{center}
\end{figure}

Now we consider the SDR scheme. As mentioned earlier, the subtracted quasi-TMDPDF matrix element contains 
logarithmic UV divergences, which shall be canceled by the short $b$ and $z$ matrix elements at zero momentum. This is illustrated in Fig.~\ref{fig:ba_z} for the choice $z=0$, where we have converted the SDR result to the $\overline{\rm MS}$ scheme using Eq.~(\ref{eq:quasiTMDcorr1}). 
As shown in the figure, the results at different lattice spacings exhibit a convergence behavior within errors. Moreover, there is an agreement with the 
perturbative $\overline{\rm MS}$ 1-loop result (dense dashed line) in the range of $b<0.4$~fm with the $\overline{\rm MS}$ scale 2 GeV. We also show sparse dashed lines for the 1-loop results with $\alpha_s(1/b)$ for comparison.

In Fig.~\ref{fig:renorm_z}, we show the renormalized $h^{\overline{\rm MS}}_{\pi,\gamma_t}$ at $b=0.12$~fm and 0.36~fm as a function of $z$. The same figures for other values of $b$ up to 0.72 fm, {and for the case of the TMD wave function matrix element} can be found in the supplemental material~\cite{supplemental}. In contrast with the $h^{\rm MOM}_{\pi,\gamma_t}$ shown in Fig.~\ref{fig:rimom}, the $h^{\overline{\rm MS}}_{\pi,\gamma_t}$ shows good convergence in the continuum limit, regardless of the value of $z$ and $b$, and agrees with the perturbative value well when $b$ is small. {Thus the renormalized TMD-PDF matrix element using the SDR scheme can actually eliminate all the UV divergence and be insensitive to the subtraction point $b_0$ (or $z_0$), and then can be used for a state-of-the-arts TMDPDF calculation on the lattice.}

{\it Summary and Outlook.} In this work, we study systematically the renormalization property of the quasi-TMDPDFs. By calculating the pion matrix elements in the rest frame at five lattice spacings and applying different renormalizations, we find that the RI/MOM renormalized matrix element has obvious residual linear divergence. In contrast, the square root of the rectangular Wilson loop can eliminate all the linear divergence in the hadron matrix element of the quasi-TMDPDF operator, and the remaining logarithmic divergence can be removed by forming the ratio with a subtracted quasi-TMDPDF matrix element at zero momentum and short $b$ and $z$, thus leading to a well-defined a continuum limit. 

In summary, this work provides a crucial test and establishes a viable solution for the renormalization of the quasi-TMD hadron matrix element on the lattice, ensuring the existence of a reliable continuum extrapolation. 
It paves the way towards the non-perturbative prediction of both TMDPDF and TMD wave functions on the lattice. 

\section*{Acknowledgment}

We thank the MILC collaboration for providing us with their gauge configurations, $\chi$QCD collaboration for sharing the overlap fermion propagators on those ensembles, and Andreas Sch\"afer, Wei Wang, and Yu-Shan Su for useful information and discussions. The calculations were performed using the Chroma software suite~\cite{Edwards:2004sx} with QUDA~\cite{Clark:2009wm,Babich:2011np,Clark:2016rdz} through HIP programming model~\cite{Bi:2020wpt}.
The numerical simulations were carried out on the SunRising-1 computing platform, and supported by Strategic Priority Research Program of Chinese Academy of Sciences, Grant No. XDC01040100, and HPC Cluster of ITP-CAS.
Y. Yang is supported by Strategic Priority Research Program of Chinese Academy of Sciences, Grant No. XDC01040100, XDB34030303, XDPB1. F. Yao and J.-H. Zhang are supported by the National Natural Science Foundation of China under grants No. 11975051. Y. Yang and J.-H. Zhang are also supported by a NSFC-DFG joint grant under grant No. 12061131006 and SCHA 458/22.

\bibliography{ref}

\clearpage

\begin{widetext}
\section*{Supplemental Materials}

\subsection{Perturbative calculation}

For perturbative calculations, we can replace the external state of the matrix element with an on-shell quark state. 
We thus have
\begin{align}\label{eq:beam}
\tilde{h}(b,z,p_z) = \frac{1}{2}\sum_{spins}\langle q(p)|\Bar{q}(0) {\cal W}(b,z,L) \gamma^t q(b,z)|q(p)\rangle
\end{align}.

For the renormalization and matching in the SDR scheme, we need to calculate the above matrix element at zero momentum. This can be done in analogy with the calculation in Ref.~\cite{Ebert:2019tvc}. 
The staple-shaped link consists of three straight links, and they can be parameterized as follows,

 \begin{align}\label{eq:link} 
 &\eta_1(s) = -sL\hat{z}, \notag \\
 &\eta_2(s) = sb\hat{x} - L\hat{z}, \notag \\
 &\eta_3(s) = b\hat{x} + [-L+s(L+z)]\hat{z}. 
 \end{align}

\begin{figure}[!th]
\begin{center}
\includegraphics[width=0.7\textwidth]{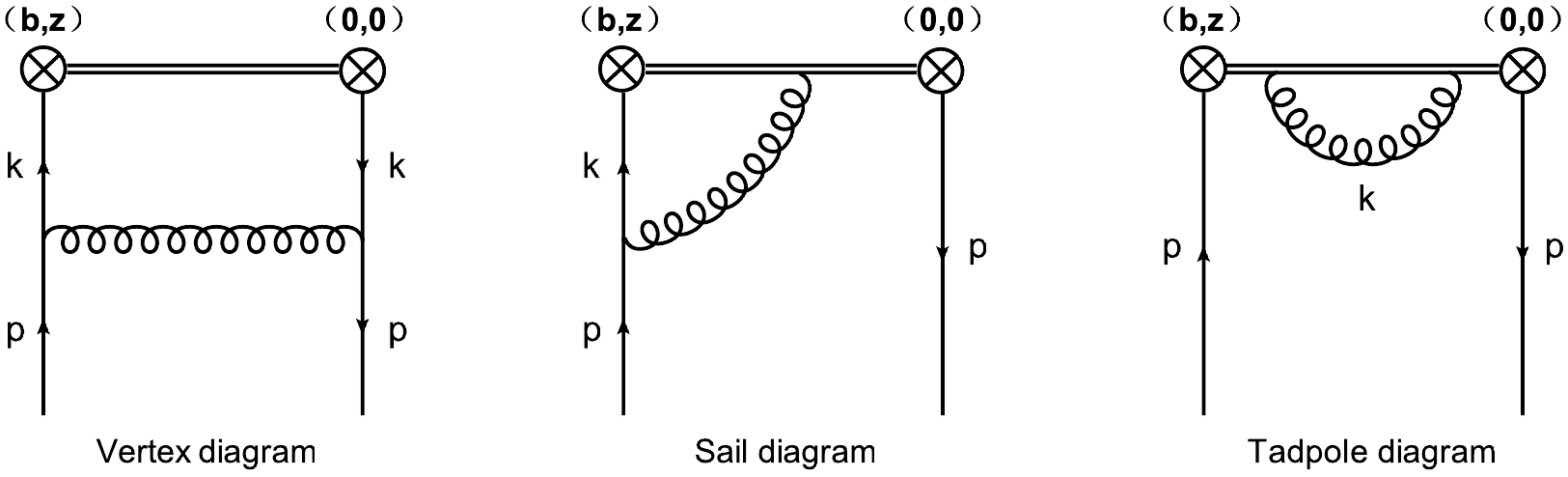}
\caption{One-loop diagrams contributing to the quasi-TMDPDF. The conjugate diagram to the second one is not shown.}\label{fig:1-loop}
\end{center}
\end{figure}

The calculation is performed in Feynman gauge with dimensional regularization $(d=4-2\epsilon)$ to regularize both the UV and IR divergences. As a result, the quark self energy diagram vanishes. 
The remaining contributing diagrams are depicted in Fig.~\ref{fig:1-loop}. they give
 
\begin{align}\label{eq:vertex} 
\tilde{h}_{\text{vertex}} &=  -i g^2C_f \frac{\mu^{2\epsilon}}{2 p_0}\frac{1}{2}\sum_s \Bar{u}(p) \int\frac{d^dk}{(2\pi)^d}\frac{\gamma^{\mu}\slashed{k}\gamma_0\slashed{k}\gamma_{\mu}}{k^4(p-k)^2}e^{-ik\cdot B}u(p), \notag \\
\tilde{h}_{sail}& = -g^2C_f\frac{\mu^{2\epsilon}}{2 p_0} \frac{1}{2}\sum_s \Bar{u}(p)\int_0^1ds\eta^{\prime}_{\mu}(s)\int\frac{d^dk}{(2\pi)^d}\frac{\gamma_0\slashed{k}\gamma^{\mu}}{k^4(p-k)^2}e^{-ip\cdot B-i(p-k)\cdot \eta(s)} u(p) \nonumber\\
&-g^2C_f\frac{\mu^{2\epsilon}}{2 p_0} \frac{1}{2}\sum_s \Bar{u}(p)\int_0^1ds\eta^{\prime}_{\mu}(s)\int\frac{d^dk}{(2\pi)^d}\frac{\gamma^{\mu}\slashed{k}\gamma_0}{k^4(p-k)^2}e^{-ik\cdot B+i(p-k)\cdot \eta(s)}u(p), \notag \\
\tilde{h}_{\text{tadpole}}
&= ig^2C_f \frac{\mu^{2\epsilon}}{2 p_0} \frac{1}{2}\sum_s \Bar{u}(p) \int_0^1ds \eta^{\prime}_{\mu}(s)\int_0^sdt \eta^{\prime}_{\nu}(t)\int\frac{d^dk}{(2\pi)^d}\frac{g^{\mu\nu}}{k^2}e^{ik \cdot [\eta(s)-\eta(t)]}\gamma_0 u(p),
\end{align}
where $B = b\hat{x} + z\hat{z}$ is the total displacement of the staple-shaped link and the external quark is massless. 
At zero momentum, each diagram gives
\begin{align}\label{eq:vertex5}
\tilde{h}_{\text{vertex}}
=& \frac{\alpha_sC_f}{4\pi}(-\frac{1}{\epsilon}-\text{ln}[\mu^2(b^2+z^2)]+1-\gamma_E-\text{ln}\pi), \notag\\
\tilde{h}_{\text{sail}}
=& \frac{\alpha_sC_f}{2\pi}(\frac{1}{\epsilon}+\text{ln}[\mu^2(b^2+z^2)]+\gamma_E+\text{ln}\pi), \notag\\
\tilde{h}_{\text{tadpole}} 
=& \frac{\alpha_sC_f}{2\pi}\big\{3(\frac{1}{\epsilon}+\text{ln}\pi+\gamma_E+2)+\text{ln}\mu^2L^2+\text{ln}\mu^2b^2 + \text{ln}\mu^2(L+z)^2 +2\frac{L}{b}\text{arctan}\frac{L}{b} \notag\\
& \quad \quad +2\frac{L+z}{b}\text{arctan}\frac{L+z}{b}-2\frac{z}{b}\text{arctan}\frac{z}{b}-\text{ln}\frac{(b^2+L^2)[b^2+(L+z)^2]}{b^2(b^2+z^2)}\big\}.
\end{align}
Add up all the contributions and take the limit $L\gg b,z$, we have the following result 
\begin{align}\label{eq:beam1}
\tilde{h}^{(1)} =& \tilde{h}_{\text{vertex}}+\tilde{h}_{\text{sail}}+\tilde{h}_{\text{tadpole}} \notag\\
=&\frac{\alpha_sC_f}{2\pi}\bigg\{\frac{7}{2}(\frac{1}{\epsilon}+\text{ln}\pi+\gamma_E)+\frac{5}{2}+\frac{3}{2}\text{ln}[\mu^2(b^2+z^2)]+2\text{ln}\mu^2b^2+\pi\frac{2L+z}{b} \notag\\
&\quad \quad -2\frac{z}{b}\text{arctan}\frac{z}{b}\bigg\}+{\cal O}(\frac{z^2}{L^2},\frac{b^2}{L^2}).
\end{align}

We then calculate the Wilson loop $Z_E(b,2L+z)$ at 1-loop level, and obtain a result that is consistent with the naive quasi soft function in Ref.~\cite{Ebert:2019tvc} with the substitution $2L$ to $2L+z$. Taking the limit $L\gg b,z$ gives
\begin{align}\label{eq:WilsonLoop01}
Z^{(1)}_E 
=& \frac{\alpha_sC_f}{\pi}\bigg\{2(\frac{1}{\epsilon}+\gamma_E+\text{ln}\pi)+4+2\text{ln}\mu^2b^2+2\frac{b}{2L+z}\text{arctan}\frac{b}{2L+z}\notag\\
&\quad \quad+2\frac{2L+z}{b}\text{arctan}\frac{2L+z}{b}+2\text{ln}\frac{(2L+z)^2}{b^2+(2L+z)^2}\bigg\},\notag\\
=&\frac{\alpha_sC_f}{\pi}\bigg\{2(\frac{1}{\epsilon}+\gamma_E+\text{ln}\pi)+2+2\text{ln}\mu^2b^2+\pi\frac{2L+z}{b}\bigg\}+{\cal O}(\frac{z^2}{L^2},\frac{b^2}{L^2}).
\end{align}
Therefore, the subtracted quasi-TMDPDF at zero momentum and short distances is

\begin{align}\label{eq:perturbativeResult}
h_{\pi,\gamma_t}(b,z,0)&\equiv \frac{\tilde{h}_{\pi,\gamma_t}(b,z,L,0)}{\sqrt{Z_E(b,2L+z)}}=1+\tilde{h}^{(1)} - \frac{1}{2}Z^{(1)}_E + {\cal O}(\alpha_s^2) \nonumber\\
&= 1+\frac{\alpha_sC_f}{2\pi}\bigg\{\frac{3}{2}(\frac{1}{\epsilon}+\gamma_E+\text{ln}\pi)+\frac{1}{2}+\frac{3}{2}\text{ln}[\mu^2(b^2+z^2)]-2\frac{z}{b}\text{arctan}\frac{z}{b}\bigg\} + {\cal O}(\alpha_s^2).
\end{align}

\clearpage

\subsection{Wilson loop extrapolation and $L$ dependence}

\subsubsection{Extrapolation of Wilson loop}

$Z_E(b,r;a)$ is the vacuum expectation value of a rectangular spacelike Wilson loop of different lattice spacings with size $b \times r$, $r\equiv2L+z$. And we need to take the square root of $Z_E(b,r;a)$ to cancel out the transverse link, pinch-pole singularity and linear divergence in $\tilde{h}$ defined in Eq.~\ref{eq:quasiTMDcorr}. 

\begin{figure}[!th]
\begin{center}
\includegraphics[width=0.45\textwidth]{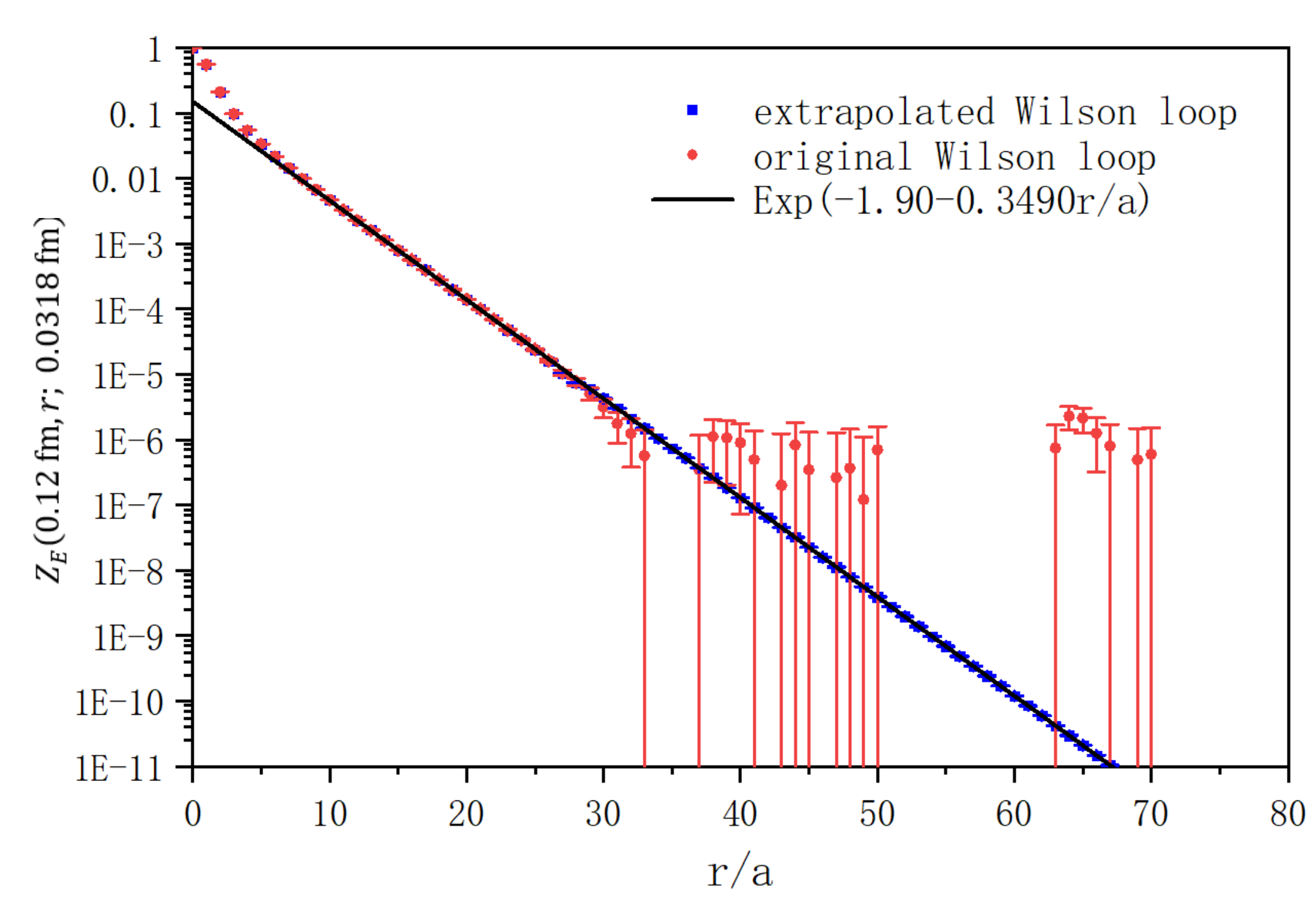}
\caption{The red dots are the original Wilson loop $Z_E(12a,r;0.0318\text{ fm})$, the blue dots are the Wilson loop after extrapolation and the black line is the fitting results.}\label{fig:extra}
\end{center}
\end{figure}

But the signal to noise ratio (SNR) of $Z_E(b,r;a)$ can be very poor at large $b$ and/or $r$ as shown in Fig.~\ref{fig:extra}. The central value of $Z_E(b, r)$ with finite statistics can even give a negative value and make $\sqrt{Z_E(b, r)}$ to be ill-defined. But since $Z_E(b, r)_{\overrightarrow{r\rightarrow \infty}} C(b)e^{-V(b)r}$ where $V(b)$ is the QCD static potential, we can fit $Z_E(b, r; a)$ by $e^{c(b;a)-V(b;a)r}$ at large $r$ and replace it with the fit result.

\begin{figure}[!th]
\begin{center}
\includegraphics[width=0.45\textwidth]{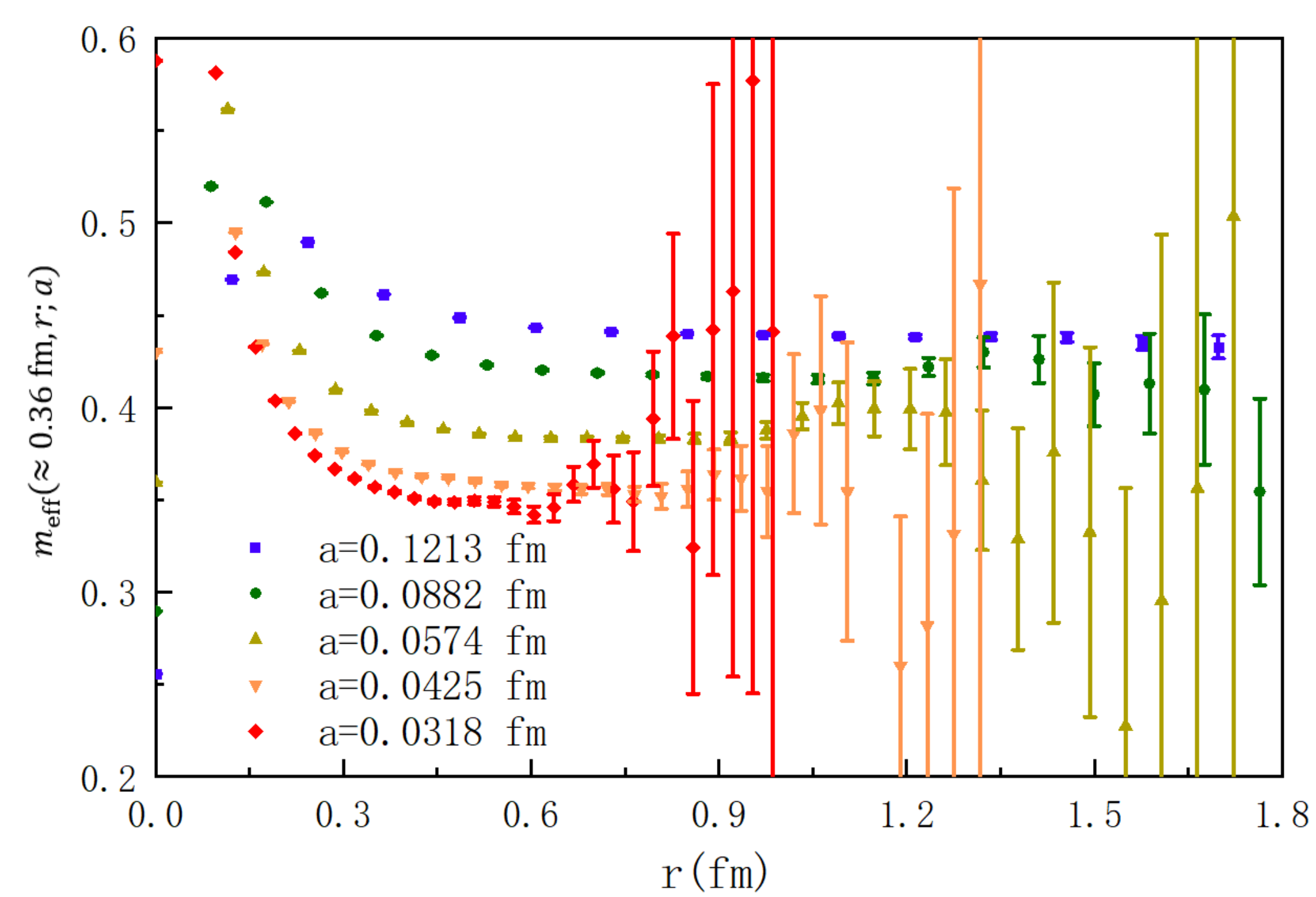}
\includegraphics[width=0.45\textwidth]{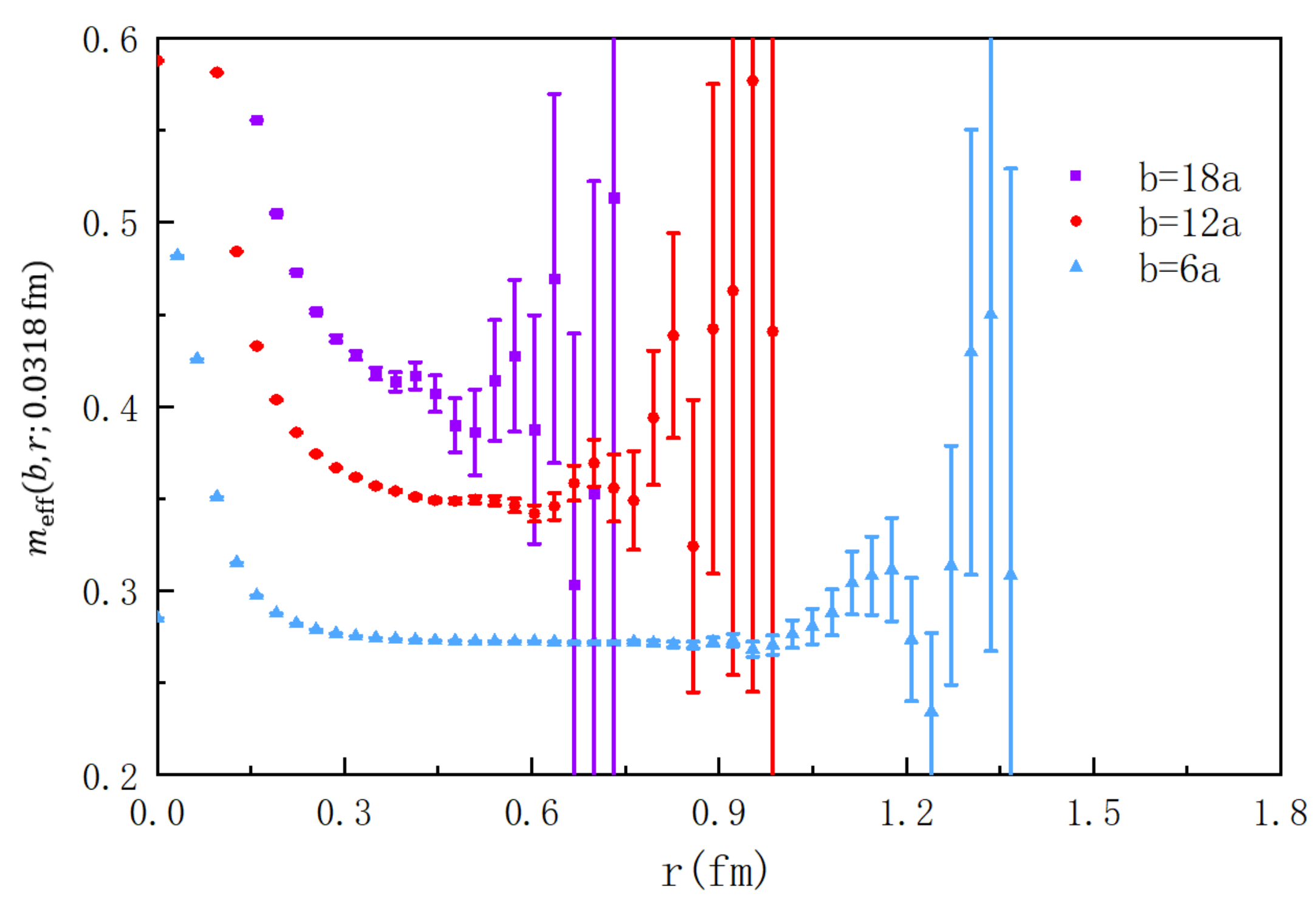}
\caption{The effective mass $m_{\text{eff}}(b,r;a)=\text{ln}\frac{Z_E(b,r;a)}{Z_E(b,r+1;a)}$ of Wilson loop as functions of r at different lattice spacing with fixed $b\simeq0.36 fm$ (left panel), and that at the smallest lattice spacing but different b (right panel). They all show a clear plateau. The plateau of smaller lattice spacing and larger $b$ is shorter.}\label{fig:eff}
\end{center}
\end{figure}

As shown in Fig~\ref{fig:eff}, the effective mass of Wilson loop $m_{\text{eff}}(b,r;a)=\text{ln}\frac{Z_E(b,r;a)}{Z_E(b,r+1;a)}$ will saturate to a plateau at large $r$, and the plateau will be shorter for smaller lattice spacing and larger $b$. For the fit window, we define the threshold $r_{\text{max}}$ as the point where SNR is more than 3, and our fitting range is chosen to be $\frac{r_{\text{max}}}{2}<r<r_{\text{max}}$. For example, with $a=0.0318\text{fm}, b=12a$, the fit window is $[15a,30a]$, and the fit gives $c(12a;0.0318\text{fm})=-1.90(0.02)$ and $V(12a;0.0318\text{fm})=0.3490(0.0015)$ with $\chi^2$/d.o.f.=0.63. 

Then we use the fitting results to replace original Wilson loop if $r>0.7r_{\text{max}}$. We fit the Wilson loop for every sample, and the statistical uncertainty comes from bootstrap re-sampling. As shown in Fig.~\ref{fig:oriLoop}, such a extrapolation is important to access the large $b$ and $z$ region, especially for small lattice spacing.

\begin{figure}[!th]
\begin{center}
\includegraphics[width=0.45\textwidth]{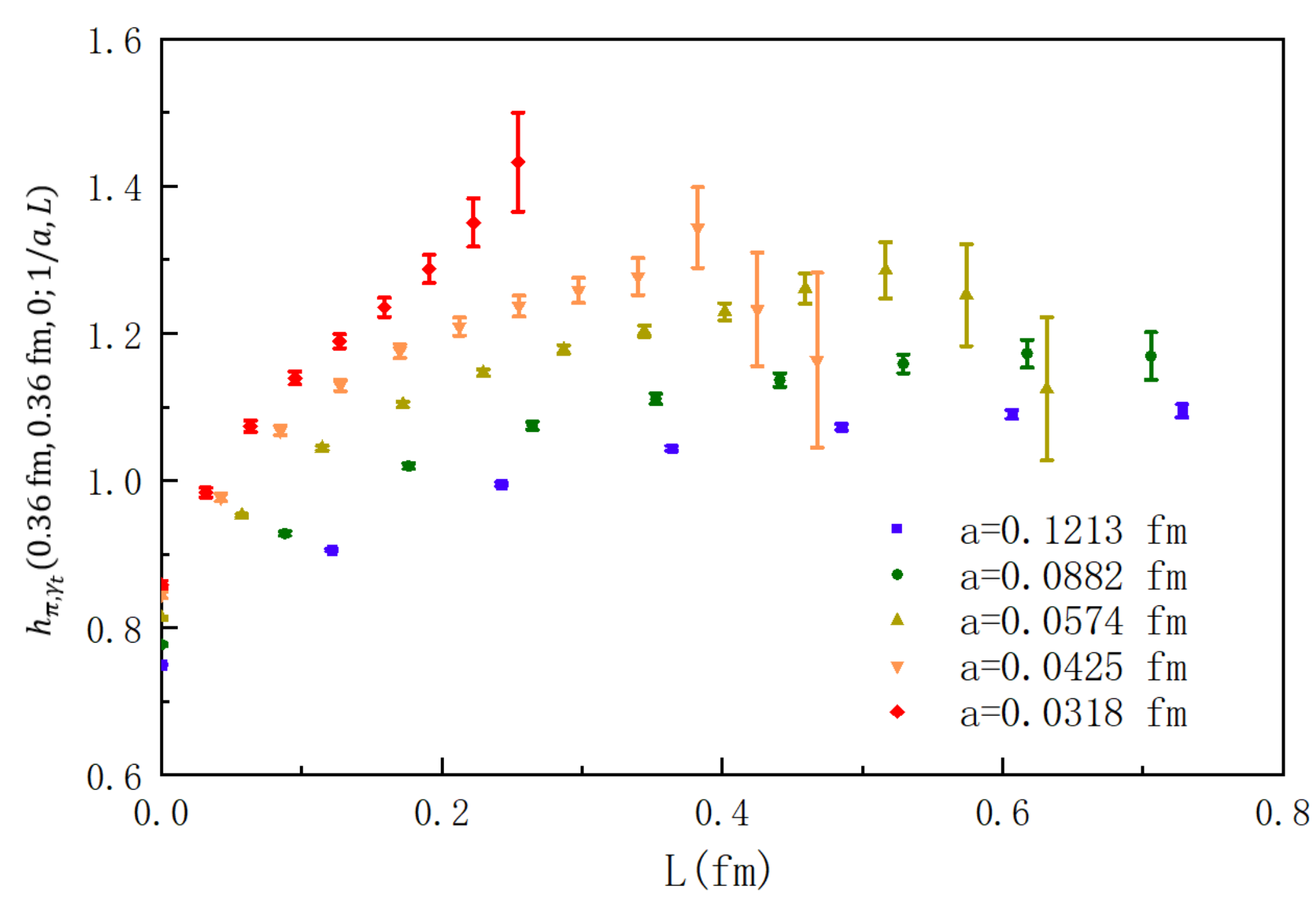}
\includegraphics[width=0.45\textwidth]{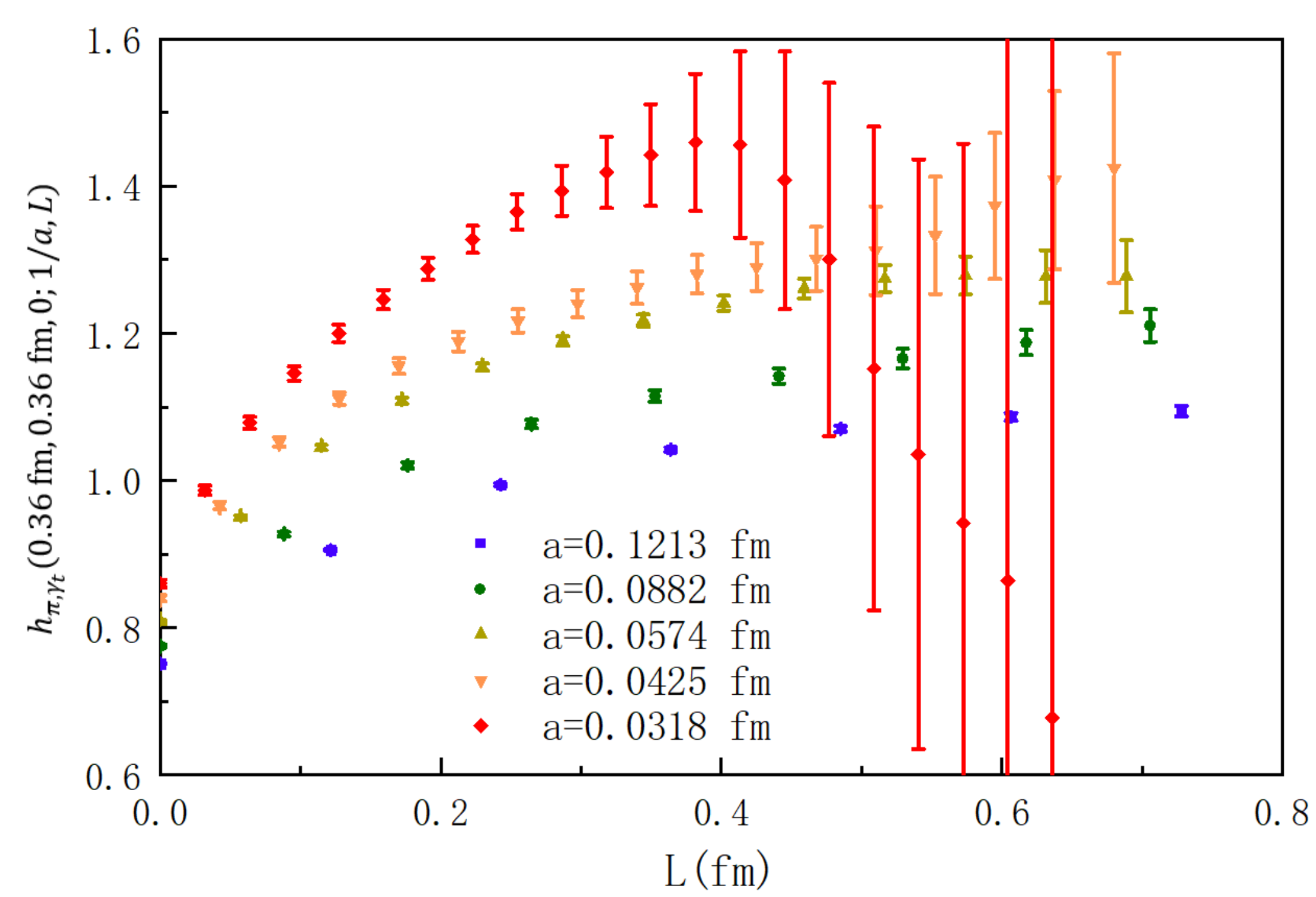}
\caption{The bare pion matrix element divided by Wilson loop, $h_{\pi,\gamma_t}(b,z,0;1/a,L)\equiv \frac{\tilde{h}_{\chi,\gamma_t}(b,z,L,P_z;1/a)}{\sqrt{Z_E(b,2L+z;1/a)}}$, at different lattice spacings. The bare matrix element is divided by the original Wilson loop (left panel), and the extrapolated Wilson loop (right panel).}\label{fig:oriLoop}
\end{center}
\end{figure}

\subsubsection{$L$ dependence of the matrix elements $h_{\pi,\gamma_t}(b,z,0;1/a)$ and $h^{\text{MOM}}_{\pi,\gamma_t}(b,z,0;p)$}

\begin{figure}[!th]
\begin{center}
\includegraphics[width=0.45\textwidth]{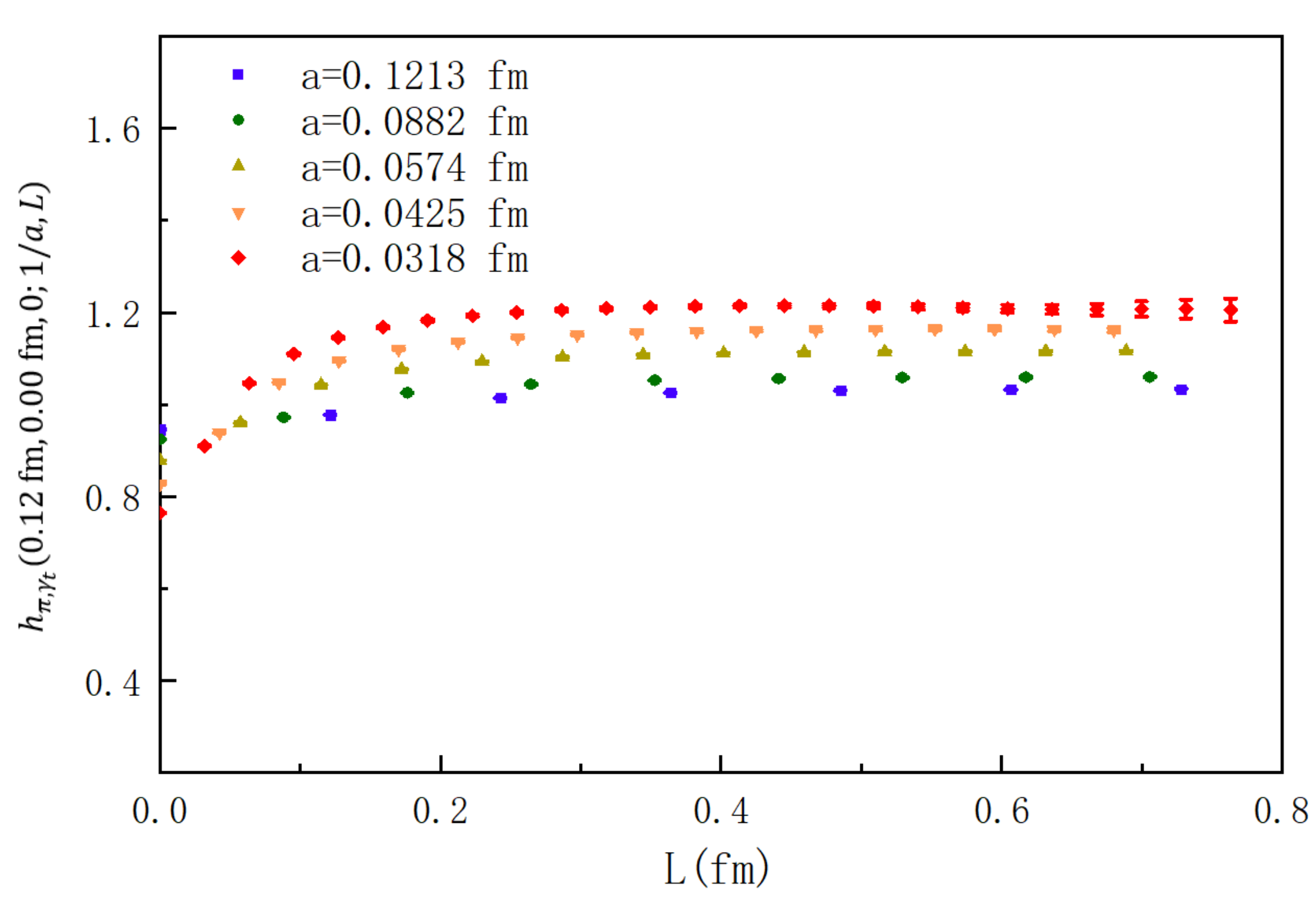}
\includegraphics[width=0.45\textwidth]{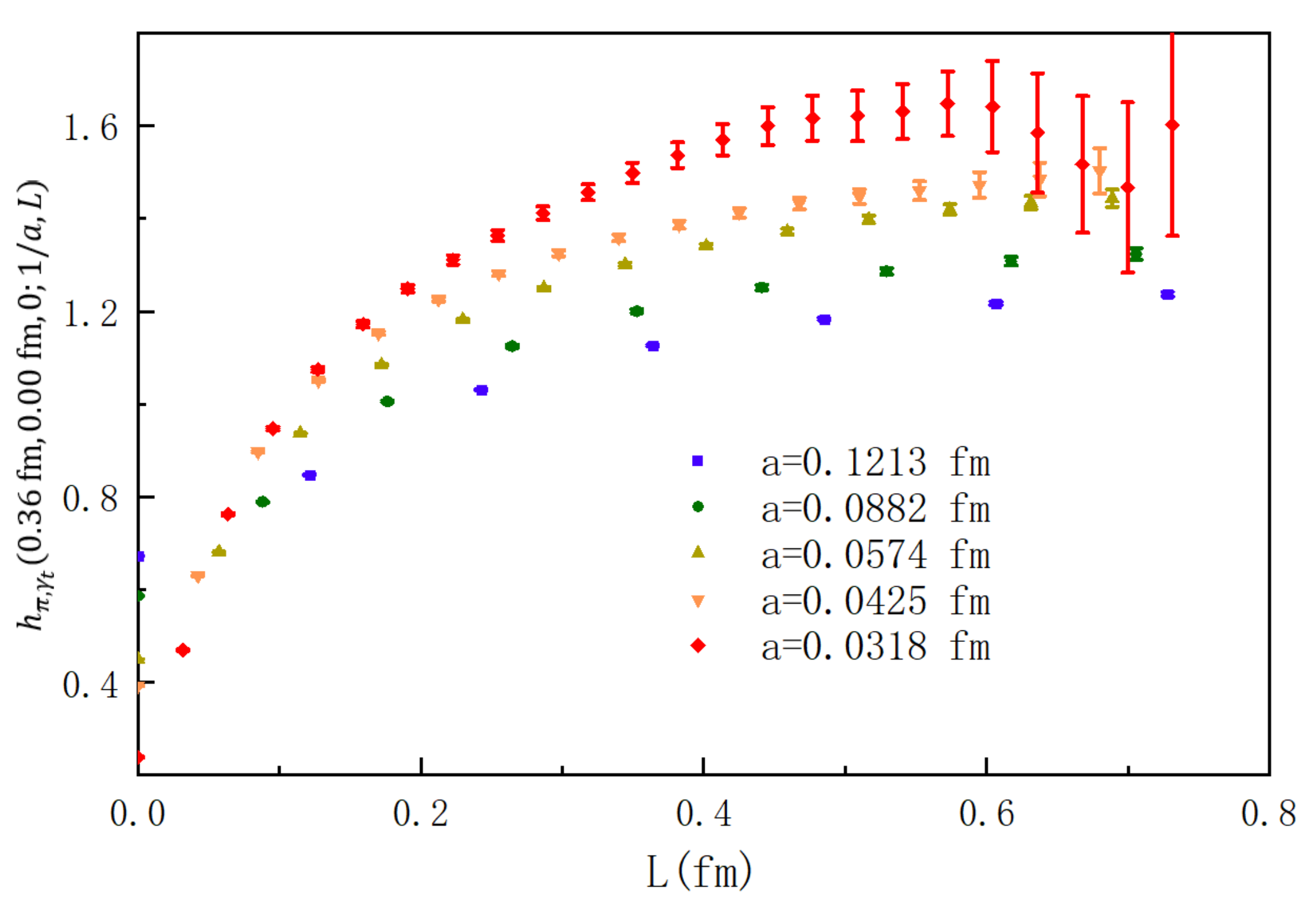}
\includegraphics[width=0.45\textwidth]{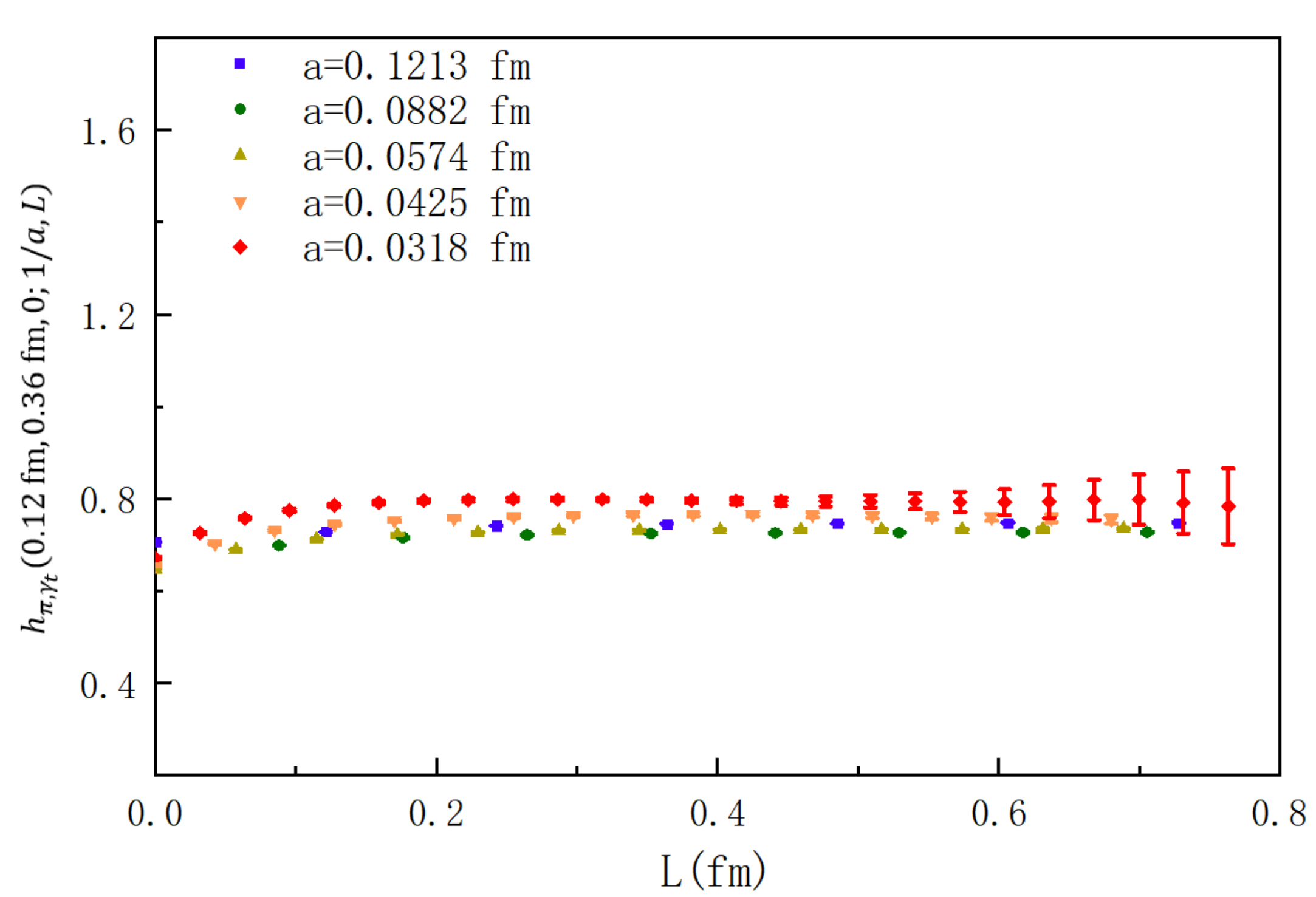}
\includegraphics[width=0.45\textwidth]{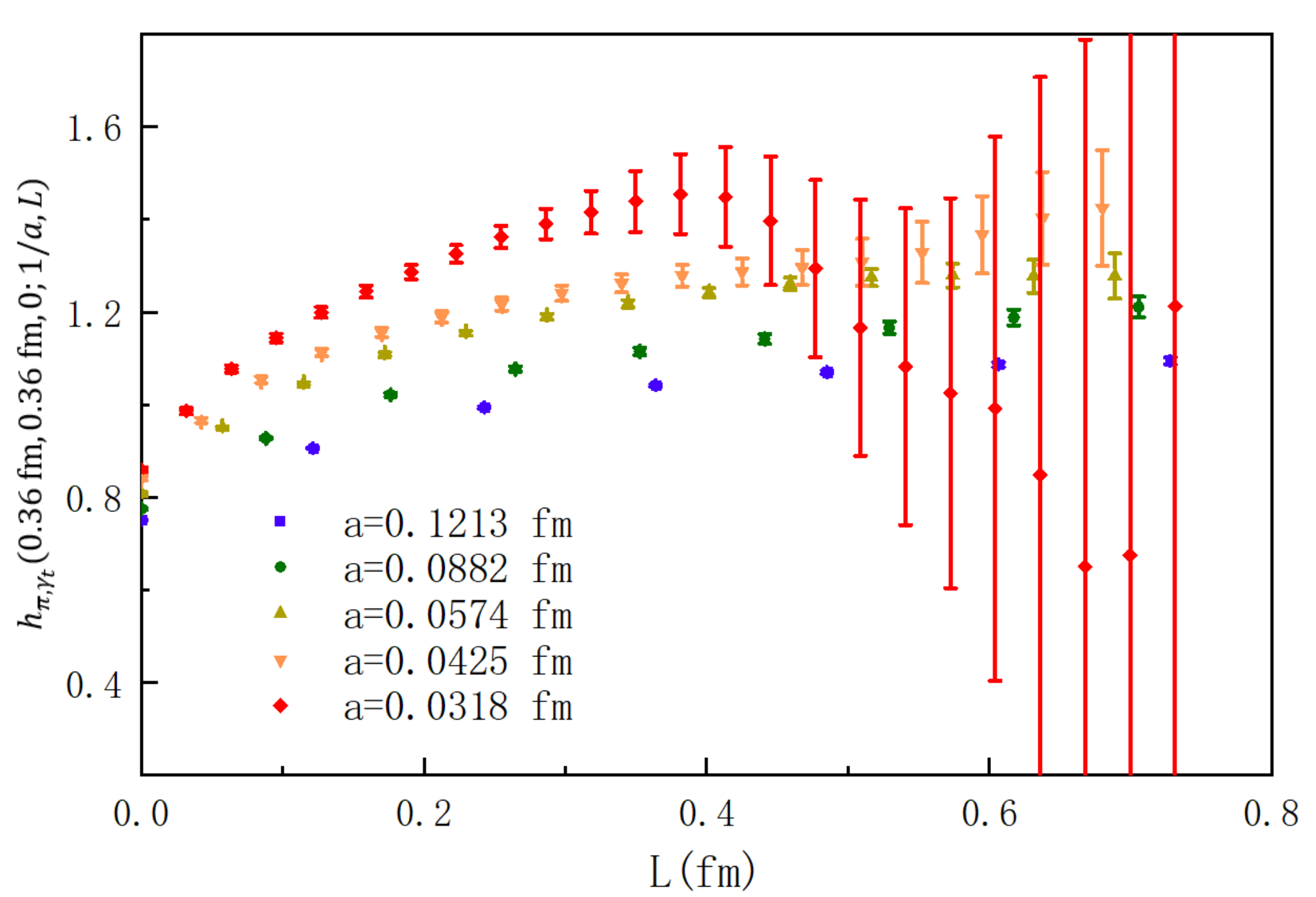}
\includegraphics[width=0.45\textwidth]{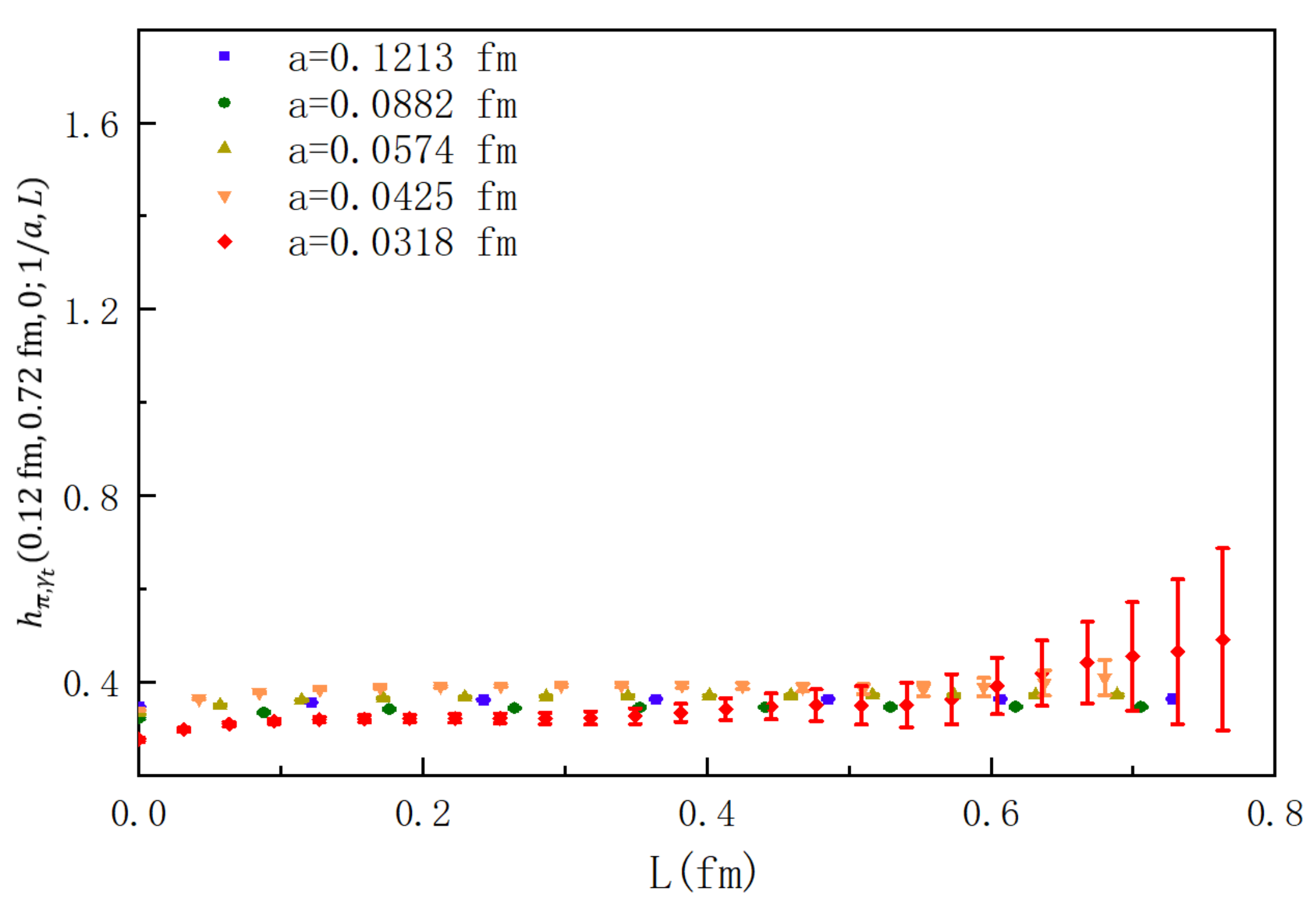}
\includegraphics[width=0.45\textwidth]{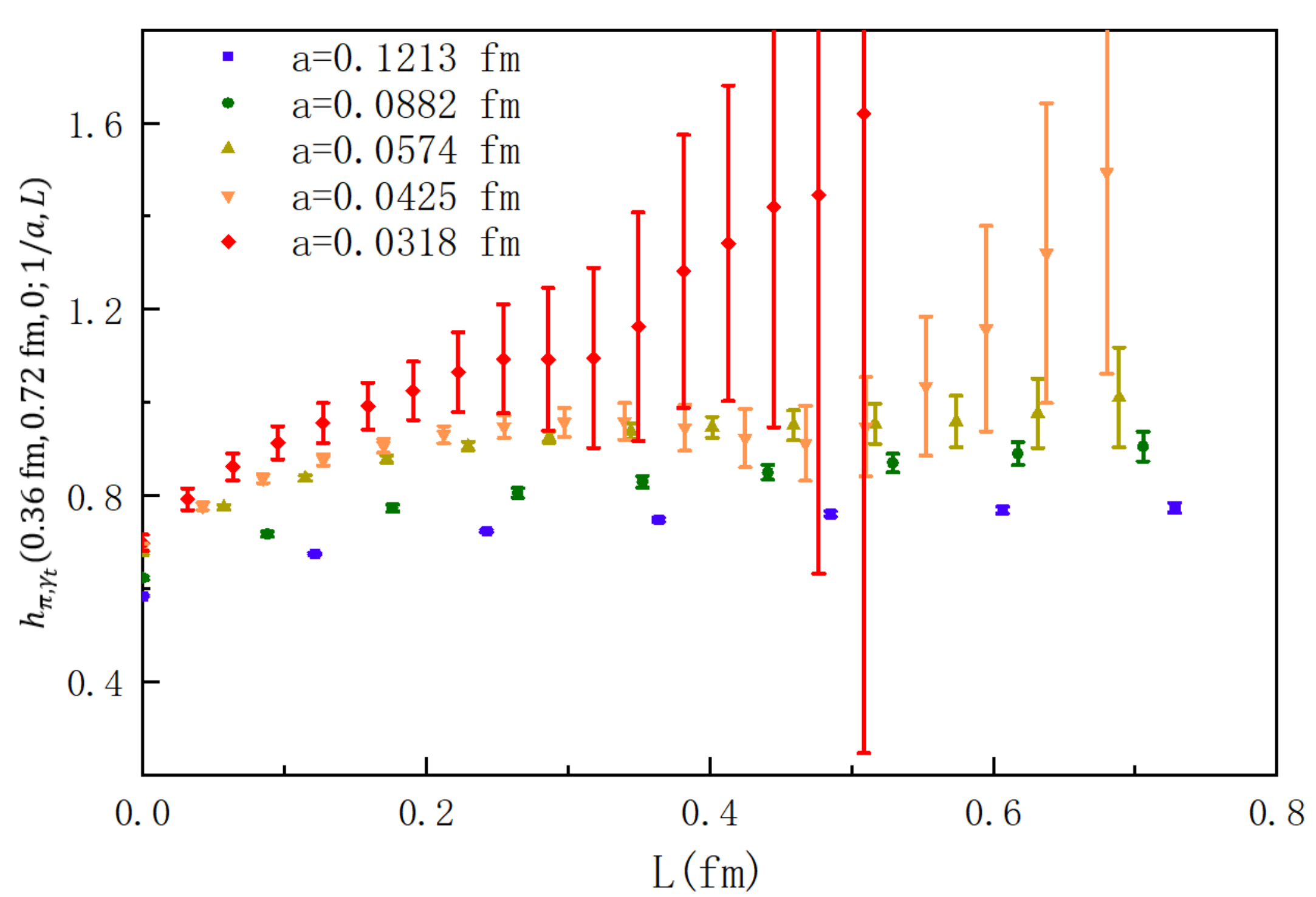}
\caption{The dependence of $L$ of the bare pion matrix element divided by Wilson loop, $h_{\pi,\gamma_t}(b,z,0;1/a,L)\equiv \frac{\tilde{h}_{\chi,\gamma_t}(b,z,L,P_z;1/a)}{\sqrt{Z_E(b,2L+z;1/a)}}$, at different lattice spacings. The statistical uncertainty comes from bootstrap re-sampling. We interpolate $b$ and $z$ to the same value for $h_{\pi,\gamma_t}$ of different lattice spacings.}\label{fig:Wilson}
\end{center}
\end{figure}

\begin{figure}[!th]
\begin{center}
\includegraphics[width=0.45\textwidth]{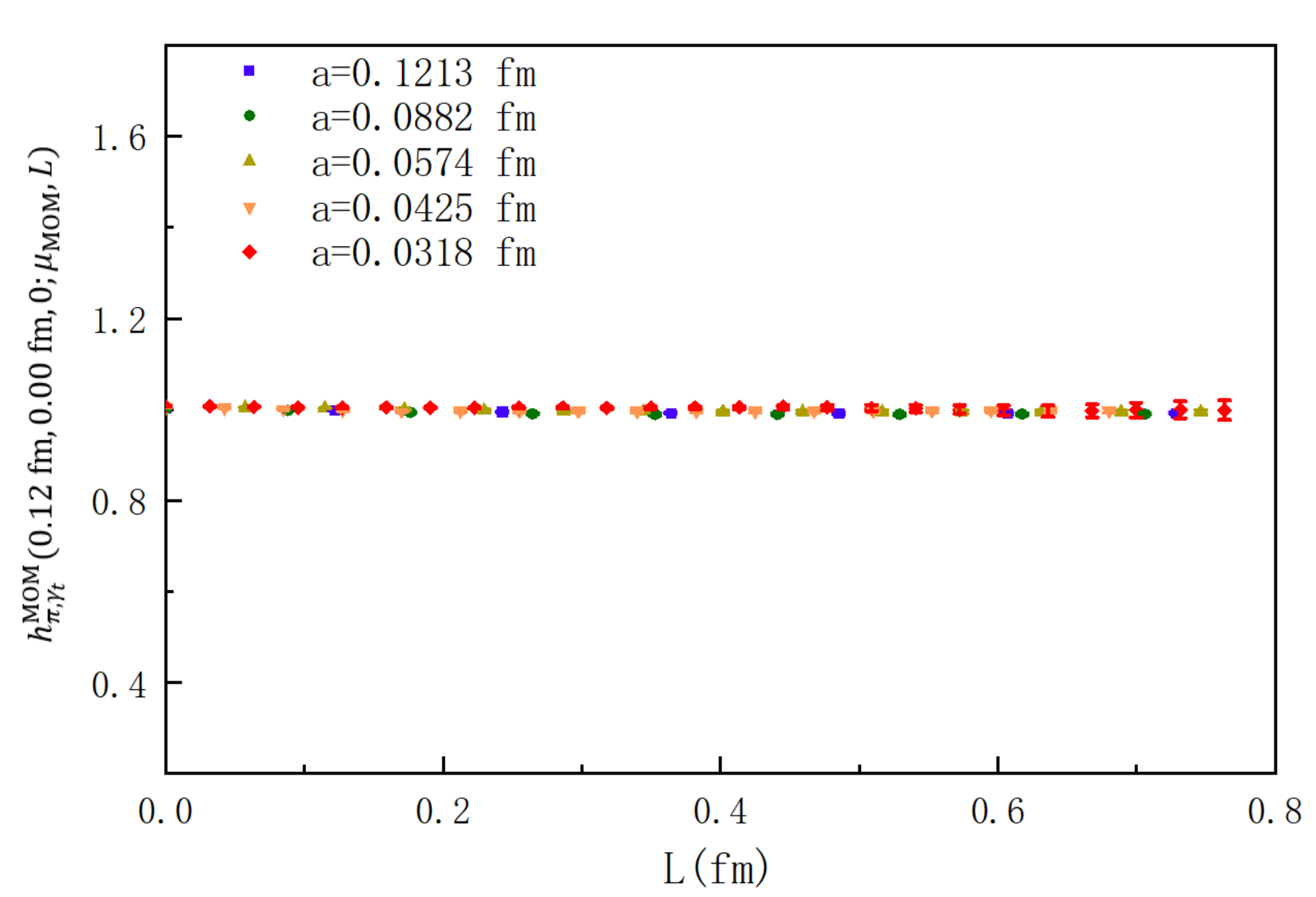}
\includegraphics[width=0.45\textwidth]{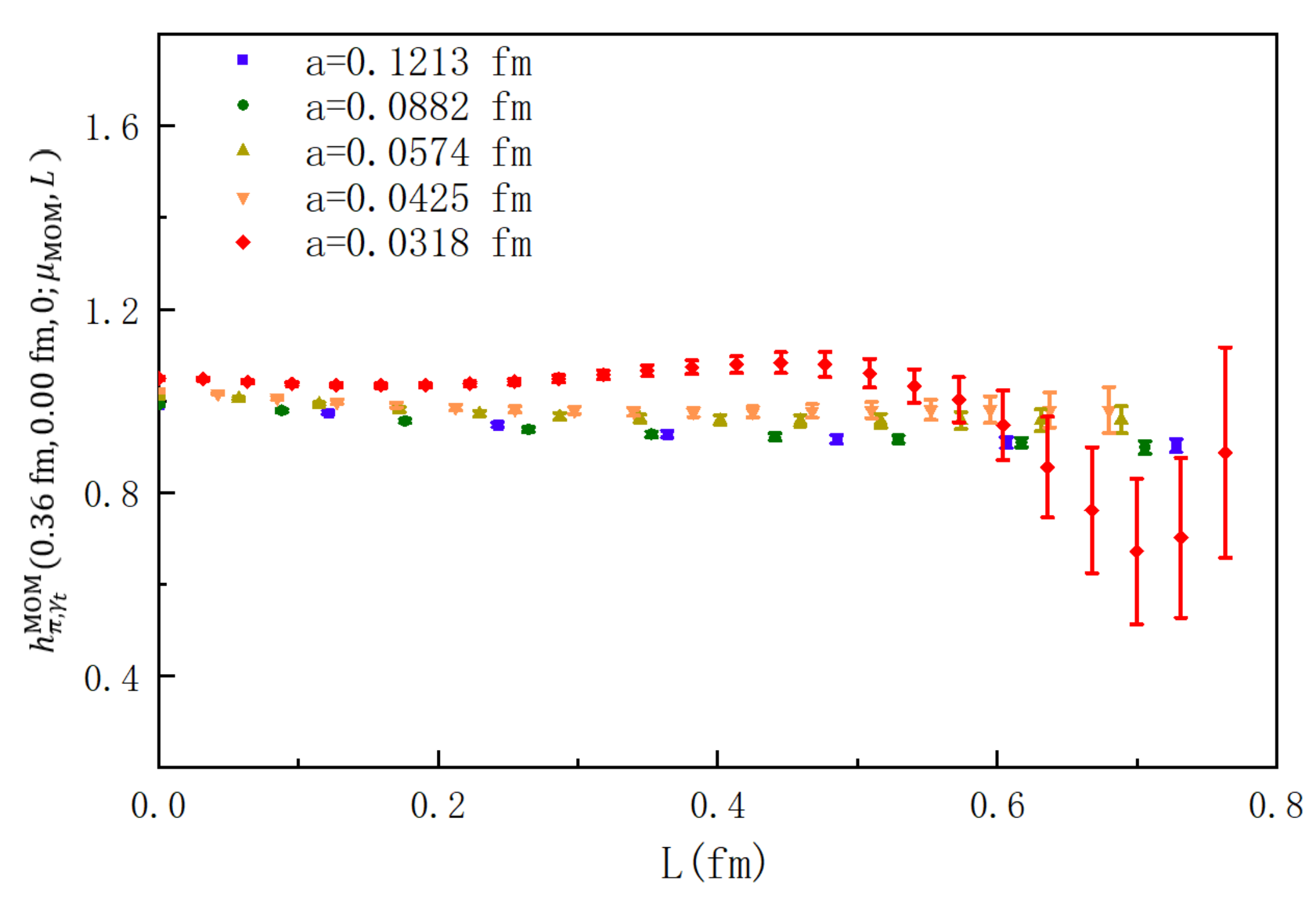}
\includegraphics[width=0.45\textwidth]{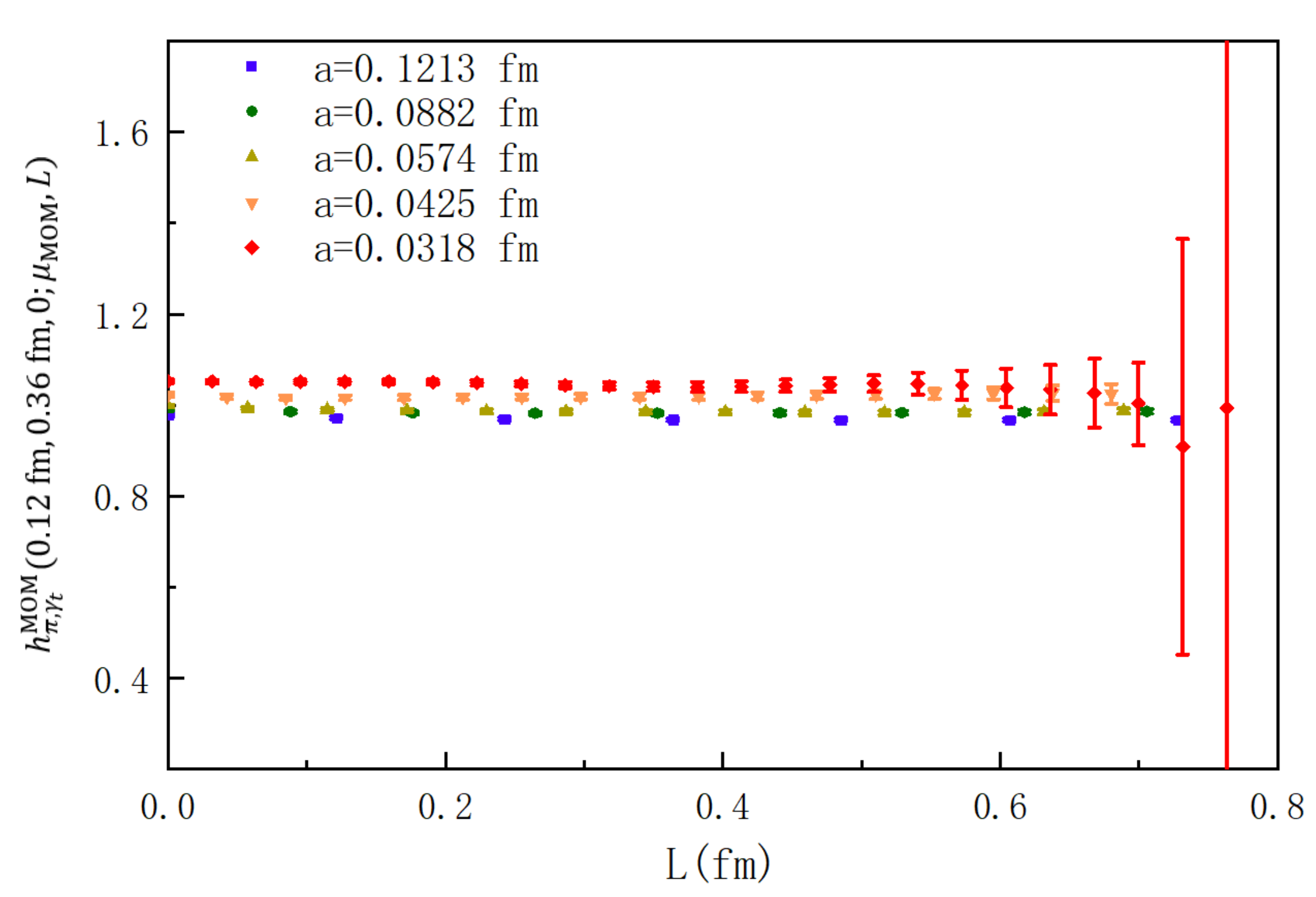}
\includegraphics[width=0.45\textwidth]{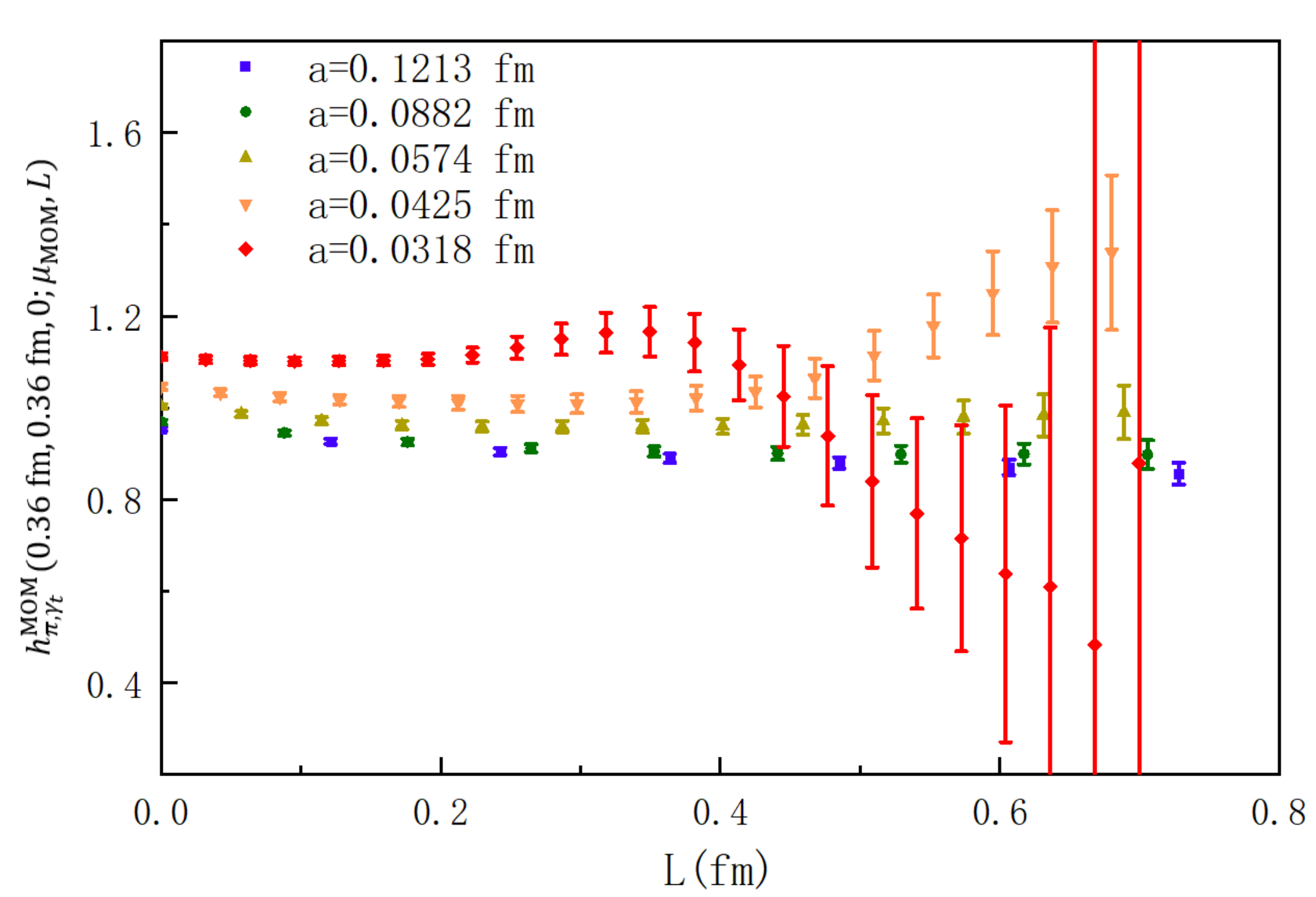}
\includegraphics[width=0.45\textwidth]{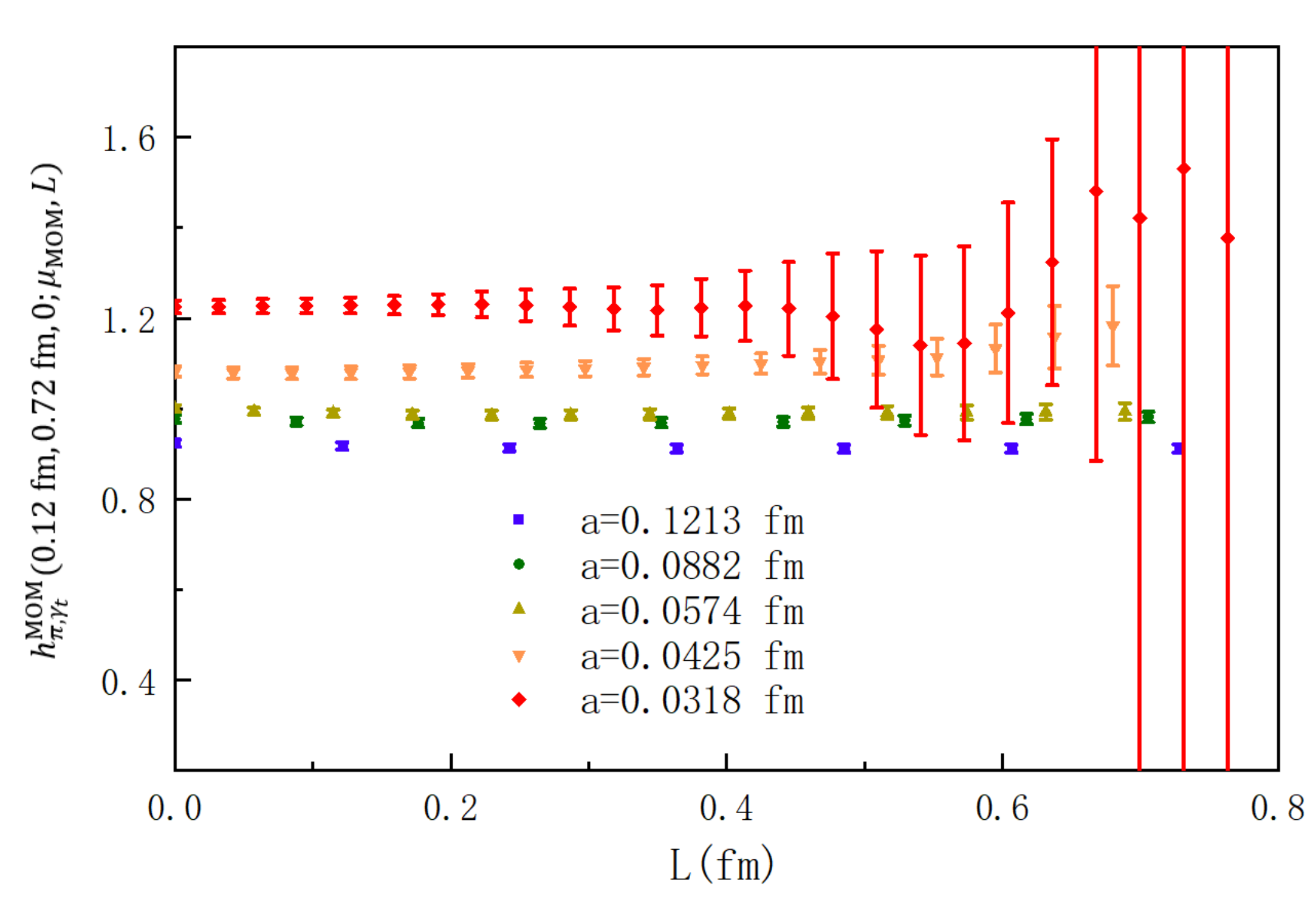}
\includegraphics[width=0.45\textwidth]{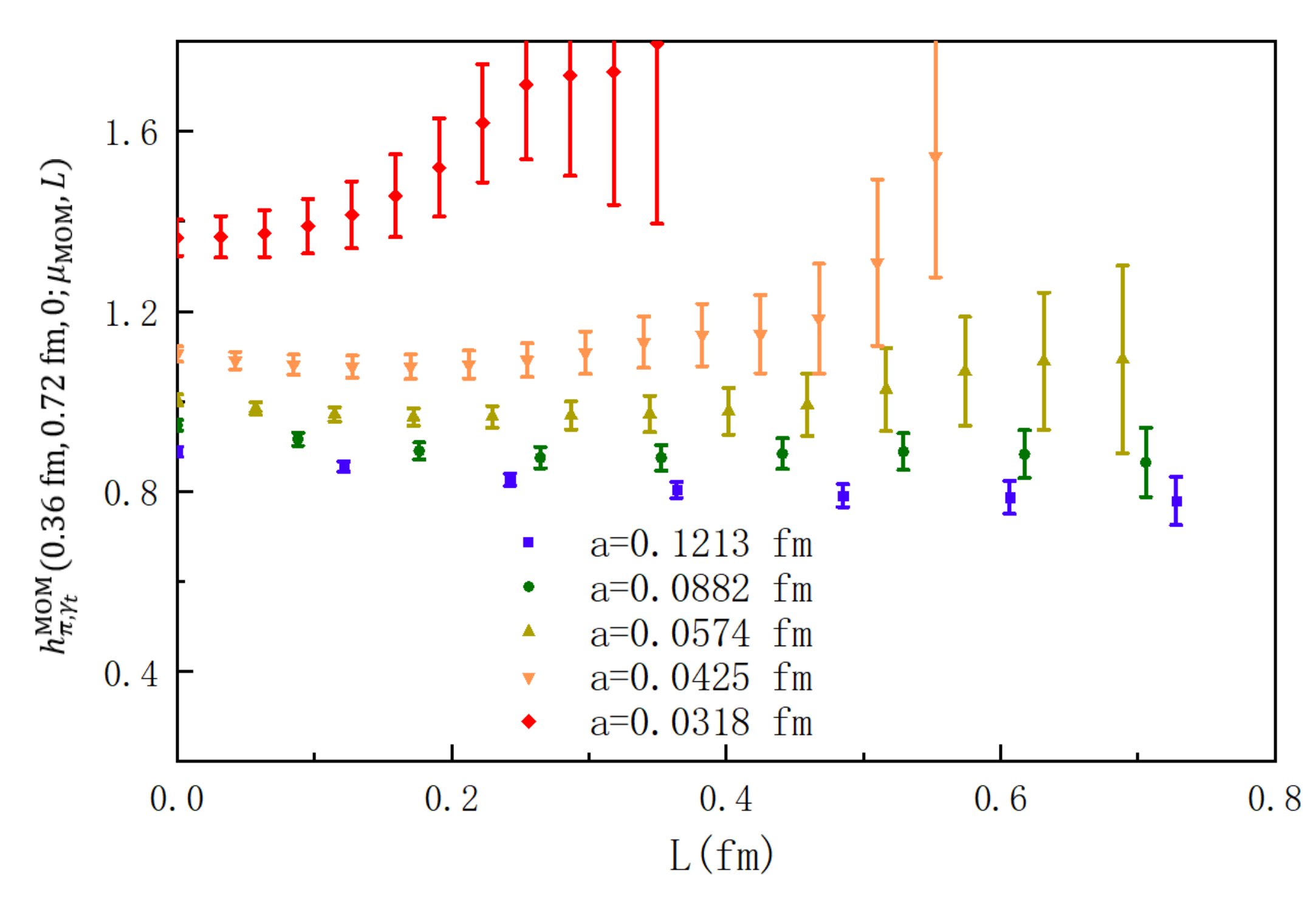}
\caption{The dependence of $L$ in pion matrix element with clover valence quark in RI/MOM scheme $h^{\text{MOM}}_{\pi,\gamma_t}(b,z,0;\mu_{\rm MOM},L)\equiv \sum_{\Gamma'}\tilde{Z}^{\rm MOM}(b,z,L,p;1/a)]^{-1}_{\Gamma\Gamma'} \tilde{h}_{\chi,\Gamma'}(b,z,L,P_z;1/a)$ defined in Eq.~\ref{eq:quasiTMDcorr_ri} at different lattice spacings. The statistical uncertainty comes from bootstrap re-sampling. We interpolate $b$ and $z$ to the same value for $h^{\text{MOM}}_{\pi,\gamma_t}$ of different lattice spacings.}\label{fig:RIMOM}
\end{center}
\end{figure}

For a well-defined quasi-TMD operator, $L$ should be large enough to ensure that the longitudinal link can extend outside the region of the parton and independent of $L$ after proper subtraction.  As shown in Fig.~\ref{fig:Wilson}, the matrix elements $h_{\pi,\gamma_t}$ all show a plateau at large $L$. To balance the statistical uncertainty and the systematic uncertainty, we take $L=6a$ for MILC12, $L=8a$ for MILC09, $L=11a$ for MILC06, $L=13a$ for MILC04, and $L=14a$ for MILC03 in most cases. For the cases with larger $b$ on the MILC03 ensemble, we set $L=b$ (more specifically, we set $L=16a$ for $b=16a$, and $L=18a$ for $b=18a$).

On the other hand, the $L$ dependence saturation in the renormalized matrix elements in RI/MOM scheme $h^{\text{MOM}}_{\pi,\gamma_t}$ is better than $h_{\pi,\gamma_t}$ as shown in Fig.~\ref{fig:RIMOM}, while the residual linear divergence is obvious. Thus practically, we take $L\simeq0.36$fm for $b=0.12$fm in the RI/MOM case to make the comparison.

\clearpage

\subsection{Action dependence and operator mixing}

\subsubsection{Matrix elements with clover valence quark and overlap valence quark}

\begin{figure}[!th]
\begin{center}
\includegraphics[width=0.45\textwidth]{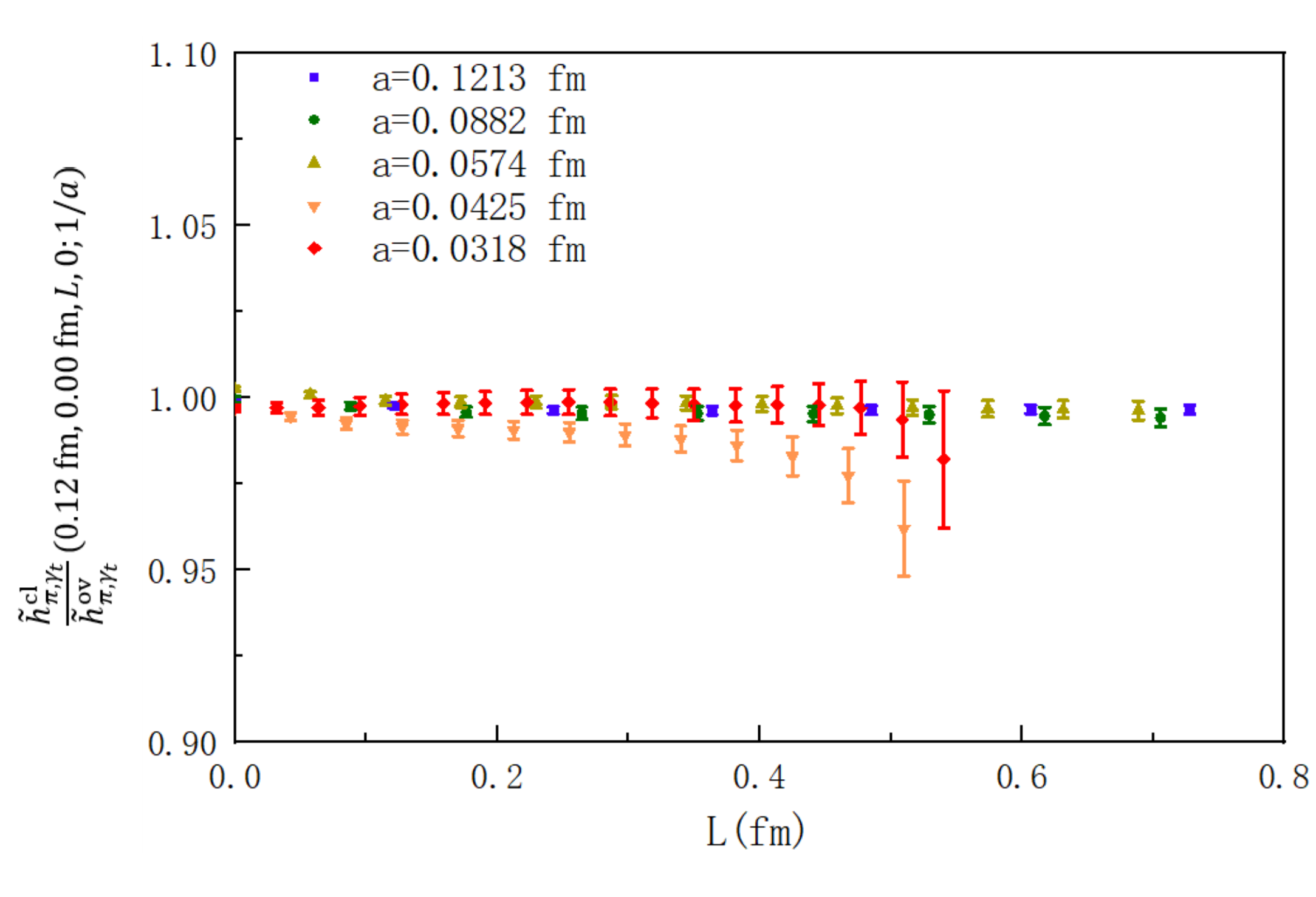}
\includegraphics[width=0.45\textwidth]{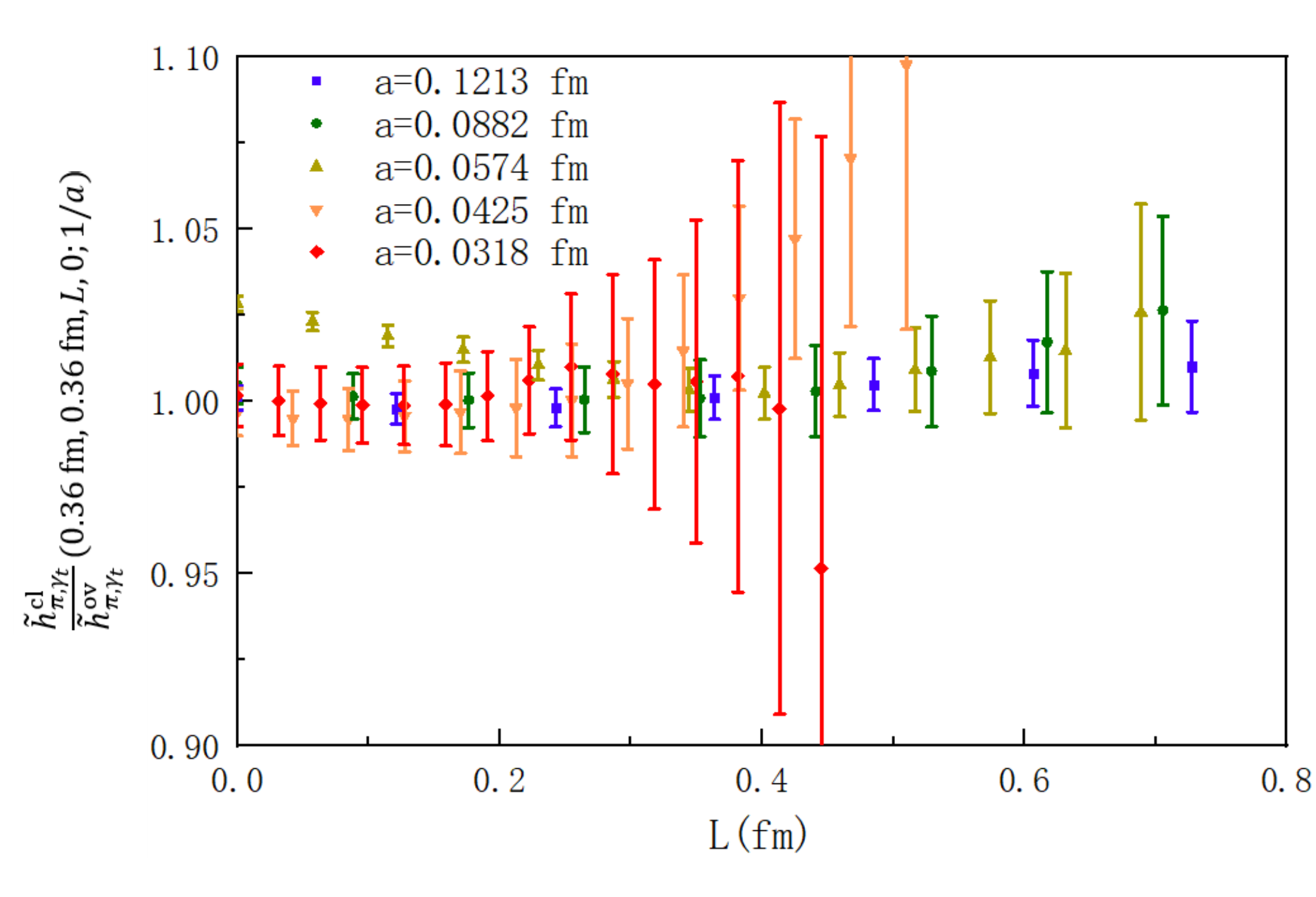}
\caption{The ratio of bare matrix elements with clover valence quark $\tilde{h}^{\text{cl}}_{\pi,\gamma_t}(b,z,L,0;1/a)$ and overlap valence quark $\tilde{h}^{\text{ov}}_{\pi,\gamma_t}(b,z,L,0;1/a)$.}\label{fig:Bare}
\end{center}
\end{figure}

As shown in Fig.~\ref{fig:Bare}, the bare pion matrix elements $\tilde{h}_{\pi,\gamma_t}(b,z,L,0;1/a)$ with clover valence quark and overlap valence quark are consistent with each other for different $b$ and $z$ at large $L$. Since Wilson loop won't be affected by the valence quark, the matrix elements $h_{\pi,\gamma_t}(b,z,0;1/a)\equiv \tilde{h}_{\pi,\gamma_t}(b,z,L,0;1/a)/\sqrt{Z_E(b,2L+z;1/a)}$ of the overlap fermion are also almost the same as that using the clover fermion. This conclusion is the same as that in the straight link case~\cite{Zhang:2020rsx}.

\begin{figure}[!th]
\begin{center}
\includegraphics[width=0.45\textwidth]{02_RIMOM_full.pdf}
\includegraphics[width=0.45\textwidth]{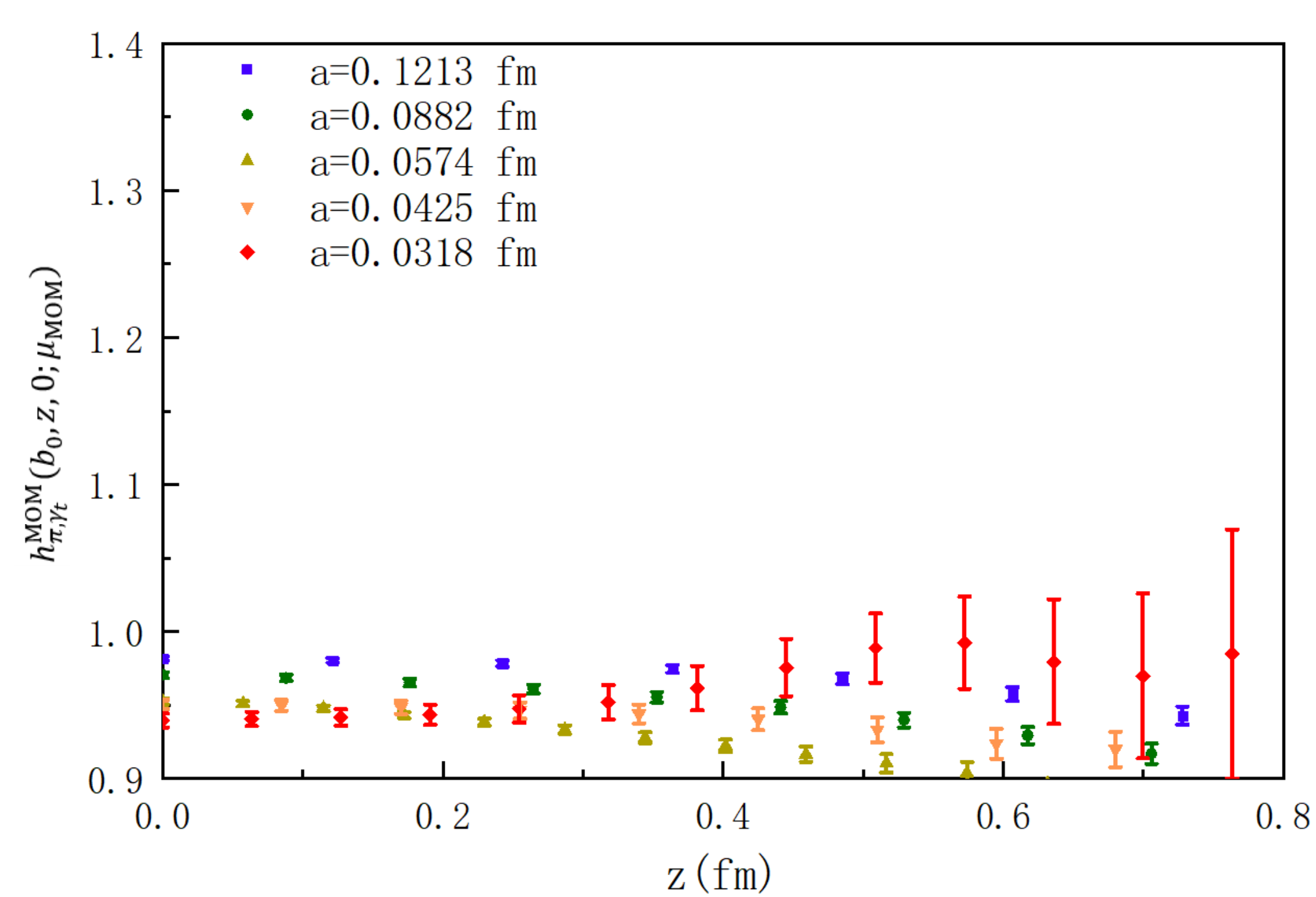}
\caption{The renormalized matrix elements $h^{\rm MOM}_{\pi,\gamma_t}(b_0,z,0;\mu_{\rm MOM})$  in RI/MOM scheme with $b_0=0.12~\text{fm}$ and $\mu_{\rm MOM}\simeq 3~\mathrm{GeV}$, using the clover fermion action (left panel) and the overlap fermion action (right panel).}\label{fig:RiValence}
\end{center}
\end{figure}

But as shown in Fig.~\ref{fig:RiValence}, the quark matrix elements of clover valence quark and overlap valence quark are quite different from each other, and make $h^{\rm MOM}_{\pi,\gamma_t}$ to be very sensitive to the quark action.

\subsubsection{Mixing, lattice spacing and chiral symmetry}

In order to estimate the operator-mixing effects, we adopt the similar method to Refs.~\cite{Shanahan:2019zcq} and calculate the nonperturbative RI/MOM renormalization/mixing factors,
\begin{align}\label{eq:mixing}
{\cal M}_{{\cal P}\Gamma}(b,z;1/a)=\frac{\textrm{Abs}\left[\textrm{Tr}\left[{\cal P}\Big\langle q(p)\Big|\overline q\Gamma {\cal W}(b,z,L)q\Big| q(p)\Big\rangle\right]\right]}{\textrm{Abs}\left[\textrm{Tr}\left[\Gamma\Big\langle q(p)\Big|\overline q\Gamma {\cal W}(b,z,L)q\Big| q(p)\Big\rangle\right]\right]}
\end{align}

\begin{figure}[!th]
\begin{center}
\includegraphics[width=0.45\textwidth]{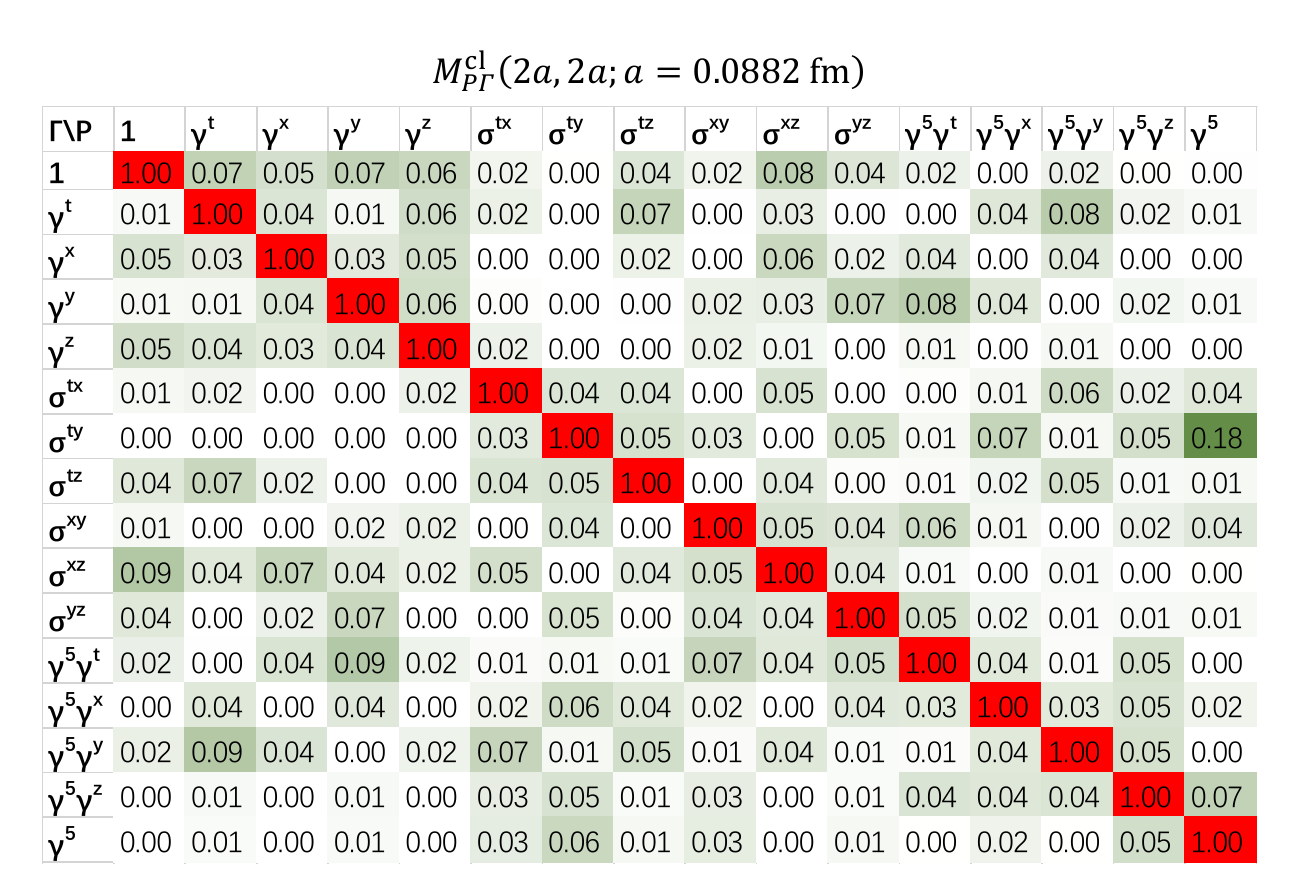}
\includegraphics[width=0.45\textwidth]{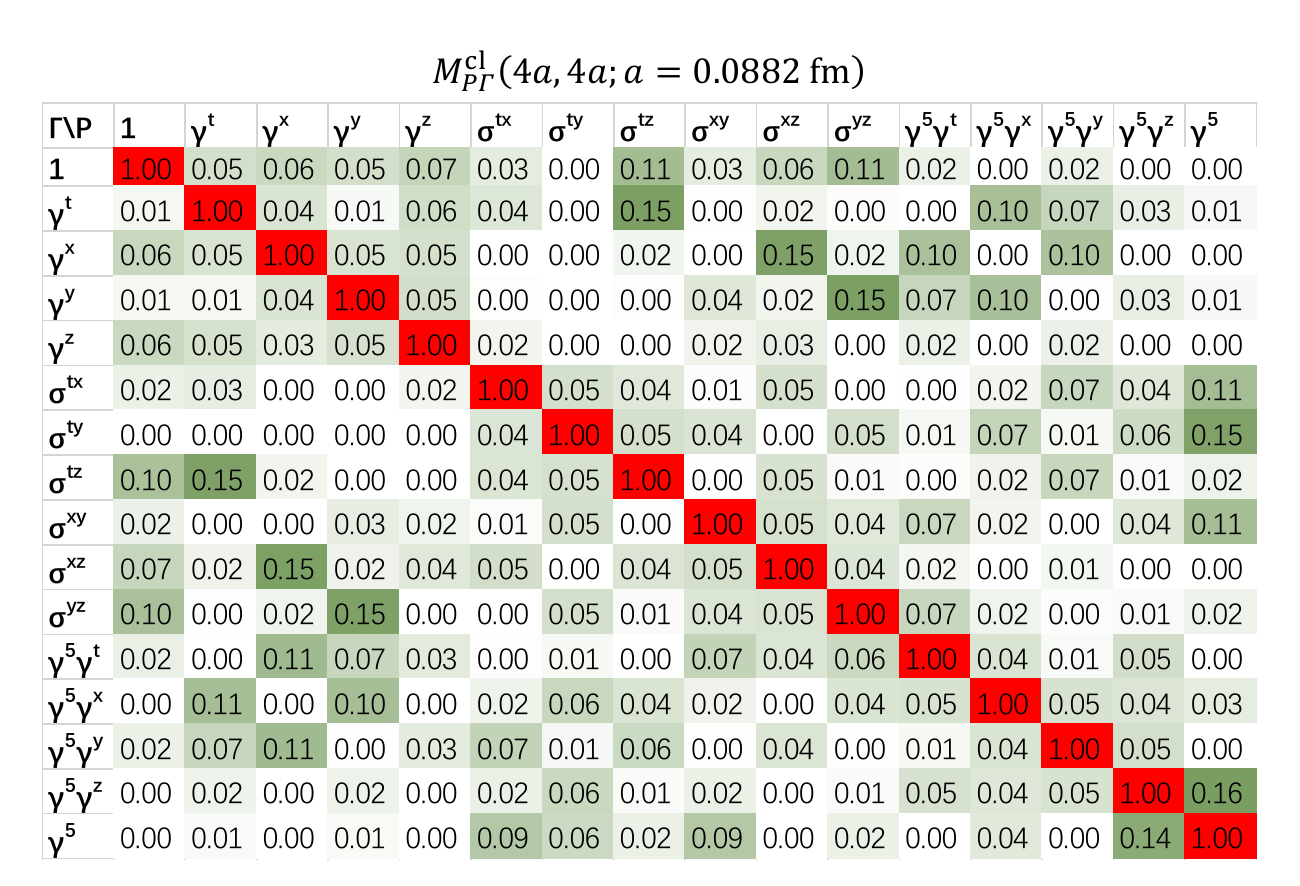}
\includegraphics[width=0.45\textwidth]{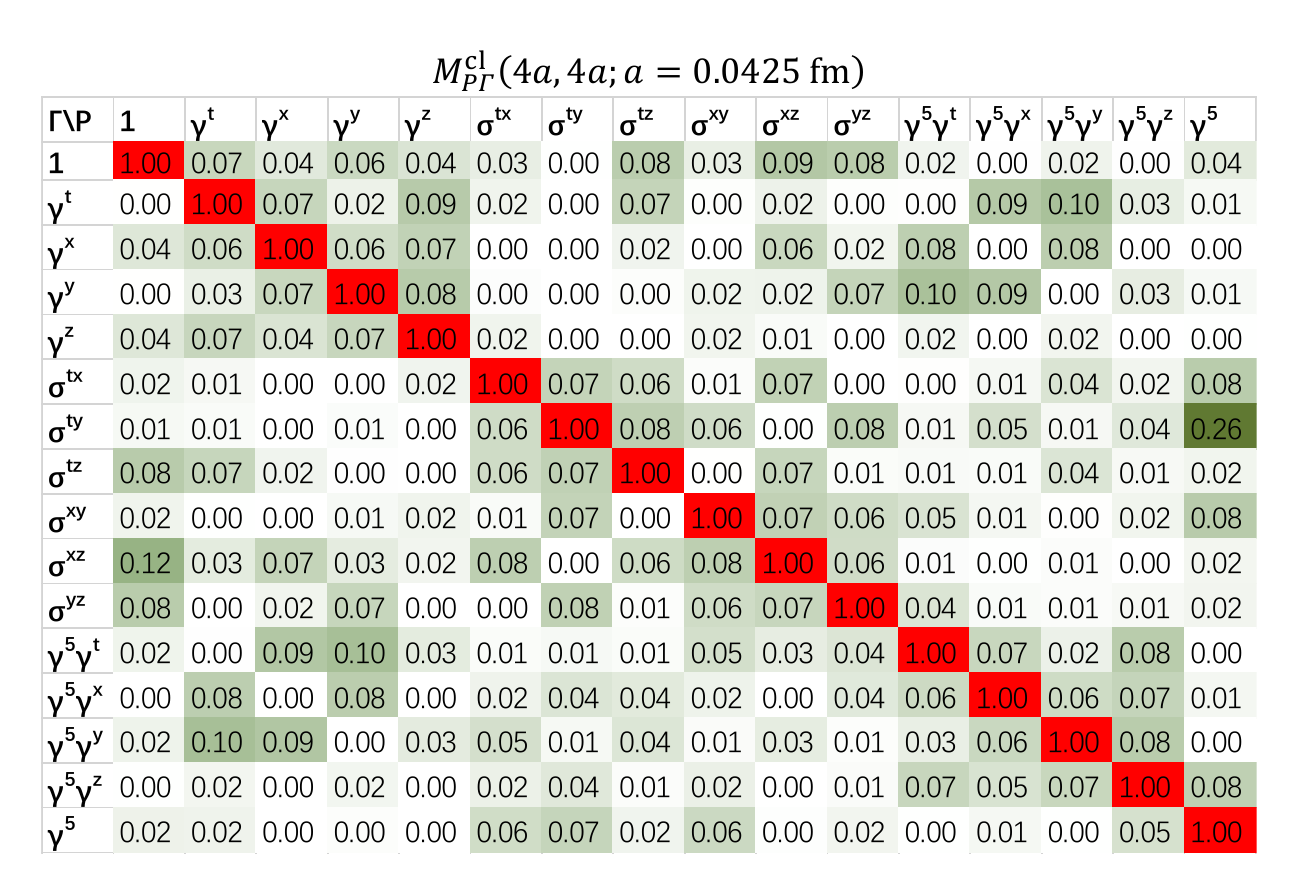}
\includegraphics[width=0.45\textwidth]{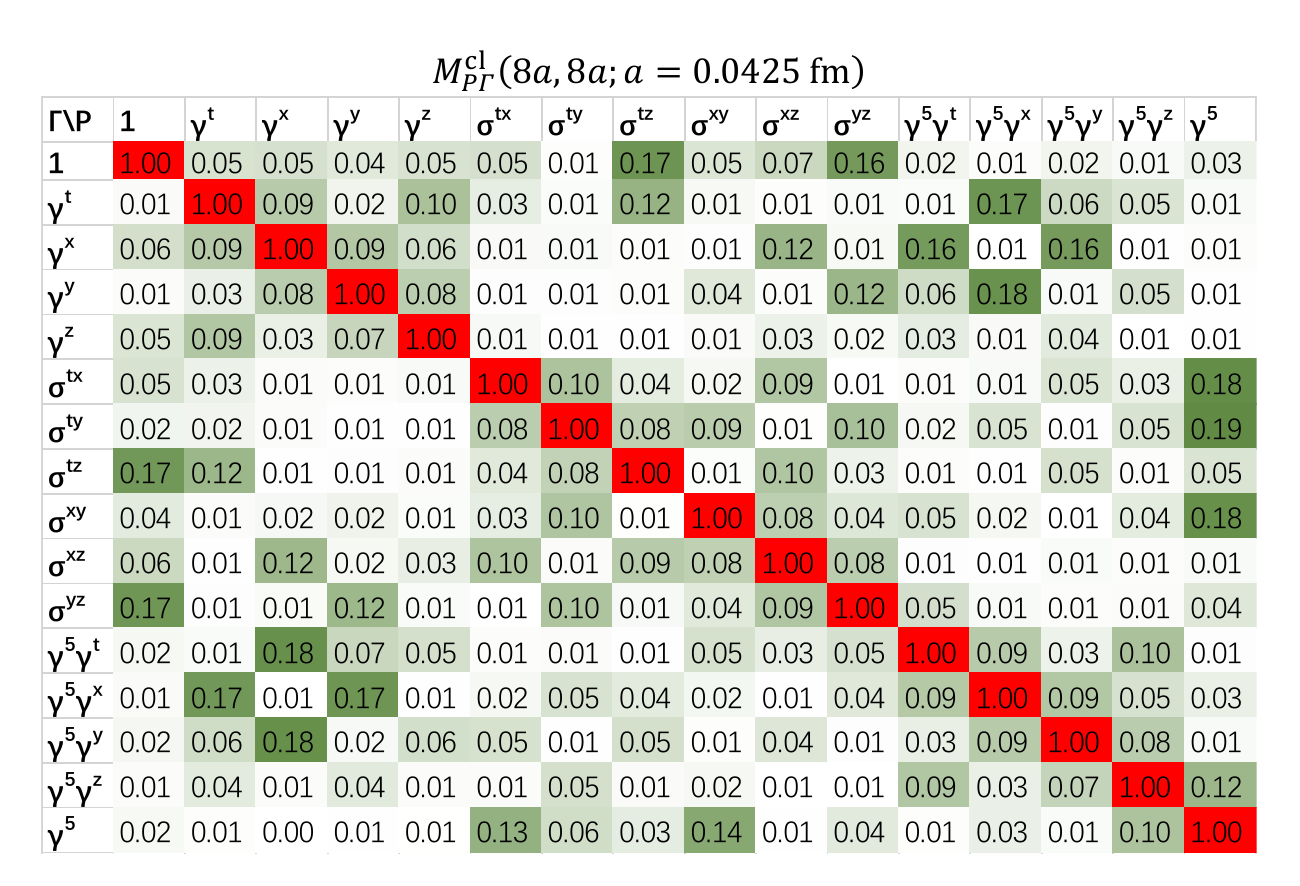}
\caption{The mixing/renormalization factors ${\cal M}_{{\cal P}\Gamma}(b,z;1/a)$ defined in Eq~\ref{eq:mixing}. They are calculated with the clover valence quark. We choose $L=7a$ for MILC09 and $L=14a$ for MILC04.}\label{fig:Mixing_cl}
\end{center}
\end{figure}

\begin{figure}[!th]
\begin{center}
\includegraphics[width=0.45\textwidth]{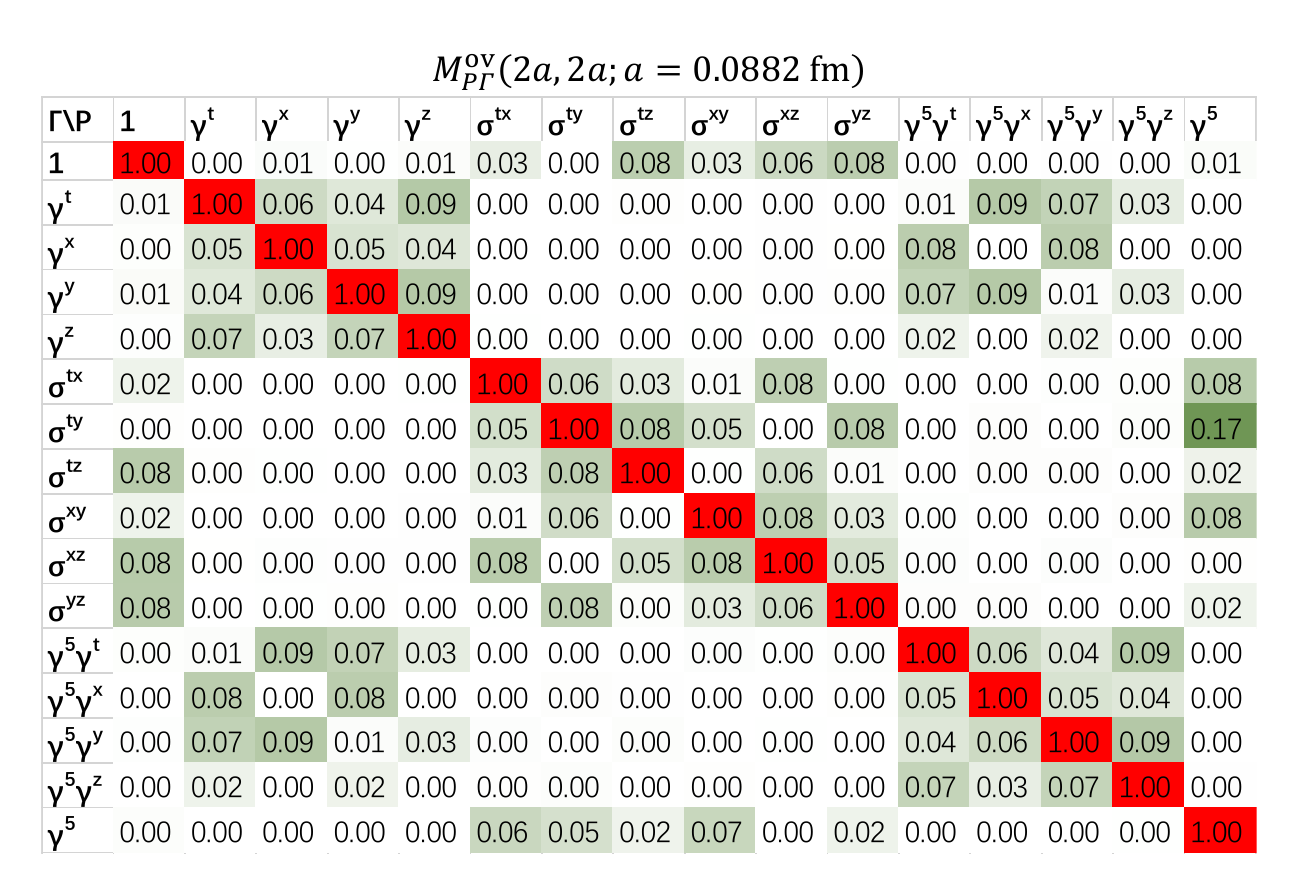}
\includegraphics[width=0.45\textwidth]{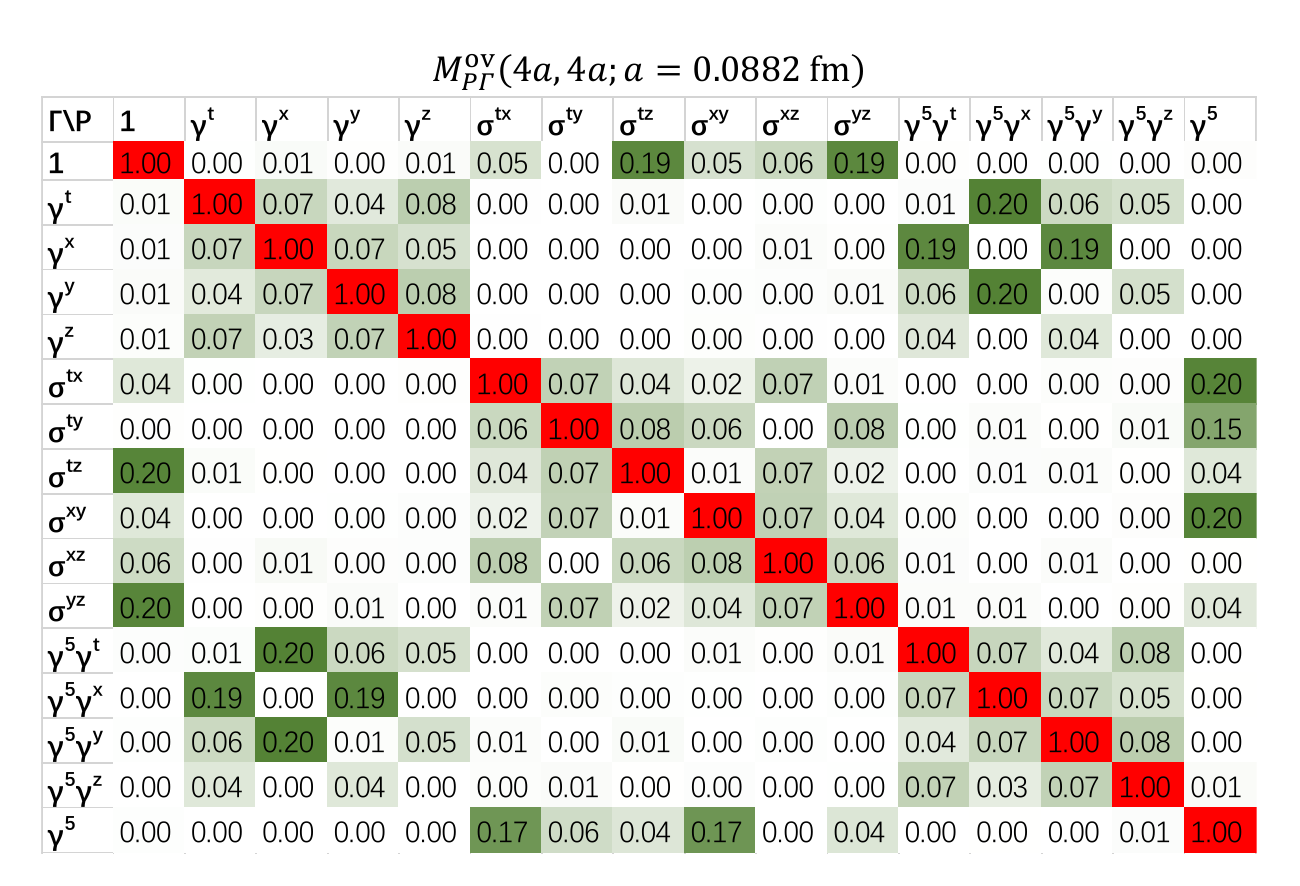}
\includegraphics[width=0.45\textwidth]{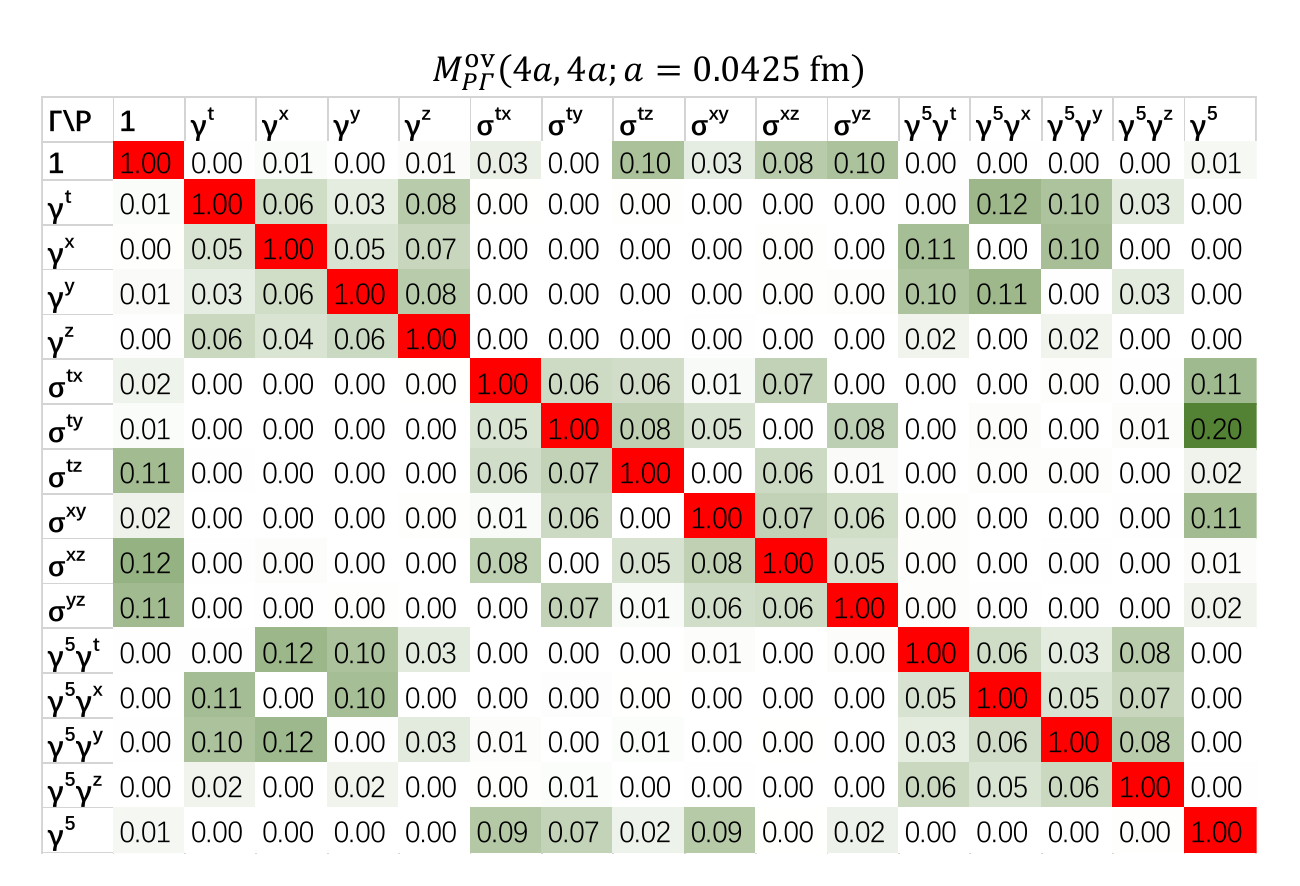}
\includegraphics[width=0.45\textwidth]{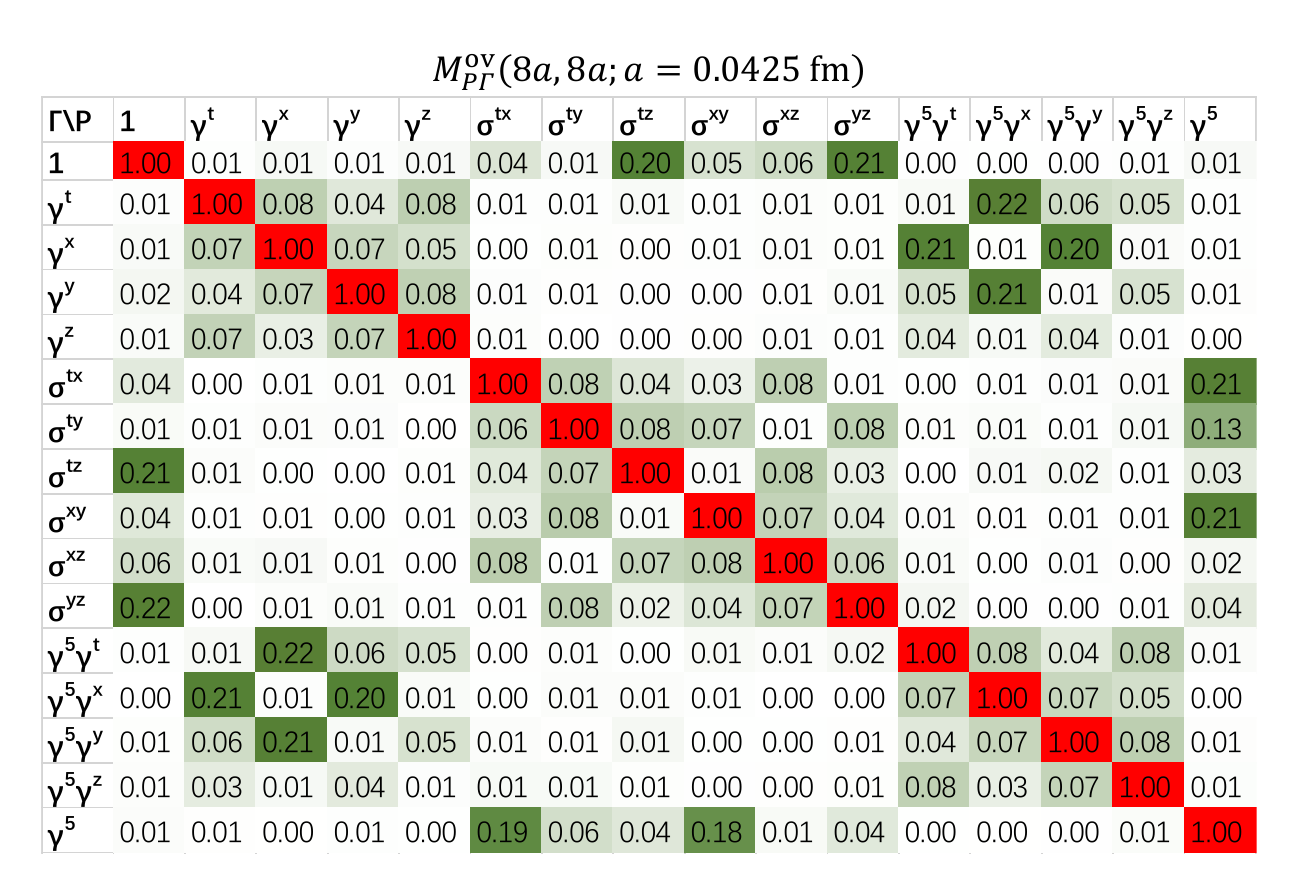}
\caption{Similar to Fig.~\ref{fig:Mixing_cl} but with the overlap valence quark.}\label{fig:Mixing_ov}
\end{center}
\end{figure}

As shown in Fig.~\ref{fig:Mixing_cl}, the mixing becomes larger with the increase of $b$ and $z$, and the decrease of $a$. This indicates that the mixing does not have a good continuum limit and is hard to deal with if we want to perform high precision calculation in the future.

Comparing to the clover case, the mixing pattern is more regular for the overlap case as shown in Fig.~\ref{fig:Mixing_ov}. Some area becomes smaller while other area becomes larger. It is quite an interesting result since the bare hadron matrix element seems to not be affected by the valence fermion action, and the same feature has been observed in the straight link case~\cite{Zhang:2020rsx}. It is well-known that overlap fermion has a much better preservation of the chiral symmetry, thus we would like to
guess that the explicit chiral symmetry of the clover fermion plays an important role to make the mixing with two quark actions to be very different. 

Compared to the Refs.~\cite{Shanahan:2019zcq}, our results are somehow different. We think the difference can come from the following reasons (from most important to least important): 1) Their external momentum has non-zero value in every direction, while our momenta is perpendicular to the staple-shaped link. We think much of the mixing comes from off-shell momentum. 2) The smearing on the gauge link used by two works are not the same, as they use Wilson flow with flow-time $t=1.0$, and we use 1-step HYP smearing; 3) They used the averaged diagonal matrix element as the denominator in the definition of ${\cal M}_{{\cal P}\Gamma}$ which makes the normalization to be slightly different from ours. 4) They use quenched QCD gauge field ensembles and we use the dynamic ensembles.

\begin{figure}[!th]
\begin{center}
\includegraphics[width=0.45\textwidth]{02_RIMOM_full.pdf}
\includegraphics[width=0.45\textwidth]{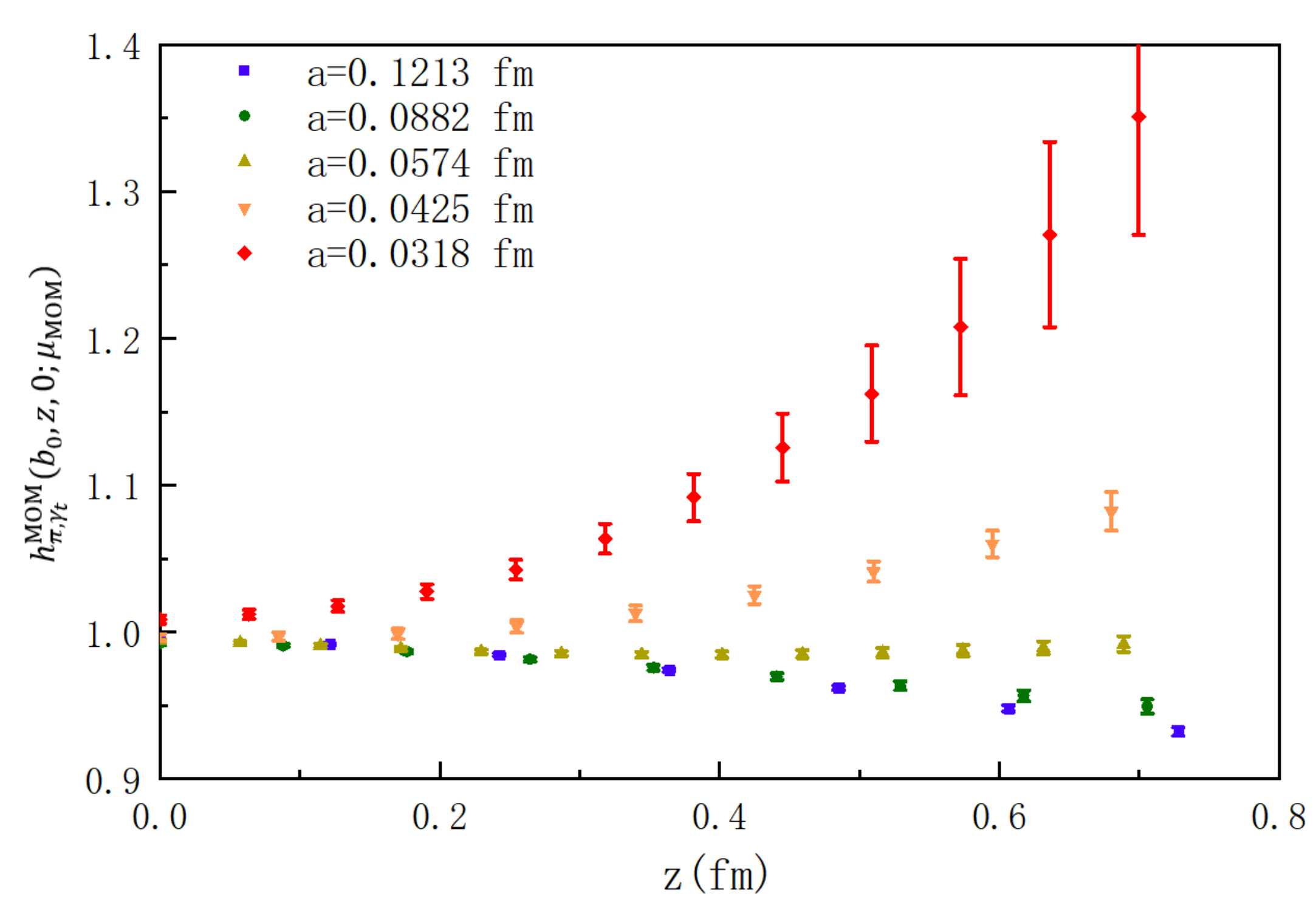}
\caption{The left panel is the renormalized matrix elements in RI/MOM scheme defined in Eq.~\ref{eq:quasiTMDcorr_ri} with the clover fermion action $h^{\rm MOM}_{\pi,\gamma_t}(b_0,z,0;\mu_{\rm MOM})$ with $b_0=0.12~\text{fm}$ and $\mu_{\rm MOM}\simeq 3~\mathrm{GeV}$. The right panel is that approximated by $h^{\rm MOM}_{\pi,\gamma_t}(b_0,z,0;\mu_{\rm MOM}) \simeq \lim_{L\to\infty}[\tilde{Z}^{\rm MOM}(b_0,z,L;\mu_{\rm MOM})]^{-1}_{\gamma_t\gamma_t} \tilde{h}_{\pi,\gamma_t}(b_0,z,L,0)$.}\label{fig:Mixing_linear}
\end{center}
\end{figure}

In Fig.~\ref{fig:Mixing_linear}, we also compare the RI/MOM renormalized result with (left panel) and that ignoring the operator mixing (right panel). As shown in the figure, the mixing has little effect on residual linear divergence. 

\clearpage

\subsection{The renormalized matrix elements $h^{\overline{\rm MS}}_{\pi,\gamma_t}(b,z,0;2\text{ GeV})$ and TMD wave function matrix element at different $b$ and $z$}

\begin{figure}[!th]
\begin{center}
\includegraphics[width=0.45\textwidth]{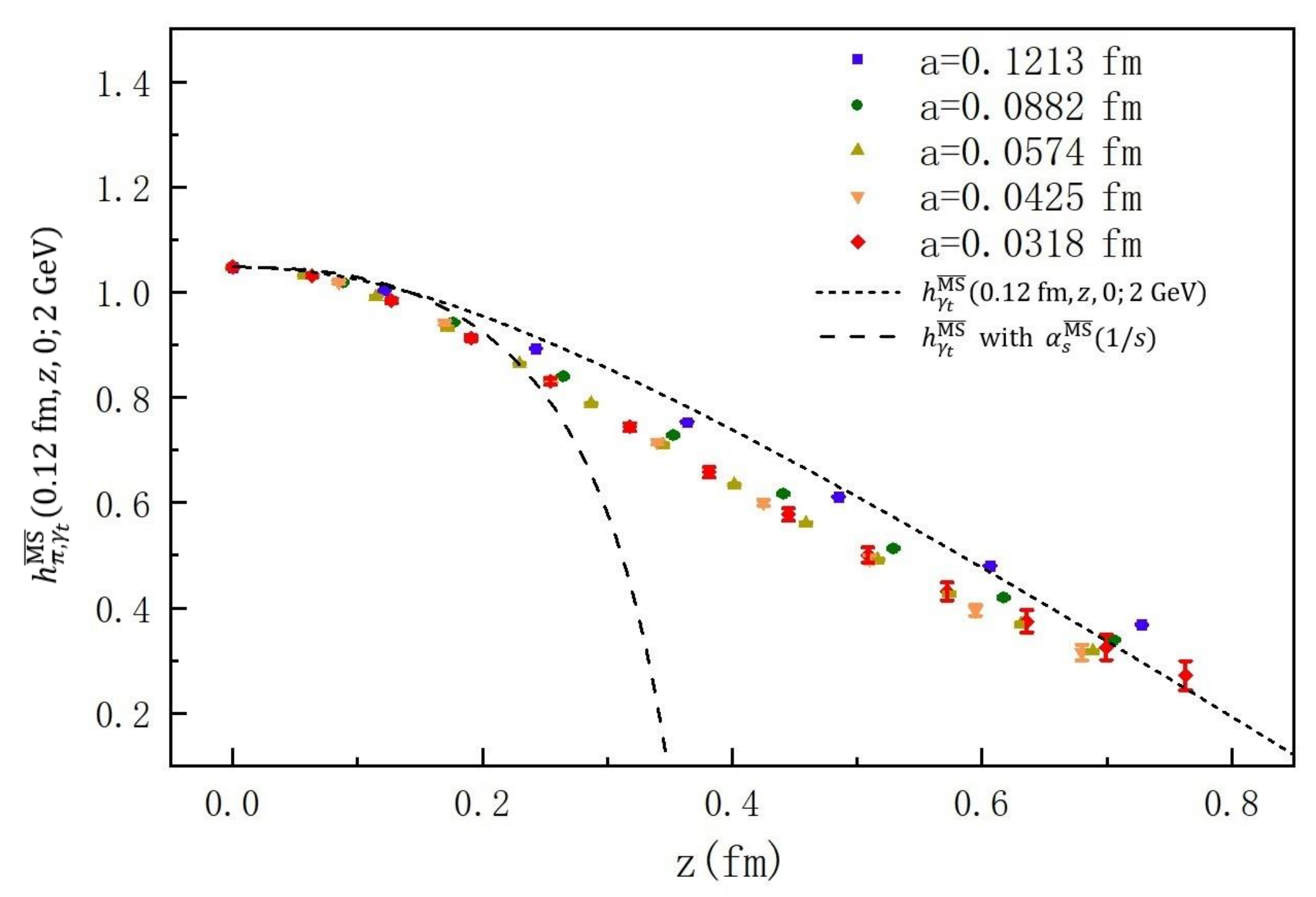}
\includegraphics[width=0.45\textwidth]{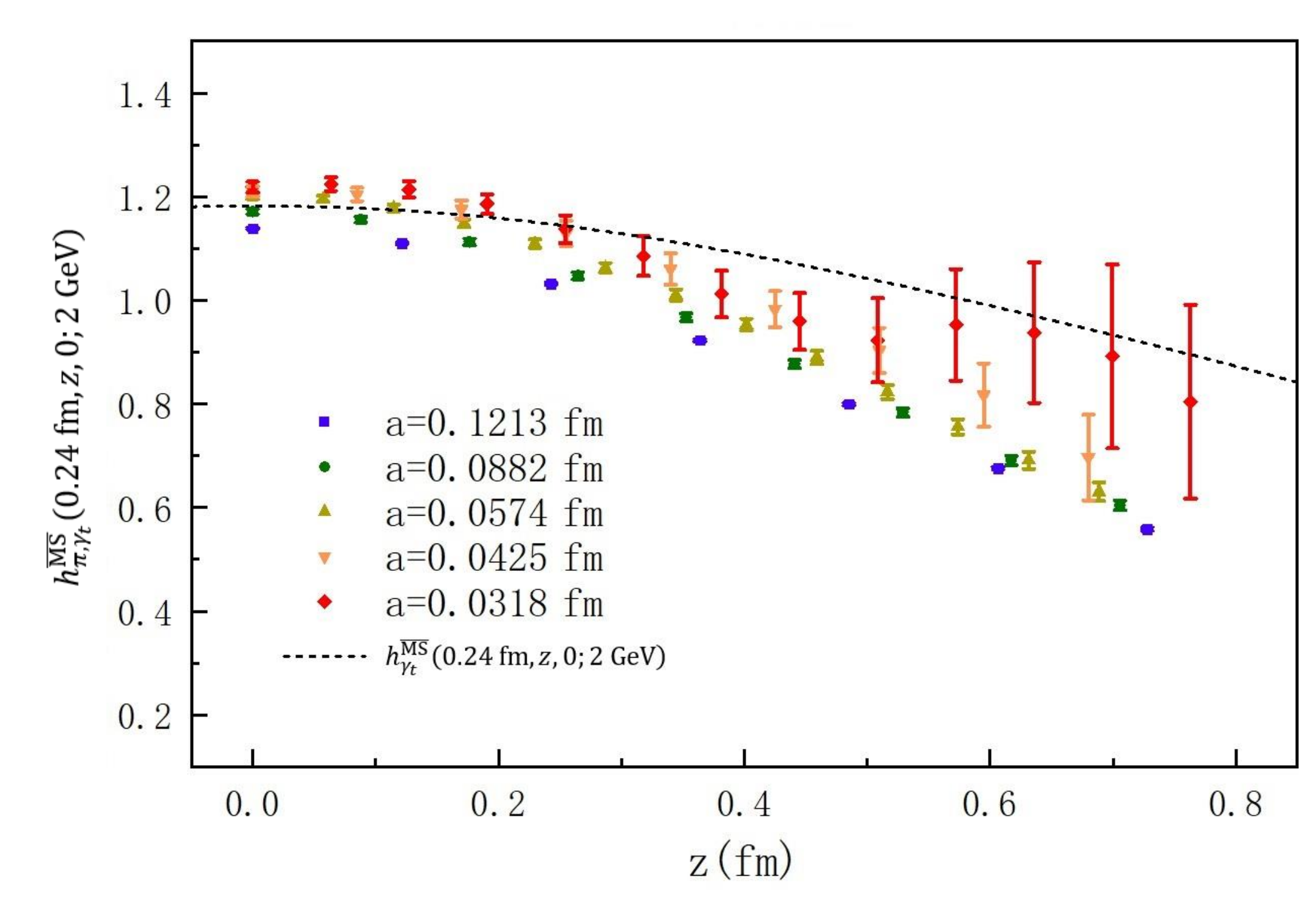}
\includegraphics[width=0.45\textwidth]{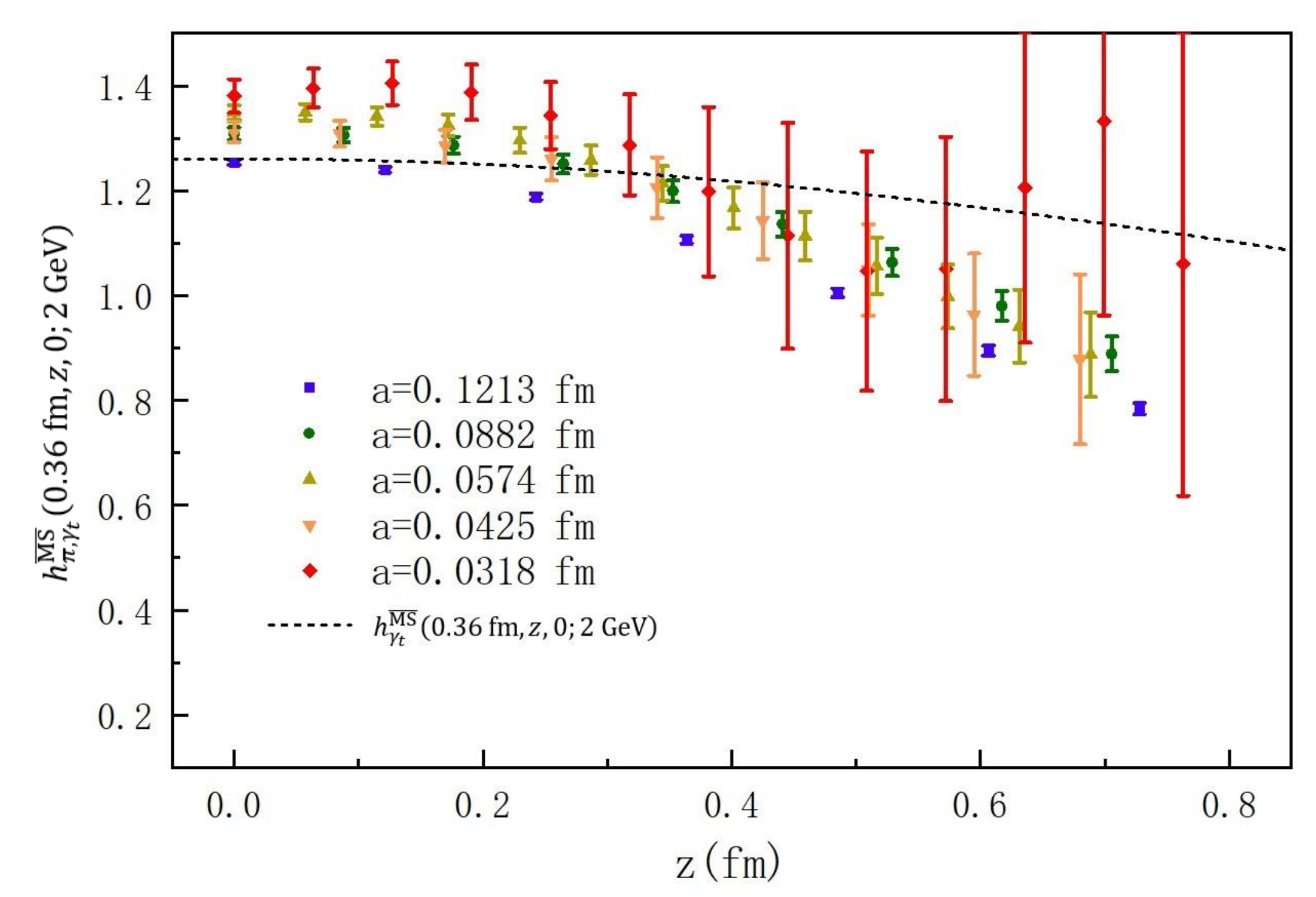}
\includegraphics[width=0.45\textwidth]{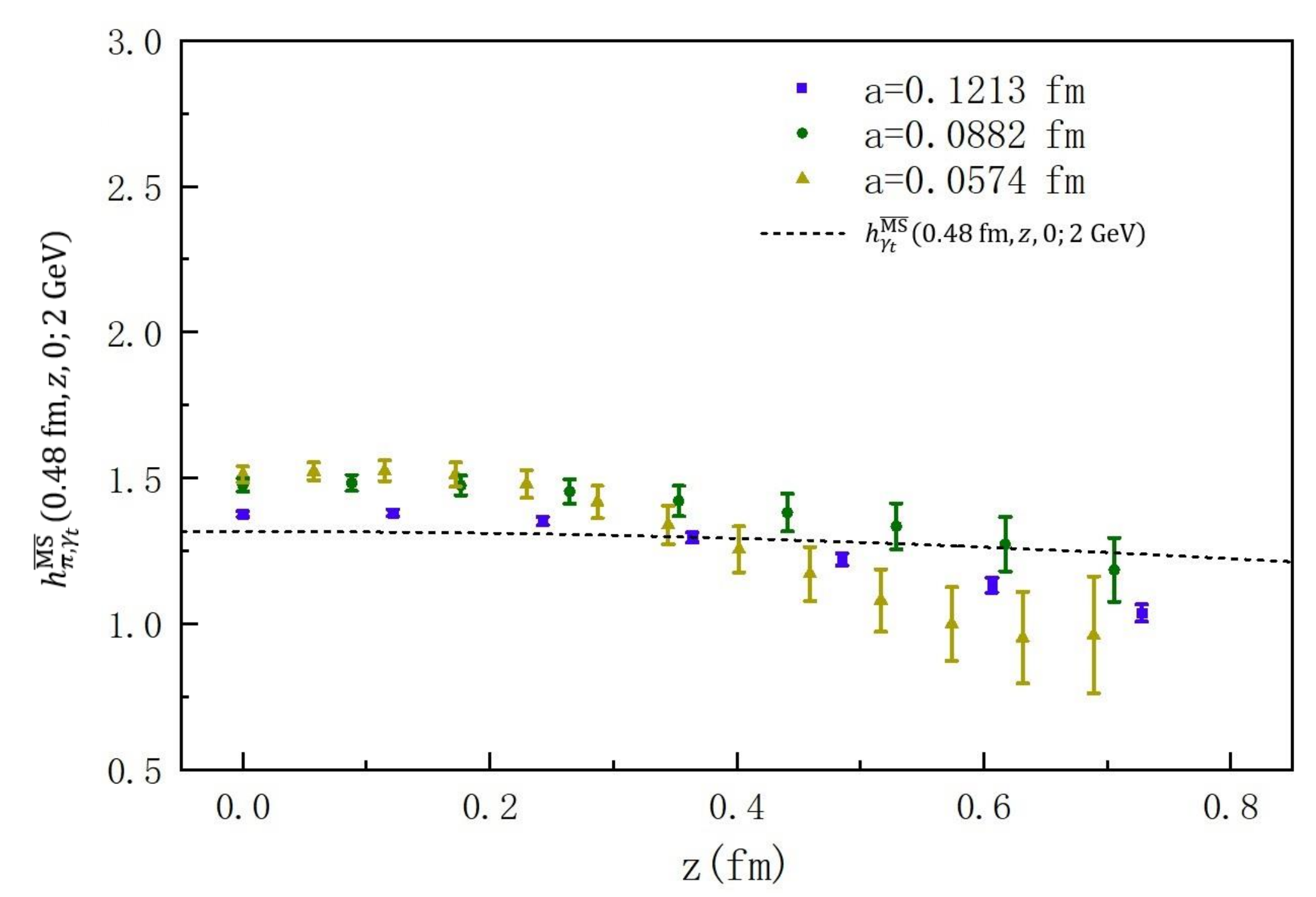}
\includegraphics[width=0.45\textwidth]{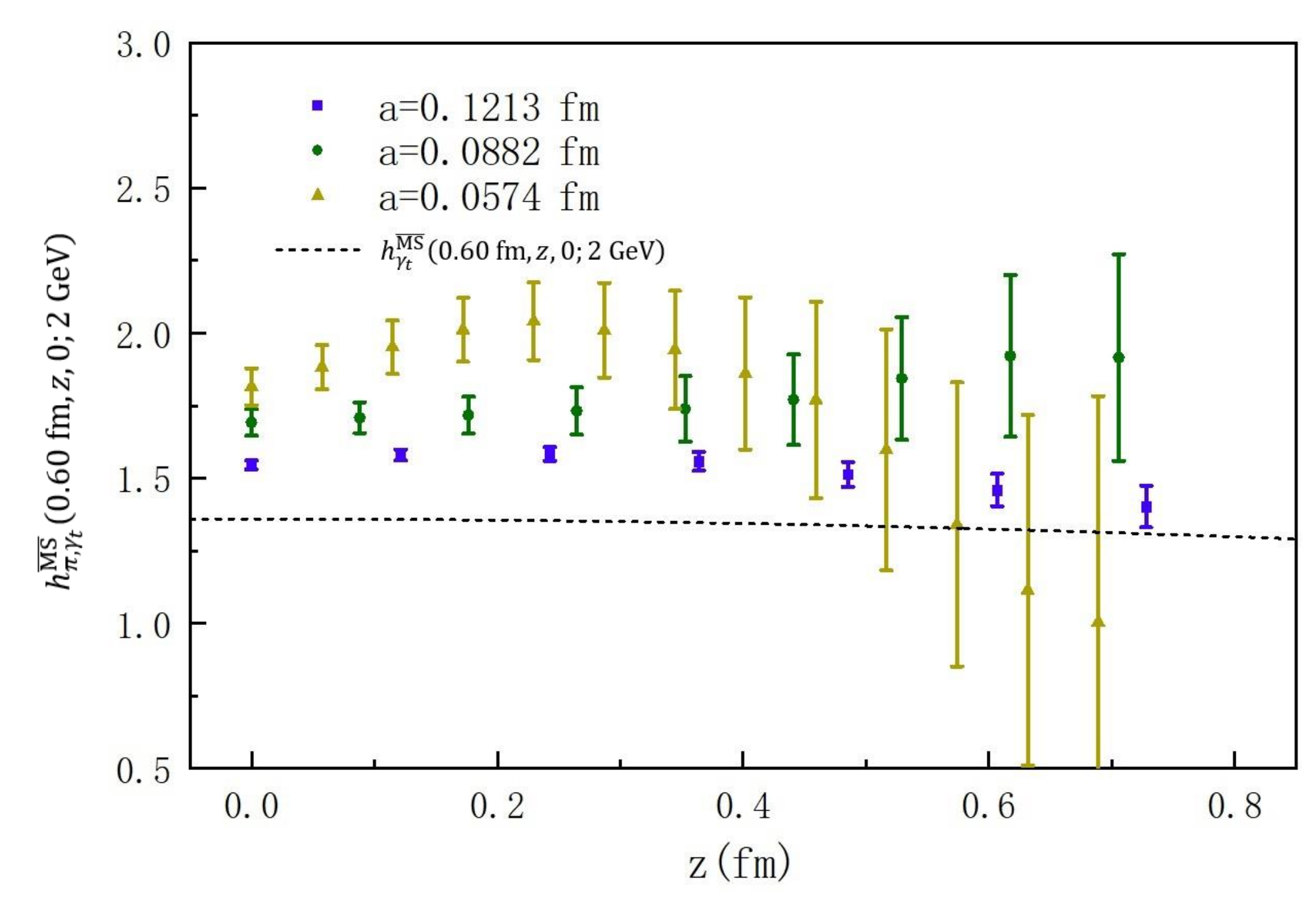}
\includegraphics[width=0.45\textwidth]{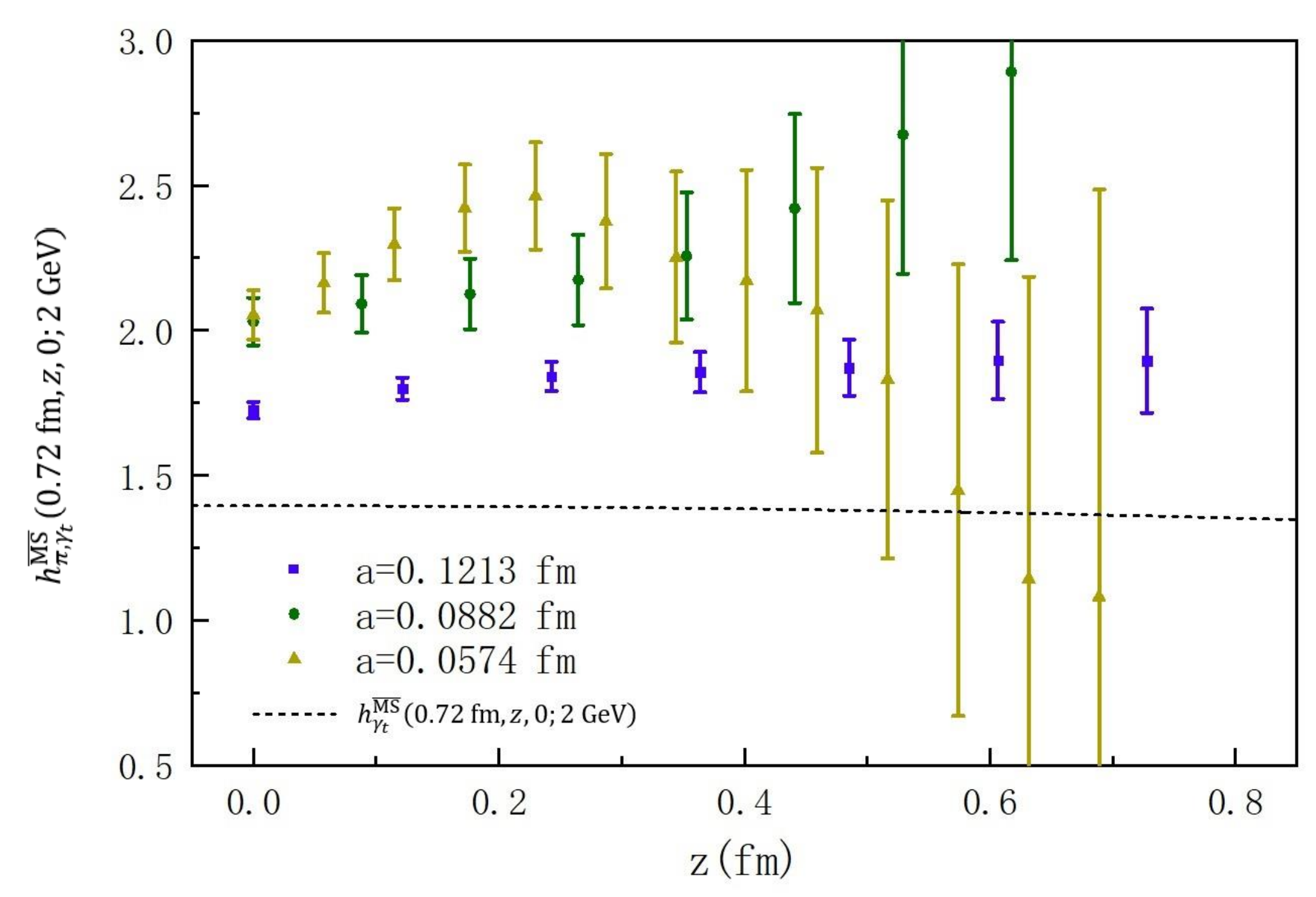}
\caption{The renormalized pion matrix elements with clover valence quark in $\overline{\rm MS}$ scheme $h^{\overline{\rm MS}}_{\pi,\gamma_t}(b,z,0;2\text{ GeV})$. The dense dashed line is the 1-loop result with $\alpha_s(2\text{ GeV})$ in $\overline{\rm MS}$ scheme. We also show a sparse dashed line for the perturbative result with $\alpha_s(1/s)$, $s=\sqrt{b^2+z^2}$ for comparison. Note that $b$ has been interpolated to the same value but $z$ are kept original. The statistical uncertainty comes from bootstrap re-sampling.}\label{fig:Cl_renorm_z}
\end{center}
\end{figure}

The renormalized pion matrix elements $h^{\overline{\rm MS}}_{\pi,\gamma_t}(b,z,0;2\text{ GeV})$ at different $b$ and $z$ are shown in Fig.~\ref{fig:Cl_renorm_z}, and there is no obvious residual linear divergence in all the cases. 

The signal of $h^{\overline{\rm MS}}_{\pi,\gamma_t}$ at small lattice spacing but large $b$ is very bad, and then we only show the results on the MILC12, MILC09, and MILC06 ensembles for $b\geq0.48$~fm. As shown in Fig.~\ref{fig:Cl_renorm_z}, the perturbative results cannot describe the lattice results for large $b$ even for small $z$, as a large $b$ (and then a small intrinsic scale) can invalidate perturbative theory of QCD. 

Similarly, we can use SDR to handle the renormalization of the TMD wave function (WF) matrix element,
\begin{align}
\Psi_{\pi,\gamma_5\gamma_t}(b,z,P_z;1/a)&=\lim_{L\to\infty}\frac{\langle O_{\gamma_5\gamma_t}(b,z,L)|\pi(P_z)\rangle}{\langle O_{\gamma_5\gamma_t}(0,0,0)|\pi(P_z)\rangle\sqrt{Z_E(b,2L+z;1/a)}},\nonumber\\
\Psi^{{\rm SDR}}_{\pi,\gamma_5\gamma_t}(b,z,P_z;\frac{1}{b_0})&=\frac{\Psi_{\pi,\gamma_5\gamma_t}(b,z,P_z;1/a)}{\Psi_{\pi,\gamma_5\gamma_t}(b_0,z_0=0,0,1/a)}.
\end{align}
Since the perturbative calculation shows that 
\begin{align}\label{eq:psiMSbar} 
&\Psi^{\overline{\rm MS}}_{\gamma_5\gamma_t}(b_0,z_0,0;\mu)=h^{\overline{\rm MS}}_{\chi,\gamma_t}(b_0,z_0,0;\mu)\nonumber\\
&=1+\frac{\alpha_s C_F}{2\pi}\bigg\{\frac{1}{2}+3\gamma_E-3\text{ln}2+\frac{3}{2}\text{ln}[\mu^2(b_0^2+z_0^2)]-2\frac{z_0}{b_0}\text{arctan}\frac{z_0}{b_0}\bigg\}+{\cal O}(\alpha_s^2),
\end{align}
the TMD WF under the $\overline{\rm MS}$ scheme can be obtained via
\begin{align}\label{eq:quasiTMDwf}
\!\! \Psi^{\overline{\rm MS}}_{\pi,\gamma_5\gamma_t}(b,z,P_z;\mu)=\psi^{\overline{\rm MS}}_{\gamma_5\gamma_t}(b_0,0,0;\mu)\Psi^{{\rm SDR}}_{\pi,\gamma_t}(b,z,P_z;\frac{1}{b_0}).
\end{align}

\begin{figure}[!th]
\begin{center}
\includegraphics[width=0.45\textwidth]{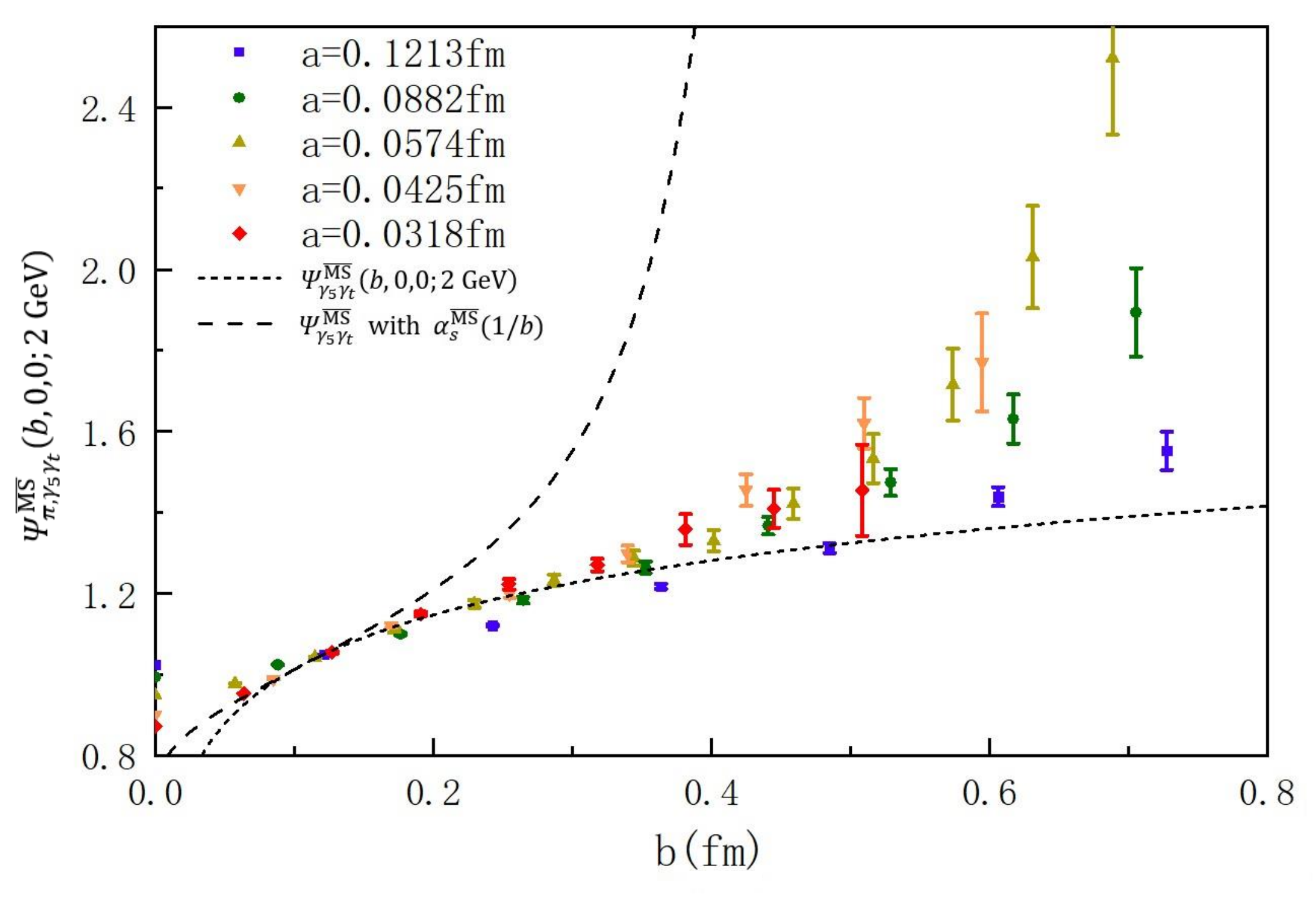}
\includegraphics[width=0.45\textwidth]{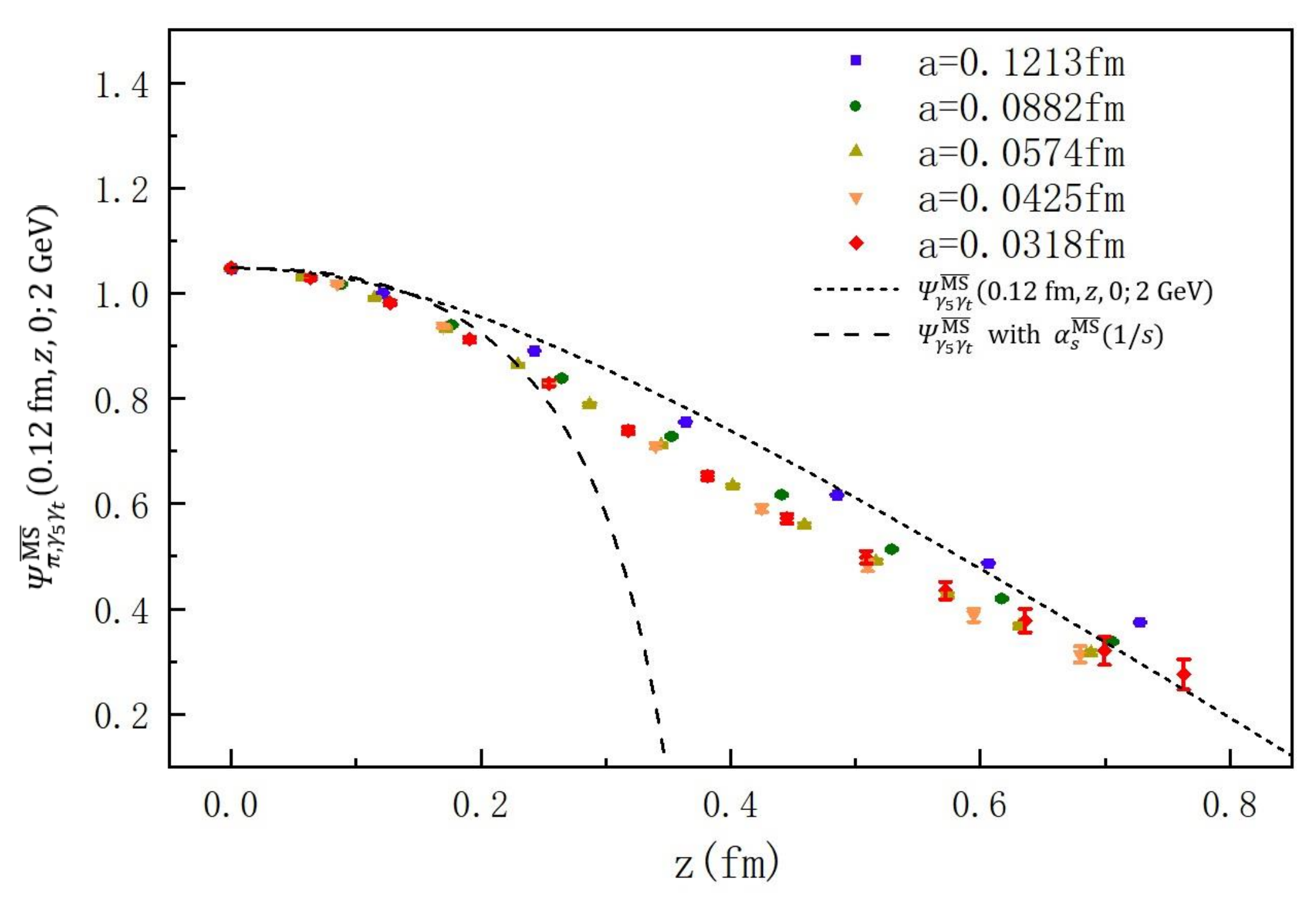}
\includegraphics[width=0.45\textwidth]{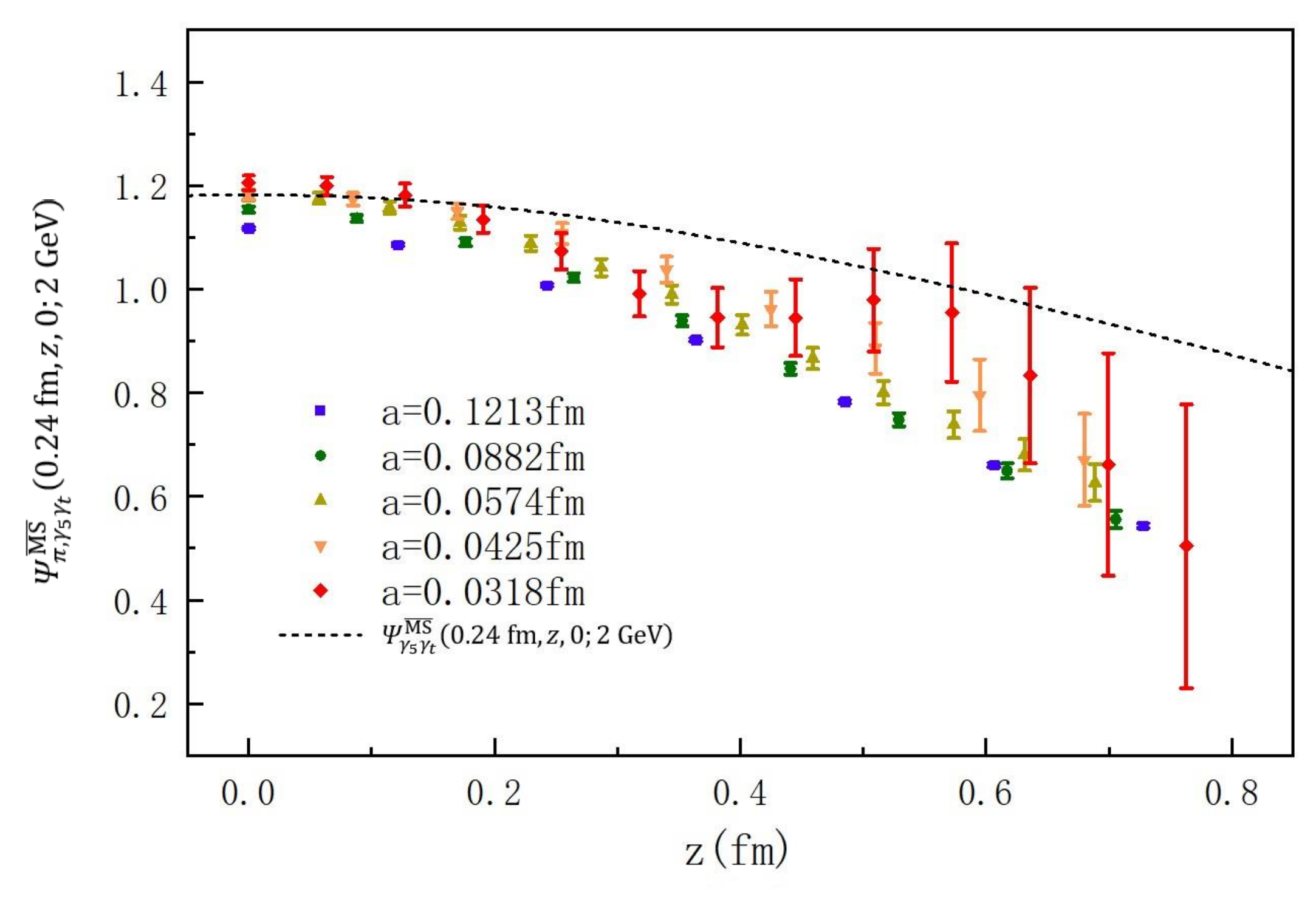}
\includegraphics[width=0.45\textwidth]{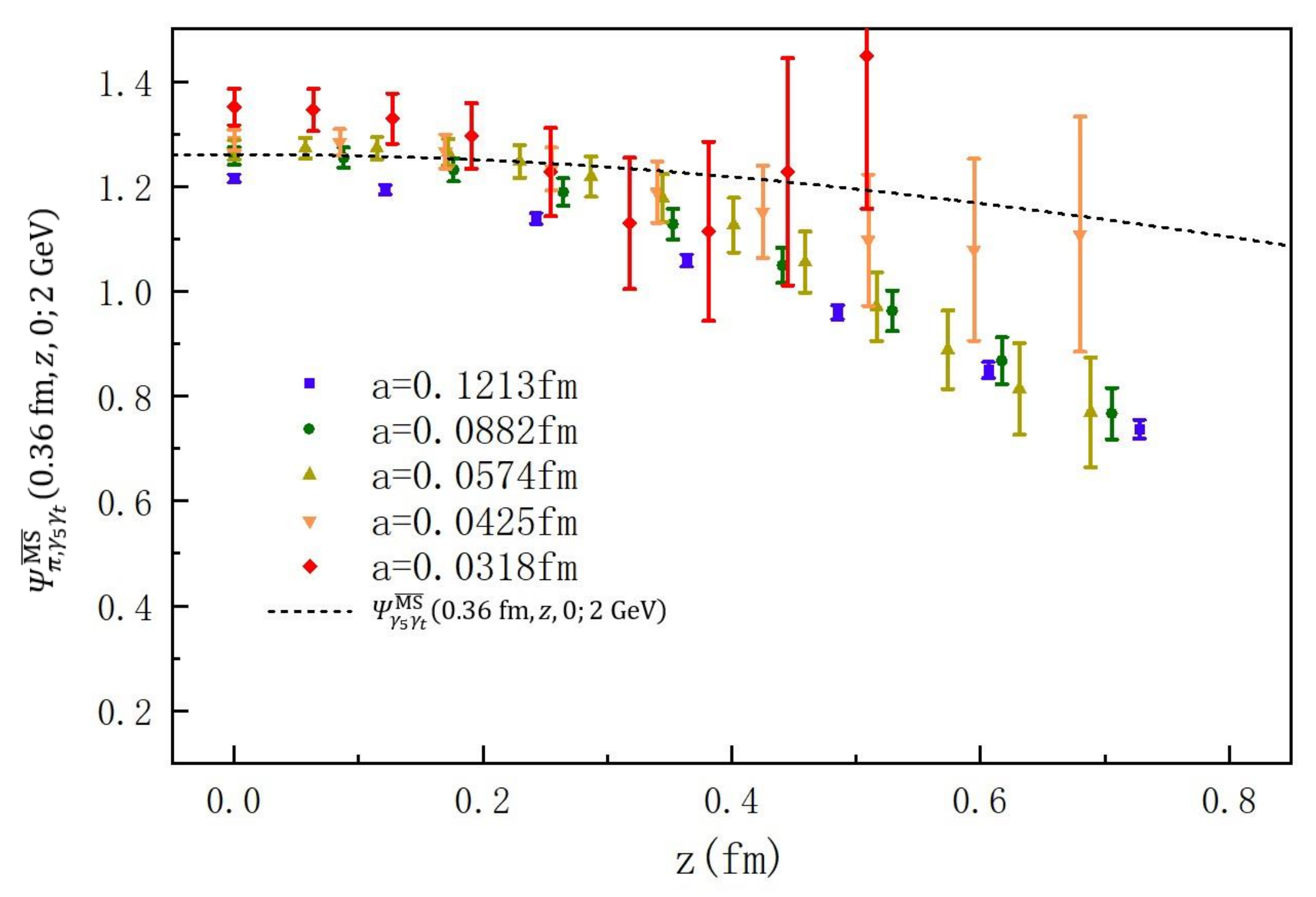}
\caption{The renormalized pion TMDWF matrix elements with clover valence quark in $\overline{\rm MS}$ scheme, $\Psi^{\overline{\rm MS}}_{\pi,\gamma_5\gamma_t}(b,0,0;2\text{ GeV})$ (left top), $\Psi^{\overline{\rm MS}}_{\pi,\gamma_5\gamma_t}(0.12~\mathrm{fm},z,0;2\text{ GeV})$ (right top), $\Psi^{\overline{\rm MS}}_{\pi,\gamma_5\gamma_t}(0.24~\mathrm{fm},z,0;2\text{ GeV})$ (left bottom) and  $\Psi^{\overline{\rm MS}}_{\pi,\gamma_5\gamma_t}(0.36~\mathrm{fm},z,0;2\text{ GeV})$ (right bottom). The dense dashed line is the 1-loop result with $\alpha_s(2\text{ GeV})$ in $\overline{\rm MS}$ scheme. We also show a sparse dashed line for the perturbative result with $\alpha_s(1/b)$ for the $z=0$ case for comparison.}\label{fig:psi_renorm_b}
\end{center}
\end{figure}

The renormalized $\Psi^{\overline{\rm MS}}_{\pi,\gamma_5\gamma_t}(b,0,0;2\text{ GeV})$ through the SDR scheme are plotted in Fig.~\ref{fig:psi_renorm_b}, and compared with the perturbative results. The situation is somehow similar to the TMD PDF matrix element case and the SDR scheme works well for the TMDWF.

\end{widetext}

\end{document}